\documentclass[namedreferences]{SolarPhysics}
\usepackage[optionalrh]{spr-sola-addons} 
\usepackage{graphicx}                    
\usepackage{courier}                    
\usepackage{amssymb}                     

\newcommand{\etal}{{\it et al.}}

\usepackage{color}                       
\usepackage{url}                         

\begin{document}
\begin{article}
\begin{opening}

\title{Solar Magnetic Carpet I: Simulation of Synthetic Magnetograms}
\author{K. A. \surname{Meyer}$^{1}$\sep D. H. \surname{Mackay}$^{1}$\sep A. A. \surname{van Ballegooijen}$^{2}$}
\author{C. E. \surname{Parnell}$^{1}$}

\runningauthor{Meyer \etal}
\runningtitle{Simulation of Synthetic Magnetograms}

\institute{ $^{1}$ School of Mathematics and Statistics, University of 
                  St Andrews, North Haugh, St Andrews, Fife, KY16 9SS, Scotland, U.K. \\
                  \url{karen@mcs.st-and.ac.uk}\\
                  $^{2}$ Harvard-Smithsonian Center for Astrophysics, 60 Garden Street, Cambridge, MA 02138}

\begin{abstract}\\
This paper describes a new 2D model for the photospheric evolution of the magnetic carpet. It is the first in a series of papers working towards constructing a realistic 3D non-potential model for the interaction of small-scale solar magnetic fields. In the model, the basic evolution of the magnetic elements is governed by a supergranular flow profile. In addition, magnetic elements may evolve through the processes of emergence, cancellation, coalescence and fragmentation. Model parameters for the emergence of bipoles are based upon the results of observational studies.
Using this model, several simulations are considered, where the range of flux with which bipoles may emerge is varied. In all cases the model quickly reaches a steady state where the rates of emergence and cancellation balance. Analysis of the resulting magnetic field shows that we reproduce observed quantities such as the flux distribution, mean field, cancellation rates, photospheric recycle time and a magnetic network. As expected, the simulation matches observations more closely when a larger, and consequently more realistic, range of emerging flux values is allowed ($4\times 10^{16}- 10^{19}$ Mx). The model best reproduces the current observed properties of the magnetic carpet when we take the minimum absolute flux for emerging bipoles to be $4\times 10^{16}$ Mx. In future, this 2D model will be used as an evolving photospheric boundary condition for 3D non-potential modelling.
\end{abstract}
\end{opening}

\section{Introduction}\label{intro} 

The solar magnetic carpet is a salt and pepper mix of small-scale opposite polarity magnetic features. These features are continually evolving in response to the photospheric flow pattern of convective cells known as supergranules. These cells range in diameter from 10\,000 km to 50\,000 km, with an average of between 13\,000 and 18\,000 km \cite{hagenaar1997}. Their flow pattern is described by an upflow at cell centre, followed by radial outflow at roughly 0.5 km s$^{-1}$ \cite{simon1964,simon2001,paniveni2004}, and downflow on the order of 0.1 km s$^{-1}$ at the cell edges \cite{wang1989}.

 Due to the constant motion of the underlying granules and supergranules, magnetic features frequently interact with one another. New magnetic flux is continually emerging from below the surface in the form of opposite polarity pairs \cite{harvey1973}. In addition, other features are formed through the coalescence of same-polarity features; large features may break apart into smaller features; and features may shrink or disappear through the cancellation of opposite polarity flux. Over the last 10 years our understanding of the solar magnetic carpet has greatly increased due to the space missions of SOHO/MDI and \textit{Hinode}/SOT. Through these missions, many observational studies of the properties of the magnetic carpet have been carried out \cite{schrijver1997,hagenaar2001,parnell2002,hagenaar2003,deforest2007,mcintosh2007,dewijn2008,hagenaar2008,lamb2008,lites2009,parnell2009,lamb2010,thornton2011}.

Three main types of small-scale magnetic features have been classified; these are ephemeral regions, network features and internetwork features. Ephemeral regions are newly emerging bipolar pairs that appear within the supergranular cells. They have an average lifetime of 4.4 h and typical diameters of 3\,000-5\,000 km \cite{harvey1973}. For the first 30 mins after appearing, the two opposite polarities are found to separate from one another at a velocity of around 4.5-5 km s$^{-1}$ \cite{harvey1993,title2000}. Later, they slow to a velocity that is on the order of the underlying supergranular flow, $\approx$ 0.4 km s$^{-1}$ \cite{hagenaar2003}. \inlinecite{schrijver1997} give the average flux of an ephemeral region to be 1.3$\times$10$^{19}$ Mx (Maxwell), while \inlinecite{chae2001} find a slightly higher value of 2.8$\times$10$^{19}$ Mx.

Network features are larger features, with diameters in the range 1\,000-10\,000 km and fluxes of 10$^{18}$-10$^{19}$ Mx, that are typically found at sites of strong downflow, \textit{i.e.} the edges of the supergranular cells \cite{martin1988}. \inlinecite{zirin1985} find them to be slow moving, with an average velocity of just 0.06 km s$^{-1}$. They do not emerge as network concentrations; rather they are produced from the residuals of other magnetic flux features. Around 90\% of their flux originates from ephemeral regions, with the remaining 10\% arising from internetwork features \cite{martin1990}.

The final type of small-scale flux features are internetwork features; they also emerge as bipolar pairs within supergranular cells. Their mean diameter is around 2\,000 km and their fluxes lie in the range 10$^{16}$-2$\times$10$^{18}$ Mx \cite{wang1995}. De Wijn \etal ~(\citeyear{dewijn2008}) find a rms velocity of 1.57$\pm$0.08 km s$^{-1}$, and an average lifetime of just 10 mins for internetwork features. \inlinecite{zhou2010} deduce an even smaller average lifespan of just 2.9$\pm$2.0 mins.

Previous theoretical models for the evolution of the magnetic carpet photosphere include \inlinecite{schrijver1997}, \inlinecite{vanballegooijen1998}, \inlinecite{simon2001}, \inlinecite{parnell2001} and \inlinecite{cranmer2010}. \inlinecite{schrijver1997} describe flux emergence, cancellation, coalescence and fragmentation through equations of magneto-chemistry. Other authors represent magnetic flux features as collections of corks that are passively advected by photospheric flows \cite{vanballegooijen1998,simon2001}, or as a continuous distribution \cite{parnell2001,cranmer2010}.

This paper describes a new model that aims to reproduce the main observational properties of this small-scale photospheric field. Many of the parameters built into it are taken from studies of observational data (such as \textit{Hinode}/SOT or SOHO/MDI magnetograms). The model also includes a supergranular flow pattern \cite{simon1964}, as well as random small-scale motions to represent the effect of granulation.

The model reproduces observed quantities, such as the power law flux distribution obtained by \inlinecite{parnell2009}. Other objectives include reproducing the rapid photospheric recycle time on the order of 1-2 h deduced by \inlinecite{hagenaar2008}; equality of the rates of emergence and cancellation of flux; and visually, the appearance of a magnetic network along the boundaries of the supergranule cells.

Our long term goal is to use the simulated magnetograms produced by this model as lower boundary data in our 3D non-linear force-free coronal field model \cite{mackay2011}. This model applies a magnetofrictional method to produce a non-potential coronal magnetic field in response to photospheric motions \cite{yeates2008}. The method will be described in Paper 2 \cite{meyer2011}, in which we simulate the non-linear force-free evolution of the coronal field for a series of simple small-scale interactions.

The paper is outlined as follows: Section~\ref{model} describes the technical aspects of the construction of the 2D magnetic carpet model and how each of the four flux evolution processes are specified. The results of the simulations are presented in Section~\ref{results} where we show that the model reproduces many observational parameters. Section~\ref{conclusion} gives the discussion, conclusions and an outline of future work we are undertaking.

\section{Model}\label{model}

The construction of the magnetic carpet model is now described. Whilst observational studies use the term `feature' to describe small-scale magnetic flux concentrations, we will refer instead to magnetic `elements'. This is to distinguish between a `magnetic element' in the model, which we choose to have a specific mathematical form. In contrast, an observational magnetic feature as would appear in a magnetogram may be composed of several of our magnetic elements. Each magnetic element within the model is treated individually as a unique discrete element. The sum of all elements then produces a synthetic magnetogram. This approach differs from other methods in several ways. In cork models, each cork represents a single intense flux tube, whereas in our model we represent our magnetic elements with Gaussian peaks, allowing features to form that are composed of many peaks and troughs in magnetic field strength. To avoid undesirable numerical effects such as numerical diffusion and pile-up at cancellation points, we move the centres of magnetic elements rather than advect their Gaussian profiles. Such a treatment also allows us to easily keep track of the number of elements and exactly which elements are involved in each of the four processes of emergence, cancellation, coalescence and fragmentation at any time. We first describe the mathematical form of the magnetic elements which produce the synthetic magnetograms. Following this, examples of the magnetograms produced over a 250 h period are shown. Finally, we discuss the rules that govern how the magnetic elements evolve.

\subsection{Synthetic Magnetograms}\label{subsec:magn}

For each discrete magnetic element we assume that the $z$-component of the element's magnetic field has a Gaussian profile,
\begin{equation}
 B_{z}=B_{0}e^{-r^2/r_{0}^2},
\end{equation}
where $B_{0}$ is the peak magnetic field strength, $r_{0}$ is the Gaussian half-width and $r$ is the distance from the centre of the Gaussian. The total flux, $\Phi$, of each element is:
\begin{equation}\label{eqn:b01}
 \Phi = \int_A B_{0} e^{-r^2/r_{0}^2} r dr d\theta = B_{0} \pi r_{0}^2.
\end{equation}

\begin{figure}
 \begin{center}
  \includegraphics[width=0.5\textwidth]{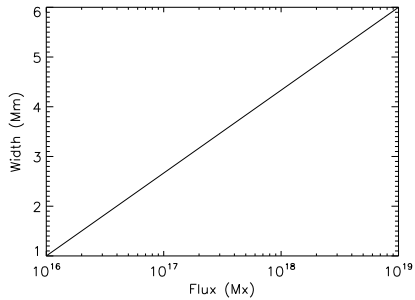}
 \end{center}
\caption{Diameter of a magnetic element versus absolute flux.}\label{fig:width}
\end{figure}

We specify the Gaussian half-width of each magnetic element be $r_{0}=d(\phi)/4$, where $d(\phi)$ is the diameter of the magnetic element and $\phi$ is its absolute flux, $\phi=|\Phi|$.
The diameter is given by
\begin{equation}\label{eqn:d}
 d(\phi)=m_{\phi}\log_{10}(\phi) + c_{\phi},
\end{equation}
where the parameters $m_{\phi}$ and $c_{\phi}$ are given by
\begin{displaymath}
 m_{\phi}=\frac{d_{\textrm{\small max}}-d_{\textrm{\small min}}}{\log_{10}(\phi_{\textrm{\small large}})-\log_{10}(\phi_{\textrm{\small small}})}, \quad \textrm{and} 
  \quad c_{\phi}=d_{\textrm{\small min}} - m_{\phi}\log_{10}(\phi_{\textrm{\small small}}).
\end{displaymath}
We let $d_\textrm{{min}}=1$ Mm, $d_{\textrm{\small max}}=6$ Mm, $\phi_{\textrm{\small small}}=10^{16}$ Mx and $\phi_{\textrm{\small large}}=10^{19}$ Mx so that the majority of magnetic elements within the simulation are confined to the range $d\in[1,6]$ Mm. These values agree with observed diameters and fluxes for magnetic carpet features such as small network features and internetwork features \cite{harvey1973,martin1988,wang1995,zhou2010}. A plot of diameter versus flux is shown in Figure~\ref{fig:width}.

Equation~(\ref{eqn:b01}) may be rearranged to give an expression for the peak magnetic field strength of a single magnetic element,
\begin{equation}
 B_{0}=\frac{16\Phi}{\pi d(\phi)^2}.
\end{equation}
The contribution of the $j$th magnetic element to the normal component of the magnetic field is then:
\begin{equation}
 B_{z,j}=B_{0,j} \exp \bigg\{\frac{-16r^2}{d(\phi_j)^2} \bigg\},
\end{equation}
\begin{displaymath}
 r^2=(x-x_j)^2+(y-y_j)^2,
\end{displaymath}
where $(x,y)$ is an arbitrary position, and $(x_j,y_j)$ is the position of the centre of the magnetic flux element.
We sum up the contribution from every magnetic element to give $B_{z}$,
\begin{equation}
 B_{z}=\sum_{j=1}^N B_{z,j}.
\end{equation}

  \begin{figure}

   \centerline{\hspace*{0.015\textwidth}
               \includegraphics[width=0.515\textwidth,clip=]{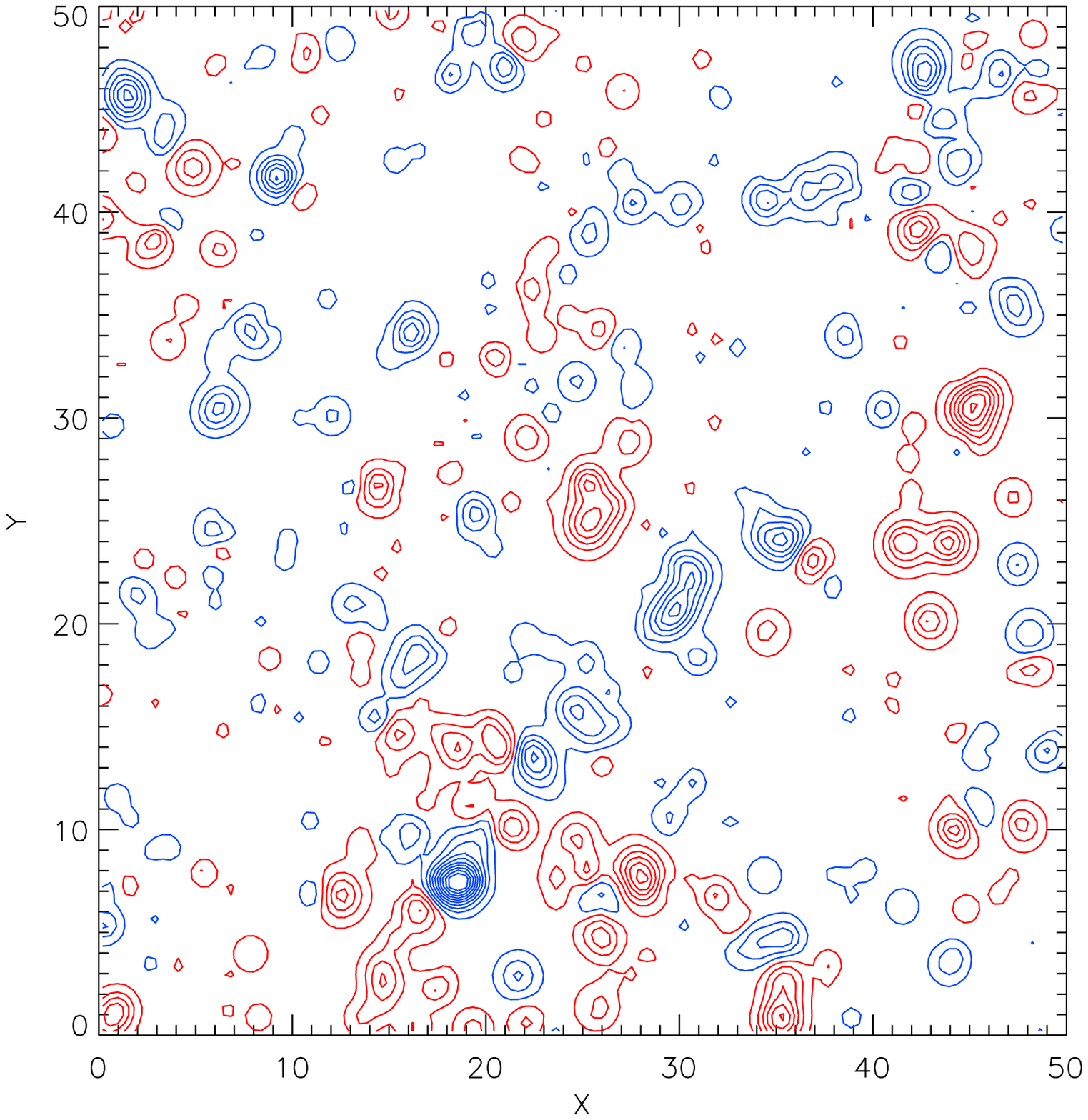}
               \hspace*{-0.03\textwidth}
               \includegraphics[width=0.515\textwidth,clip=]{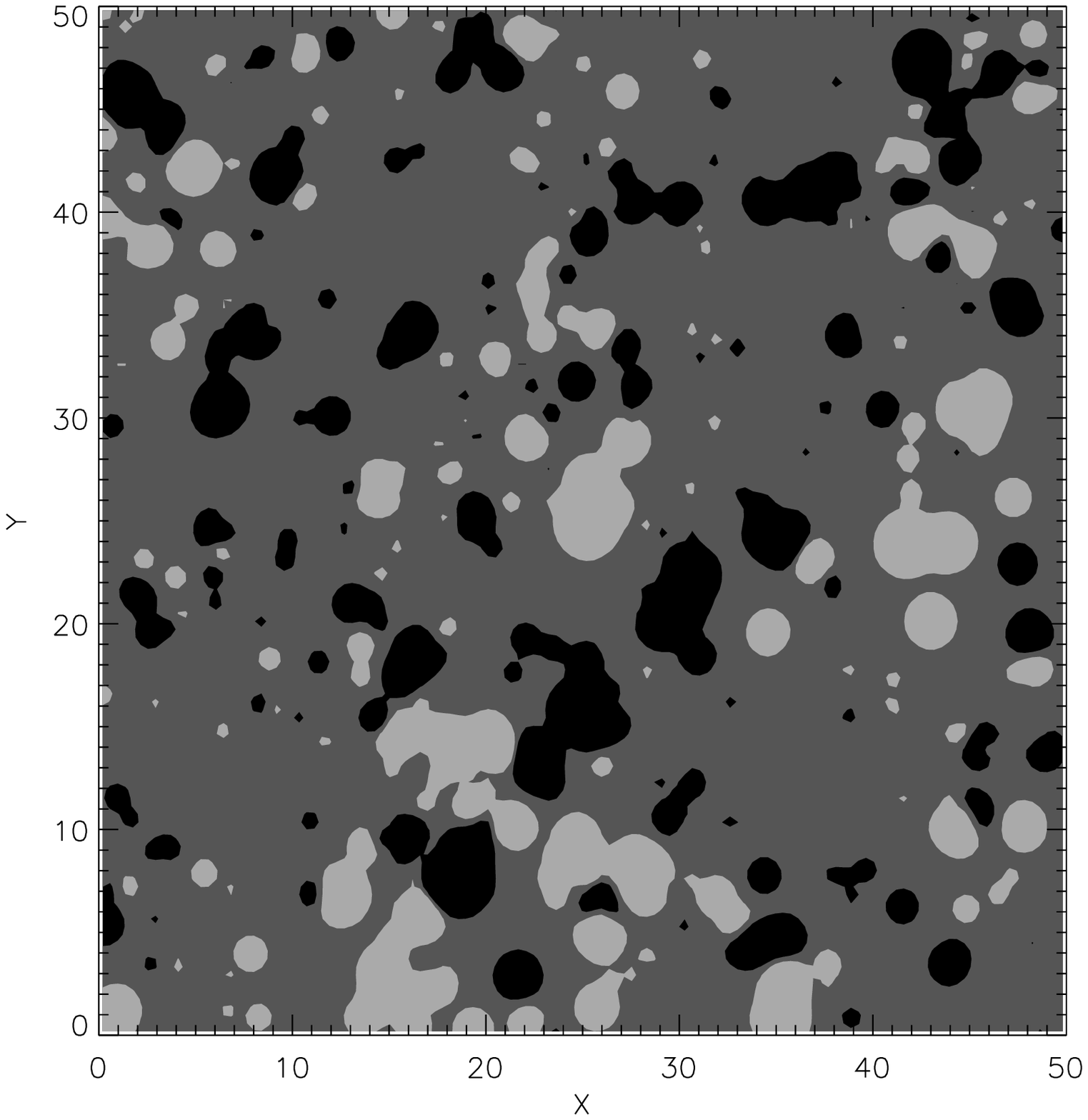}
              }
     \vspace{-0.48\textwidth}   
     \centerline{ \bf     
      \hspace{-0.02 \textwidth}  \color{black}{(a)}
      \hspace{0.44\textwidth}  \color{black}{(b)}
         \hfill}
     \vspace{0.43\textwidth}    

\caption{Synthetic magnetograms for a simulation with a flux emergence range of $8\times 10^{16}$ - $1\times 10^{19} $ Mx. Both are taken at $t=20$ h. (a) Red contours represent positive magnetic field, blue represents negative magnetic field, where 10 contour levels are shown for each polarity with an absolute peak value of 70 G (Gauss). (b) The same as image (a), with contours filled in to simulate a magnetogram.}\label{fig:mag_ex}
   \end{figure}

Figure~\ref{fig:mag_ex} shows contour plots of $B_{z}$ for a simulation in which newly emerging bipoles have a total flux in the range $8\times 10^{16}$ - $1\times 10^{19} $ Mx. This simulation is described in more detail in Section~\ref{results}. A black and white image of the same region is shown to simulate a magnetogram.

Throughout the simulation, the motion of the magnetic elements is determined by the underlying supergranular flow, which is described next. In addition, details of the flux emergence and interaction processes are given in Sections~\ref{subsec:emerge}, \ref{subsec:fragment} and \ref{subsec:cancel}. Readers not wishing to read the technical details of how these processes are implemented may jump to Section~\ref{results}.

\subsection{Steady Flow Profile}\label{subsec:flow}

\begin{figure}
   \centerline{\hspace*{0.015\textwidth}
               \includegraphics[width=0.95\textwidth,clip=]{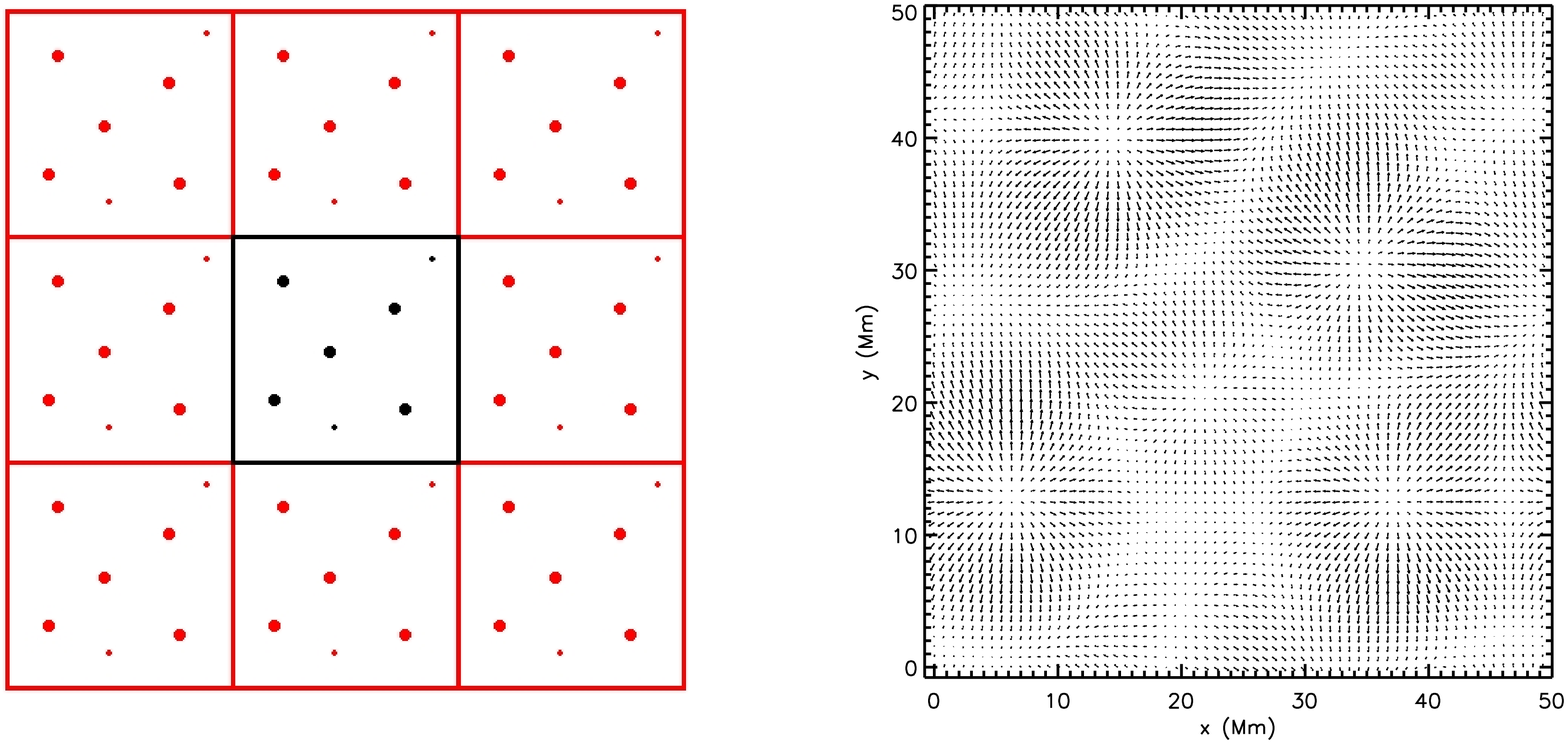}
              }
     \vspace{-0.01\textwidth}   
     \centerline{ \bf     
      \hspace{0.215 \textwidth}  \color{black}{(a)}
      \hspace{0.46\textwidth}  \color{black}{(b)}
         \hfill}
   
\caption{(a) The positions of the supergranule cell centres are translated eight times to surround the originals. This simulates the influence of outside flows on the domain. (b) A simulated supergranular flow profile in which seven cells have been specified in the computational region.}\label{fig:sgflow}
   \end{figure}

We construct a supergranular velocity profile for our model that is similar to that of \inlinecite{simon2001}, except that the diameter of each cell varies from supergranule to supergranule. For simplicity our flow profile is currently steady throughout the simulation, although in reality, supergranules evolve in time. Various authors have estimated lifetimes of supergranules to be anywhere from 10 hours to 2 days, depending on the technique used \cite{rieutord2010}. However, we will show that our steady profile does not lead to an unphysical buildup of magnetic flux at the cell boundaries. The flow profile of a single supergranule is given by
\begin{equation}\label{eqn:sgflow}
 v_{R}=A_{0}R\exp\bigg\{\frac{-R^2}{R_{0}^2}\bigg\},
\end{equation}
where $v_{R}$ is the velocity from cell centre in the $x$-$y$ plane. $R$ is the distance from cell centre, and $A_{0}$ and $R_{0}$ are normalised values that determine the flow strength and radius of the supergranule. We set $R_0=0.173$ for all supergranules, which corresponds to 8.7 Mm in our simulations. $A_0$ is a number between 0 and 1 that determines the strength of each cell. We choose the positions $(x_{\textrm{\small c}},y_{\textrm{\small c}})$ of the centres of $n$ supergranules in the simulation. To introduce the influence of cells outside the computed domain, we translate the positions of these $n$ cells eight times to surround the original pattern, as illustrated by Figure~\ref{fig:sgflow}(a). The original domain is shown in black, and spans the range $[x_{\textrm{\small min}},x_{\textrm{\small max}}] \times [y_{\textrm{\small min}},y_{\textrm{\small max}}]$, where we choose $x_{\textrm{\small min}}=y_{\textrm{\small min}}=0$ Mm and $x_{\textrm{\small max}}=y_{\textrm{\small max}}=50$ Mm. Translation of the cells means that the supergranular flow matches through the side boundaries, which are periodic.

To produce the steady flow pattern, the velocity at any point is found by summing over the contributions of $9n$ supergranular cells. This includes the $n$ original supergranules which describe the central region of Figure~\ref{fig:sgflow}(a), outlined in black, and the eight sets of translations. Once the contribution of all cells has been calculated, the resulting velocity profile is scaled so that the maximum value of $v_{R}$ is 0.5 km s$^{-1}$ \cite{simon1964,paniveni2004,rieutord2010}.

The flow profile chosen for the simulation illustrated in Figure~\ref{fig:mag_ex} is shown in Figure~\ref{fig:sgflow}(b). This flow profile is used for all of the simulations described in this paper. The positions of the cell centres were selected to produce an irregular pattern, as would be seen in real observations of the solar surface. One can also see flows contributing from the supergranules through the boundaries, as a result of the translation previously described.

Contributions from granular motions are also added to each magnetic element's velocity at each time step. The granular velocity contribution is randomly chosen to be between 0 and 0.1 km s$^{-1}$ if an element is within $0.75R_{\textrm{\small sg}}$ of the centre of the supergranular cell, and between 0 and 0.2 km s$^{-1}$ if it is further out than this, where $R_{\textrm{\small sg}}$ is the radius of the supergranule. The direction of this velocity is also chosen as random. A slightly higher value for the granular velocity contribution is chosen if the magnetic element is near the boundary of the supergranule, since the contribution from the supergranular flow profile is small at these locations. This prevents elements from becoming stationary once they reach the network between supergranules.

One time step within the model is equal to $1$ min. This is an arbitrary choice within the model and may be chosen to be higher or lower. We have chosen 1 min as the current cadence of magnetogram data from instruments such as SOHO/MDI and SDO/HMI are of a similar size. Full disc MDI magnetograms are typically of cadence 1 min at best \cite{scherrer1995}, while HMI line-of-sight magnetograms are of cadence 45 s (http://hmi.stanford.edu/hmi-specs.html). Many authors studying the observational properties of the solar magnetic carpet have used magnetogram data sets of cadence roughly 1 min, occasionally averaged over a longer time period to reduce noise, for example \inlinecite{hagenaar2001}, \inlinecite{parnell2002}, \inlinecite{zhou2010} and \inlinecite{thornton2011}. As higher cadences become available for observational data, we may choose a smaller time step in our model. The simulations in Section~\ref{results} are each run for 250 h, which corresponds to 15\,000 time steps.

In the next three sections, we discuss the methods used to implement the processes of emergence, fragmentation, cancellation and coalescence.

\subsection{Emergence}\label{subsec:emerge}

Each emerging bipole is made up of two separate magnetic elements that are of equal flux and opposite polarity. We use the term `bipole' only when referring to a newly emerging pair of magnetic elements, at all other times we consider single magnetic elements.

To simplify computations and prevent infinitesimally small flux elements from arising, we set the minimum unit of flux that a single polarity may have to be equal to $\phi_{0}=10^{16}$ Mx, and define all magnetic elements to have flux that is an integer multiple of $\phi_{0}$ \cite{parnell2001}.

A bipole of absolute flux $\phi_{\textrm{\tiny bp}}$ consists of two magnetic elements of equal absolute flux, $\phi=\phi_{\textrm{\tiny bp}}/2$, but opposite polarity. The total signed flux of the bipole is therefore $0$.

\subsubsection{Parameters for Newly Emerging Bipoles}

\begin{figure}
 \begin{center}
  \includegraphics[width=0.6\textwidth]{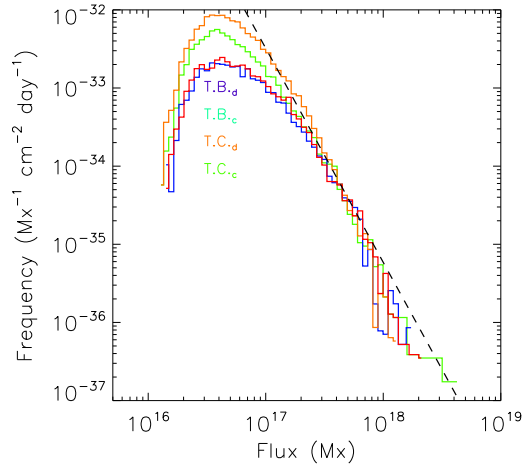}
 \end{center}
\caption{Log-log plot of the frequency of emergence against flux emerged in a 5 h long set of \textit{Hinode}/SOT magnetograms. Results were obtained using a tracked bipolar (TB) and tracked cluster (TC) method applied to clumping (subscript `c') and downhill (subscript `d') data. See Thornton and Parnell (2010) for definitions of tracking methods used. The dashed line is a power-law fit as described by Equation~(\ref{eqn:nall}), with $n_{0}=1.77\times10^{-14}$ cm$^{-2}$ day$^{-1}$ and $\alpha=2.74$.}\label{fig:em_hinode}
\end{figure}

Flux emergence within the simulation is determined by the probability distribution for emerging bipoles from \inlinecite{thornton2011}. This is determined from \textit{Hinode}/SOT high resolution magnetograms and is in the form of a power law:
\begin{equation}\label{eqn:nall}
 N(\phi_{\textrm{\tiny bp}})=\frac{n_{0}}{\phi_{0}}\bigg( \frac{\phi_{\textrm{\tiny bp}}}{\phi_{0}} \bigg)^{-\alpha},
\end{equation}
where $n_{0}=1.77\times10^{-14}$ cm$^{-2}$ day$^{-1}$ and $\alpha=2.74$. The values for $n_0$ and $\alpha$ come from a feature tracking study of \textit{Hinode}/SOT data that is described by \inlinecite{thornton2011}. Figure~\ref{fig:em_hinode} shows a log-log plot of their results for the frequency of emergence versus flux emerged. TB and TC indicate emergence detected by a \emph{tracked bipolar} and a \emph{tracked cluster} method respectively. These methods are explained fully in the original paper. The subscripts `d' and `c' stand for the \emph{downhill} and \emph{clumping} methods of magnetic feature identification, these are described by \inlinecite{deforest2007}.

The quantity $N(\phi_{\textrm{\tiny bp}})d\phi_{\textrm{\tiny bp}}$ is the total number of bipoles that emerge with total absolute flux in the range $[\phi_{\textrm{\tiny bp}},\phi_{\textrm{\tiny bp}}+d\phi_{\textrm{\tiny bp}}]$, where $d\phi_{\textrm{\tiny bp}}$ is very small.
The flux emergence rate for bipoles with flux in the range $[\phi_{a},\phi_{b}]$ in Mx cm$^{-2}$ day$^{-1}$ may be computed as:
\begin{equation}\label{eqn:fluxemerge}
 F_{\textrm{\small emer}}(\phi_a,\phi_b)=\int_{\phi_a}^{\phi_b} N(\phi_{\textrm{\tiny bp}}) \phi_{\textrm{\tiny bp}} d\phi_{\textrm{\tiny bp}}
 = \frac{n_0\phi_0}{2-\alpha}\bigg[ \bigg( \frac{\phi_{\textrm{\tiny bp}}}{\phi_{0}} \bigg)^{2-\alpha} \bigg]_{\phi_a}^{\phi_b}.
\end{equation}
Correspondingly, the number of bipoles emerging in the range $[\phi_{a},\phi_{b}]$ in units of cm$^{-2}$ day$^{-1}$ is then
\begin{equation}\label{eqn:nbipole}
 \int_{\phi_{a}}^{\phi_{b}} N(\phi_{\textrm{\tiny bp}})d\phi_{\textrm{\tiny bp}} = \int_{\phi_{a}}^{\phi_{b}} \frac{n_{0}}{\phi_{0}}\bigg( \frac{\phi_{\textrm{\tiny bp}}}{\phi_{0}} \bigg)^{-\alpha} d\phi_{\textrm{\tiny bp}} = \frac{n_{0}}{1-\alpha} \bigg[ \bigg( \frac{\phi_{\textrm{\tiny bp}}}{\phi_{0}} \bigg)^{1-\alpha} \bigg]_{\phi_{a}}^{\phi_{b}}.
\end{equation}
Let $\phi_{\textrm{\small min}}$ and $\phi_{\textrm{\small max}}$ be the minimum and maximum flux allowed for emerging bipoles in our simulation. They must both be integer multiples of $2\phi_{0}$, as each individual polarity will then have an absolute flux that is an integer multiple of $\phi_{0}$. In addition to this, we must have $\phi_{\textrm{\small min}}\geq 2\phi_0$. For our simulations, we choose to emerge only discrete values of flux from $\phi_{\textrm{\small min}}$ to $\phi_{\textrm{\small max}}$ in steps of size $d\phi=2\phi_{0}$. If $m=(\phi_{\textrm{\small max}}-\phi_{\textrm{\small min}})/2\phi_{0}$ then the set of emerging flux values of bipoles is
\begin{displaymath}
 \{ \phi_{k}=\phi_{\textrm{\small min}}+2k\phi_{0}, k=0,1,...,m \}.
\end{displaymath}
For each discrete value $\phi_{k}$, we integrate over the range $[\phi_{k}-\phi_{0},\phi_{k}+\phi_{0}]$ to approximate the number of bipoles with that absolute flux that will emerge during the simulation:
\begin{equation}
  N^{\prime}_{k} = \int_{\phi_{k}-\phi_{0}}^{\phi_{k}+\phi_{0}}\frac{n_{0}}{\phi_{0}}\bigg( \frac{\phi_{\textrm{\tiny bp}}}{\phi_{0}} \bigg)^{-\alpha} d \phi_{\textrm{\tiny bp}} \times A \times D,
\end{equation}
where $A$ is the area of the domain in cm$^2$ and $D$ is the number of days of the simulation. We cannot allow a non-integer number of bipoles to emerge, so $N_{k}^{\prime}$ must be converted to an integer, $N_k$, by rounding up or down randomly. The $N=\sum_{k=0}^{m}N_{k}$ bipoles are then randomly assigned a value from $0$ to $t_{\textrm{\small max}}=15\,000$ using a uniform distribution, which will be the time step in which they emerge. This means that a random number of bipoles emerge each time step, and also a random total quantity of flux.

In addition to flux, diameter and sign, several other parameters must be chosen for each newly emerging bipole. For each bipole, a random uniform integer between $0$ and $n-1$ is chosen. This is the index of the supergranule cell that it will emerge in. Its location within the cell is then randomly chosen. Since ephemeral regions have been observed to emerge with a preference towards the edge of a supergranule \cite{wang1988}, we also build this into our model. We allow emergence to occur in the range $[0.5R_{\textrm{\small sg}},0.75R_{\textrm{\small sg}}]$, where $R_{\textrm{\small sg}}$ is the radius of the supergranule.
The final parameter that must be chosen for an emerging bipole is its tilt angle, $\theta$. This is the angle of the axis along which the two polarities of the bipole will separate from one another and it is simply a random uniform number between $0$ and $2\pi$ radians. The separation velocities of newly emerging magnetic elements are discussed below.

\subsubsection{Appearance of Newly Emerging Bipoles}

\begin{figure}
 \begin{center}
  \includegraphics[width=0.8\textwidth]{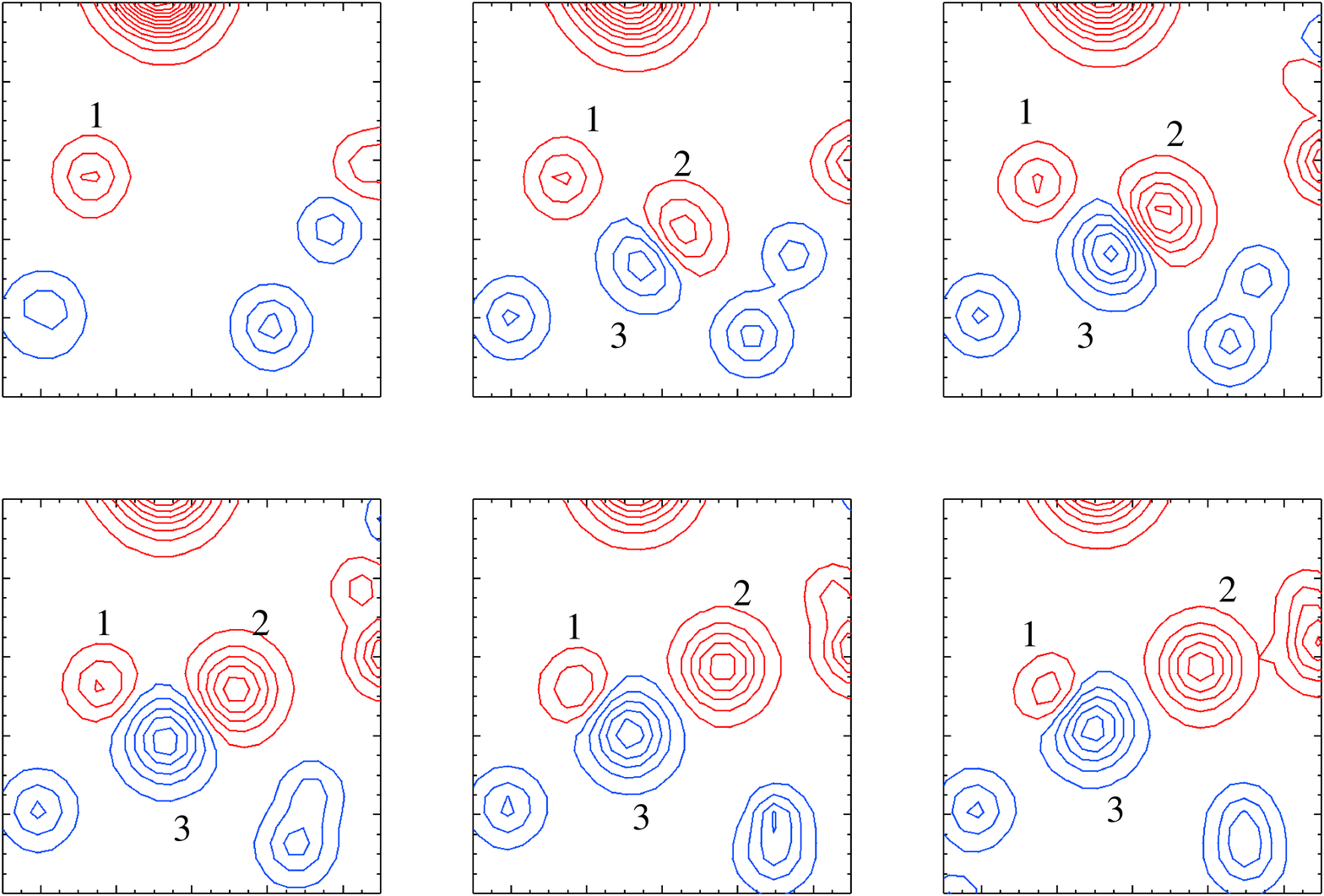}

     \vspace{-0.31\textwidth}   
     \centerline{ \bf     
      \hspace{0.2 \textwidth}  \color{black}{\small{(a)}}
      \hspace{0.211\textwidth}  \color{black}{\small{(b)}}
      \hspace{0.211\textwidth}  \color{black}{\small{(c)}}
         \hfill}
     \vspace{0.253\textwidth}    
     \centerline{ \bf     
      \hspace{0.2 \textwidth}  \color{black}{\small{(d)}}
      \hspace{0.211\textwidth}  \color{black}{\small{(e)}}
      \hspace{0.211\textwidth}  \color{black}{\small{(f)}}
         \hfill}

      \vspace{-0.01\textwidth}
 \end{center}
\caption{A sequence of still images (left to right) showing a newly emerging bipole (elements 2 and 3). Red contours represent positive magnetic field, blue contours represent negative magnetic field. 20 contour levels are shown for each polarity, with an absolute peak value of 70 G. The region is 10 Mm $\times$ 10 Mm in area. The images are taken 4 mins apart, and elements of interest have been numbered from 1 to 3, where elements 2 and 3 illustrate emergence. This event may be seen in the accompanying movie, `mag2\_em.mpg'.}\label{fig:em1}
\end{figure}

At each time step, newly emerging bipoles are added into the simulation. Each magnetic element undergoes the specified emergence process outlined below until it has travelled $e_{\textrm{\small rad}}\times d(\phi)=1.5d(\phi)$ from its initial position. $e_{\textrm{\small rad}}=1.5$ is a constant that defines the separation that a magnetic element must reach from its emergence point before supergranular flows or other processes may take over its evolution. The element's velocity $v_{\textrm{\small sep}}$ depends upon how long it has been emerging for and its direction is given by its tilt angle $\theta$. Observational studies show that the two polarities of a newly emerging bipole initially separate at several km s$^{-1}$, later slowing to velocities that are of a similar order to those of the underlying supergranular flows \cite{harvey1993,title2000,hagenaar2003}. Within our simulations, two elements will initially separate with a velocity of $3$ km s$^{-1}$, slowing to $1.0-1.3$ km s$^{-1}$ after 30 mins and later to $0.5$ km s$^{-1}$ which is on the order of the underlying supergranular flow. Initially, the positive and negative magnetic elements within the bipole move in opposite directions along the axis of their tilt angle.

Figure~\ref{fig:em1} (and the accompanying movie, `mag2\_em.mpg') shows a sequence of images taken from a small region of the synthetic magnetograms produced by one of the simulations. The region is $10\times 10$ Mm$^2$ in area, and the images are taken 4 mins apart. One can see that between images (a) and (b), a new bipole begins to emerge in the centre of the box, the positive and negative polarities are marked `2' and `3' respectively. The two polarities grow in flux as they move apart. Towards the end of the sequence, magnetic element 3 begins to interact with a pre-existing magnetic element, indicated by `1'. Notice also that in addition to the two polarities moving in opposite directions to one another, there is a slight drift of the bipole towards the upper left due to the underlying supergranular flow.

\subsection{Fragmentation}\label{subsec:fragment}

Fragmentation within our model is based upon the process described by \inlinecite{parnell2001}, which depends upon both the flux of the magnetic element and time. Every element is checked for fragmentation at each time step. The fragmentation rate $R_{\textrm{\small f}}$ is an input parameter for the simulation. We assume that every element of sufficient flux will fragment within $T_{\textrm{\small f}}=1/R_{\textrm{\small f}}$ s. \inlinecite{parnell2001} suggests that a fragmentation rate of $R_{\textrm{\small f}}> 1.2\times 10^{-4}$ s$^{-1}$ is required to reproduce the correct absolute flux density and flux distribution, so we take $R_{\textrm{\small f}}=1.5\times 10^{-4}$ s$^{-1}$. This means that a magnetic element of sufficiently large flux will fragment roughly once every $1/R_{\textrm{\small f}}=6667$ s $\approx$ 1 h 50 mins. For an element $j$ to have \emph{sufficient flux} to fragment, it must satisfy the inequality
\begin{equation}\label{eqn:frag1}
  \phi_{j} > \psi \Big( 1-\frac{k_{\textrm{\small f}}}{q} \Big),
\end{equation}
where $q$ is a random number such that $k_{\textrm{\small f}}<q<1$, and $0<k_{\textrm{\small f}}\leq 1$. $k_{\textrm{\small f}}$ is fixed at the start of the simulation, but a random $q$ is chosen for every magnetic element every time we check whether it will fragment. \inlinecite{parnell2001} take a value of $\psi$ that is at the large end of their range of expected flux values, $\psi=8 \times 10^{18}$ Mx, and $k_{\textrm{\small f}}=0.75$. Without observational evidence to suggest differently, we take the same values for our simulations. With these parameters, the largest value that the right hand side of Equation~(\ref{eqn:frag1}) can take is
\begin{displaymath}
 \psi\bigg( 1-\frac{k_{\textrm{\small f}}}{\textrm{max}(q)} \bigg)=8 \times 10^{18}\bigg(1-\frac{0.75}{1.0}\bigg)=2 \times 10^{18} \textrm{ Mx}.
\end{displaymath}

Similarly, the minimum value that the right hand side of Equation~(\ref{eqn:frag1}) can take is $0$. However, elements are not allowed to fragment if they are of unit flux, $\phi_{0}$. This means that all magnetic elements of absolute flux $\geq 2\phi_{0}$ may fragment, but elements of greater flux have a higher probability of fragmentation.

Time dependence is built into the fragmentation process as follows. If $\phi_{j}$ satisfies Equation~(\ref{eqn:frag1}), we then choose a random $s$ such that $0<s<1$. Element $j$ will only fragment during the current time step if
\begin{displaymath}
 s < \frac{T_{j}}{T_{\textrm{\small f}}}.
\end{displaymath}
$T_{j}$ is the `age' of magnetic element $j$ in s. An element's age is reset to $0$ every time it fragments, coalesces or cancels with another element. Clearly once $T_{j}>T_{\textrm{\small f}}$ the right hand side will be greater than 1, and the inequality will be satisfied.
Therefore, all magnetic elements of flux greater than $2 \times 10^{18}$ Mx are guaranteed to fragment within $T=T_{\textrm{\small f}}/60$ time steps, unless some other process takes over their evolution before then.

\subsubsection{New Elements Arising From Fragmentation}

Within our simulations, elements are allowed to split into just two new elements at a time. In reality, most fragmenting magnetic features only split into two \cite{thornton2011a}. Once it has been determined that element $j$ will fragment, a new element $k$ is introduced to represent the element resulting from the fragmentation. The flux and diameter of $j$ and $k$ must now be recomputed. As in \inlinecite{parnell2001}, the original flux $\phi_{j}$ is split into two new fluxes, $\phi_{1}$ and $\phi_{2}$. If $p$ is a random number between 0.55 and 0.95, then
\begin{displaymath}
 \phi_{1}=p \phi_j.
\end{displaymath}
Since all fluxes within the simulation must be integer multiples of $\phi_{0}$, we round $\phi_1/\phi_{0}$ to the nearest integer. We then take
\begin{displaymath}
\phi_{2}=\phi_j - \phi_{1}.
\end{displaymath}
Elements of flux $\phi_{0}$ are not allowed to fragment. If $\phi_{j}=2\phi_{0}$, then $\phi_{1}=\phi_{2}=\phi_{0}$, and if $\phi_{j}=3\phi_{0}$, then $\phi_{1}=2\phi_{0}$ and $\phi_{2}=\phi_{0}$. We may now set $\phi_{j}=\phi_{1}$ and $\phi_{k}=\phi_{2}$, then compute their new diameters $d(\phi_{j})$ and $d(\phi_{k})$.

\subsubsection{Motion of Fragmenting Elements}

\begin{figure}
 \begin{center}
  \includegraphics[width=1.0\textwidth]{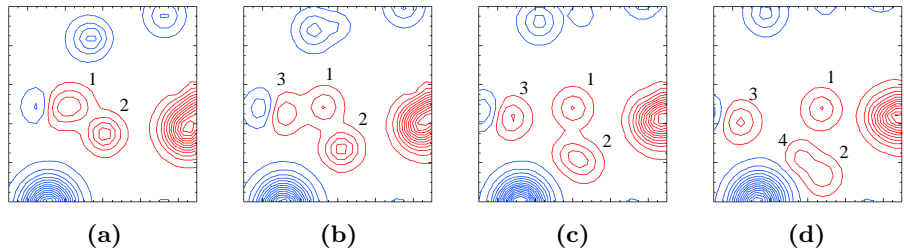}

     \vspace{-0.01\textwidth}  
     \centerline{ \bf     
      \hspace{0.091 \textwidth}  \color{black}{\small{(a)}}
      \hspace{0.191\textwidth}  \color{black}{\small{(b)}}
      \hspace{0.191\textwidth}  \color{black}{\small{(c)}}
      \hspace{0.191\textwidth}  \color{black}{\small{(d)}}
         \hfill}

    \vspace{-0.02\textwidth}

 \end{center}
\caption{A sequence of still images (left to right) that features fragmentation of magnetic elements. Red contours represent positive magnetic field, blue contours represent negative magnetic field. 15 contour levels are shown for each polarity, with an absolute peak value of 70 G. The region is 10 Mm $\times$ 10 Mm in area. The images are taken 2 mins apart, and elements of interest have been numbered from 1 to 4. These interactions may be seen in the accompanying movie, `mag3\_frag.mpg'.}\label{fig:frag1}
\end{figure}

An element's behaviour after it has fragmented depends upon the process that the original element was undergoing before the fragmentation occurred. If it was undergoing emergence, fragmenting from yet another element, or simply being advected by supergranular flows, then its velocity $v_{R}$ returns to being determined by this process\footnote{If the underlying supergranular velocity is very small ($<0.1$ km s$^{-1}$), the new elements may be given a `push' of $0.1-0.2$ km s$^{-1}$ to help them separate.}. In the case of emergence, the two new elements produced by fragmentation are also treated as emerging. Their velocities therefore still decrease with time, though the directions of motion of the two new elements $j$ and $k$ are different. If $\theta$ is the direction of motion of the original element, then
\begin{displaymath}
 \theta_{1}=\theta+0.3 \pi q \quad \textrm{and} \quad \theta_{2}=\theta-0.3 \pi q,
\end{displaymath}
where $0<q\leq 1$ is randomly chosen for each individual fragmentation \cite{parnell2001}. Therefore the two magnetic elements move in the same direction with slightly offset motions.

If the element determined to fragment was undergoing cancellation or coalescence with another element $l$ before it fragmented, the treatment is slightly different. The fluxes and diameters of the two new elements $j$ and $k$ are computed. The new element $j$ then continues to cancel or coalesce with $l$, while $k$ moves off in a different direction at the velocity of the underlying supergranular flow\footnote{Again, the element may be given a `push' if the underlying supergranular velocity is very small.}.
The new direction of motion for $k$ is given by
\begin{displaymath}
\theta_{2}=\theta \pm 0.6 \pi q,
\end{displaymath}
with $0<q\leq 1$ random.

The parameter $f_{\textrm{\small rad}}$ defines the separation that a magnetic element will reach from its point of fragmenting. An element $j$ will continue to move away from its fragmentation site until it has reached a distance of $f_{\textrm{\small rad}}d(\phi_{j})=1.5d(\phi_{j})$.

Figure~\ref{fig:frag1} (and the accompanying movie, `mag3\_frag.mpg') shows examples of fragmentation occurring within one of the simulations. Element 1 splits to form a new element, 3, element 2 then begins to fragment to form element 4. Fragmentation may also be seen in the sequence in Figure~\ref{fig:canc1} of the following section, where element 2 fragments into a large and a small element.

\subsection{Coalescence and Cancellation}\label{subsec:cancel}
Every time step, each element that is not currently undergoing the emergence process is checked to see if it will cancel or coalesce with another element. In the current model, an element $j$ may only cancel or coalesce with one single element at any given time. First we determine how many magnetic elements are within interaction range of $j$ (including those through the periodic boundaries). An element $k$ is defined to be within interaction range if $d_{\textrm{\small sep}}$, the separation distance, satisfies
\begin{displaymath}
 d_{\textrm{\small sep}}(j,k)\leq c_{\textrm{\small rad}}(d(\phi_{j})+d(\phi_{k})),
\end{displaymath}
where
\begin{displaymath}
 d_{\textrm{\small sep}}(j,k)=|(x_j,y_j)-(x_k,y_k)| \quad \textrm{and} \quad c_{\textrm{\small rad}}=0.5.
\end{displaymath}
The constant $c_{\textrm{\small rad}}$ defines the interaction range between two magnetic elements. It is important that $c_{\textrm{\small rad}}<0.75$ to prevent elements that have just separated from one another due to emergence from immediately cancelling with one another again. There are also several other conditions for the cancelling or coalescing of $j$ with some element $k$:
\begin{itemize}
 \item $k$ must not currently be undergoing emergence,

 \item $k$ must not currently be cancelling or coalescing with another element,

 \item If $j$ and $k$ are in the process of fragmenting from one another, then they are not allowed to immediately coalesce. If they are still within range after the fragmentation process has ended, they may then coalesce together (see Section~\ref{subsec:fragment}).
\end{itemize}
If these conditions are all satisfied by more than one element within range of $j$, then $j$ will cancel or coalesce with the closest element.

Once it has been determined that two elements will cancel or coalesce, they move together at a constant velocity of $v_{\textrm{\small c}}=1.0$ km s$^{-1}$ until their centres meet. Their direction of motion is along the axis defined by their centres.

\begin{figure}
 \begin{center}
  \includegraphics[width=1.0\textwidth]{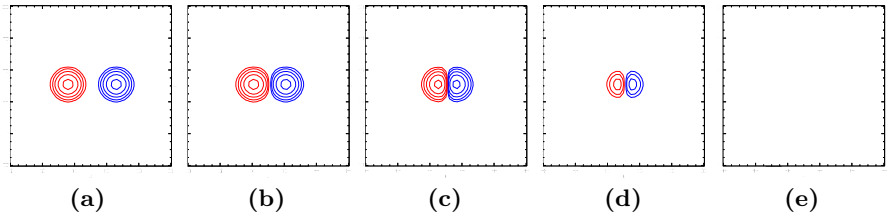}

     \vspace{-0.01\textwidth}  
     \centerline{ \bf     
      \hspace{0.084 \textwidth}  \color{black}{\small{(a)}}
      \hspace{0.131\textwidth}  \color{black}{\small{(b)}}
      \hspace{0.131\textwidth}  \color{black}{\small{(c)}}
      \hspace{0.131\textwidth}  \color{black}{\small{(d)}}
      \hspace{0.131\textwidth}  \color{black}{\small{(e)}}
         \hfill}

    \vspace{-0.02\textwidth}

 \end{center}
\caption{A cancellation between two magnetic elements of equal flux and opposite polarity. The bipole shrinks as the elements move together and their Gaussian profiles overlap. By the last frame the magnetic elements overlap completely and no contours of flux remain.}\label{fig:canc2}
\end{figure}

Since each magnetic element is given a Gaussian profile, their profiles overlap as they move towards one another. In the case of cancellation, this results in both magnetic elements shrinking as they move towards one another. If they are of equal flux and opposite polarity, they will completely cancel one another out. This is illustrated in Figure~\ref{fig:canc2}. Within the magnetic carpet model, both magnetic elements are removed at this stage. In the case of partial cancellation or coalescence, one of the elements is removed when the peaks of their Gaussian profiles meet. The remaining element's flux is then updated to be the difference between the two original flux values. This new element's motion is now determined by supergranular flows until another process takes over.

\begin{figure}
 \begin{center}
  \includegraphics[width=1.0\textwidth]{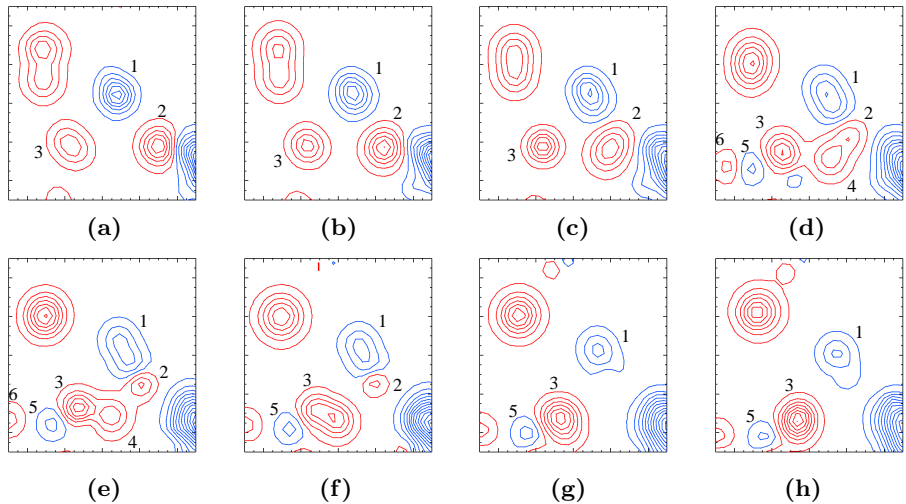}

     \vspace{-0.29\textwidth}   
     \centerline{ \bf     
      \hspace{0.091 \textwidth}  \color{black}{\small{(a)}}
      \hspace{0.191\textwidth}  \color{black}{\small{(b)}}
      \hspace{0.191\textwidth}  \color{black}{\small{(c)}}
      \hspace{0.191\textwidth}  \color{black}{\small{(d)}}
         \hfill}
     \vspace{0.25\textwidth}    
     \centerline{ \bf     
      \hspace{0.091 \textwidth}  \color{black}{\small{(e)}}
      \hspace{0.191\textwidth}  \color{black}{\small{(f)}}
      \hspace{0.191\textwidth}  \color{black}{\small{(g)}}
      \hspace{0.191\textwidth}  \color{black}{\small{(h)}}
         \hfill}
     \vspace{-0.02\textwidth}
 \end{center}
\caption{A sequence of still images (left to right) showing magnetic elements coalescing and cancelling. Red contours represent positive magnetic field, blue contours represent negative magnetic field. 15 contour levels are shown for each polarity, with an absolute peak value of 70 G. The region is 10 Mm $\times$ 10 Mm in area. The images are taken 4 mins apart, and elements of interest have been numbered from 1 to 6. These interactions may be seen in the accompanying movie, `mag1\_canc.mpg'.}\label{fig:canc1}
\end{figure}

The processes of cancellation and coalescence occur frequently within the model, and can easily be detected by eye in the synthetic magnetograms. Figure~\ref{fig:canc1} (and the accompanying movie, `mag1\_canc.mpg') shows a sequence of eight images taken from one of the simulations, in which examples of both cancellation and coalescence of magnetic elements can be seen. We follow the evolution of a negative polarity element 1, and two positive elements, 2 and 3. In frame (d) an emergence occurs in the bottom left (5 and 6) and element 2 begins to fragment, creating a new element, 4. As the sequence progresses, a cancellation occurs between 1 and 2, and elements 3 and 4 coalesce. By the end of the sequence, 2 and 4 have disappeared, and 3 has begun cancelling with 5.

\subsection{Summary of Photospheric Evolution}\label{subsec:whole}

After the parameters for newly emerging bipoles have been determined, we compute the magnetic field on the photosphere for each time step. At each time step, we first loop over all magnetic elements and determine their current velocities. In addition to the four flux evolution processes described in Sections~\ref{subsec:emerge},~\ref{subsec:fragment} and~\ref{subsec:cancel}, we add a contribution from the underlying supergranular flow to each element's velocity. This prevents the motion of elements from being too linear, particularly when undergoing emergence. It seems likely that supergranular flows would continually have an effect on magnetic elements on the real solar surface. We also add a random motion to every magnetic element that represents the effect of granulation. The main purpose of this is to prevent elements from becoming stationary once they reach the network between supergranules, as our supergranular flow profile does not currently evolve in time.

We now update the positions of every element according to its velocity:
\begin{displaymath}
 \mathbf{x}=\mathbf{x} + \mathbf{v_{x}} dt,
\end{displaymath}
\begin{displaymath}
 \mathbf{y}=\mathbf{y} + \mathbf{v_{y}} dt.
\end{displaymath}

The next stage is to check which elements have just started or finished emerging, cancelling, coalescing or fragmenting. Elements are added and removed from the simulation as necessary, and parameters updated. We also check that no elements have moved out of the computational domain. If they have, their coordinates are translated so that they reappear at the opposite side of the box due to the periodic boundaries.

To construct synthetic magnetograms for the line-of-sight component of the magnetic field, we recompute $B_{z}$ analytically at each time step from the information stored for each magnetic element. This simulated magnetogram is output to a file every time step. The method is as described in Section~\ref{subsec:magn}.

\section{Results} \label{results}

Using the techniques described above, nine simulations are run keeping all interaction and evolution parameters fixed at values determined to produce the most realistic evolution. The only parameter that is varied is the lowest flux value taken by emerging bipoles. For the model, we consider a quiet region of the solar surface which has no contribution from active regions. The parameters common to each model are given in Table~\ref{tab:param}. The locations of supergranular cells and the parameters $A_0$ and $R_0$ that specify their strengths are also kept fixed, as described in Section~\ref{subsec:flow}.
\begin{table}
\begin{center}
\begin{tabular}{|ccc|}
\hline
Parameter & Value & Description  \\
\hline
 $x_{\textrm{\small min}}$, $x_{\textrm{\small max}}$ & 0, 50 Mm & $x$-range for computational box. \\
 $y_{\textrm{\small min}}$, $y_{\textrm{\small max}}$ & 0, 50 Mm & $y$-range for computational box. \\
 $t_{\textrm{\small max}}$ & 15\,000 & Length of simulation.  \\
 $step\_length$ & 1 min & Number of mins 1 time step represents.  \\
 $\phi_{\textrm{\small max}}$ & $10^{19}$ Mx & Maximum absolute flux for newly \\
              &              & emerging bipoles. \\
 $\phi_0$ & $10^{16}$ Mx & Minimum unit of flux. \\
 $e_{\textrm{\small rad}}$ & 1.5 & Defines distance a magnetic element will travel \\
           & & from its initial location under emergence. \\
 $c_{\textrm{\small rad}}$ & 0.5 & Defines interaction distance for cancellation \\
           &     & and coalescence. \\
 $f_{\textrm{\small rad}}$ & 1.5 & Defines distance a magnetic element will travel \\
           & & from its initial location under fragmentation. \\
 $v_{\textrm{\small sg}}$  & 0.5 km s$^{-1}$ & Peak value for supergranular flow profile. \\
 $R_{\textrm{\small f}}$  & $1.5\times10^{-4}$ s$^{-1}$ & Fragmentation rate. \\
 $\psi$  & $8\times10^{18}$ Mx & Fragmentation parameter. \\
\hline
\end{tabular}
\caption{Key parameters within the magnetic carpet model, along with their values specified.}\label{tab:param}
\end{center}
\end{table}
In the nine simulations, the varying values taken for the minimum absolute flux of our distribution of emerging bipoles are:
\begin{displaymath}
 \phi_{\textrm{\small min}}=10^{16}\times[4,\: 8,\: 10,\: 20,\: 30,\: 40,\: 60,\: 80,\: 100] \textrm{ Mx}.
\end{displaymath}
The results of the magnetic carpet simulations and the effect of varying $\phi_{\textrm{\small min}}$ are considered in the following four sections. Two movies showing the full field of view of the synthetic magnetograms accompany this paper. The movies show the $\phi_{\textrm{\small min}}=8\times 10^{16}$ Mx (`mag4.mpg') and $\phi_{\textrm{\small min}}=10^{17}$ Mx (`mag1.mpg') simulations between the 50th and 60th hour. By this time, both simulations have reached an equilibrium state in which the rate of emergence roughly equals the rate of cancellation of magnetic flux. Section~\ref{subsec:syn} shows some still images from the simulation with $\phi_{\textrm{\small min}}=8\times 10^{16}$ Mx.

\subsection{Example Synthetic Magnetograms}\label{subsec:syn}

  \begin{figure}

   \centerline{\hspace*{0.015\textwidth}
               \includegraphics[width=0.515\textwidth,clip=]{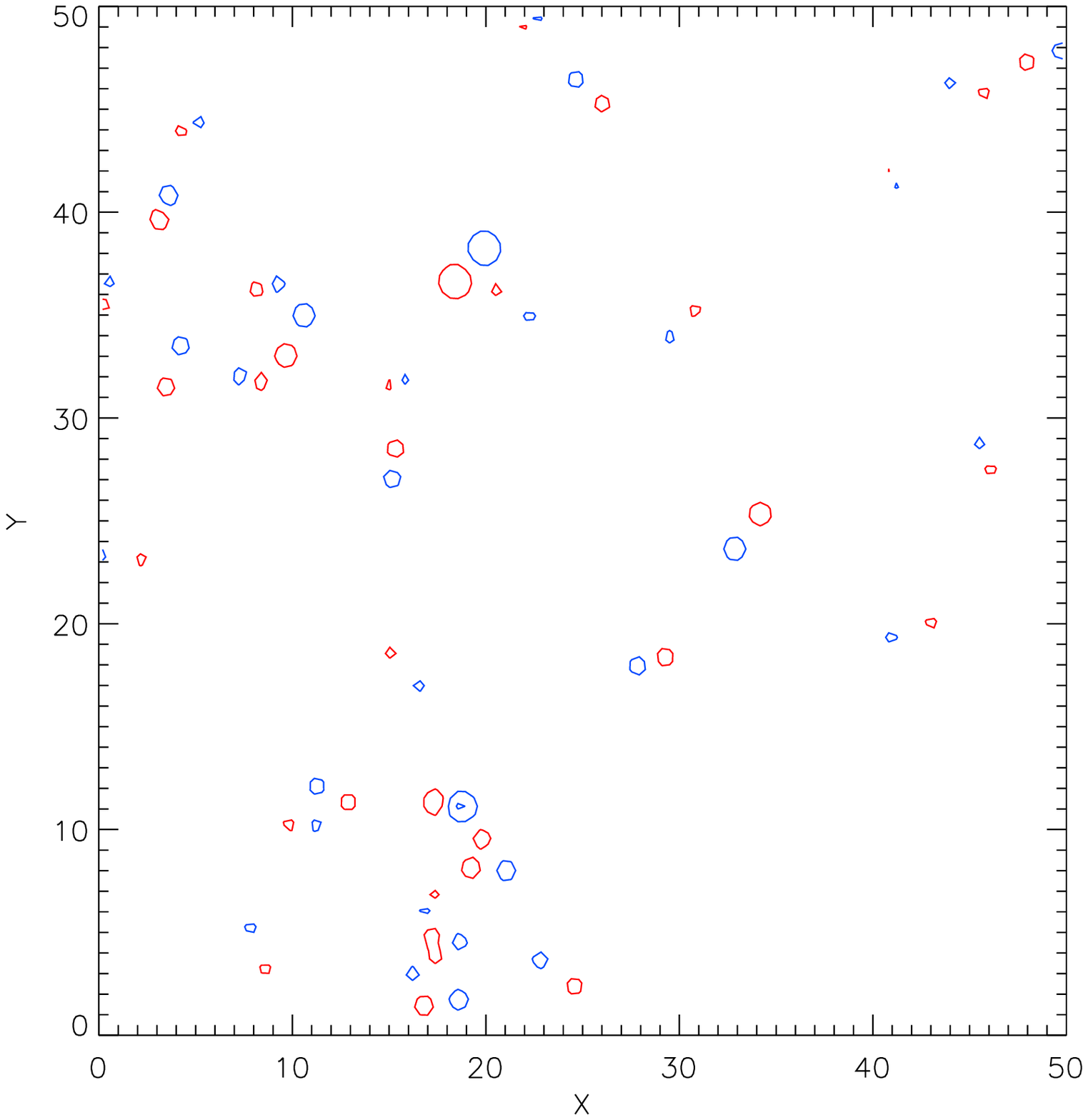}
               \hspace*{-0.03\textwidth}
               \includegraphics[width=0.515\textwidth,clip=]{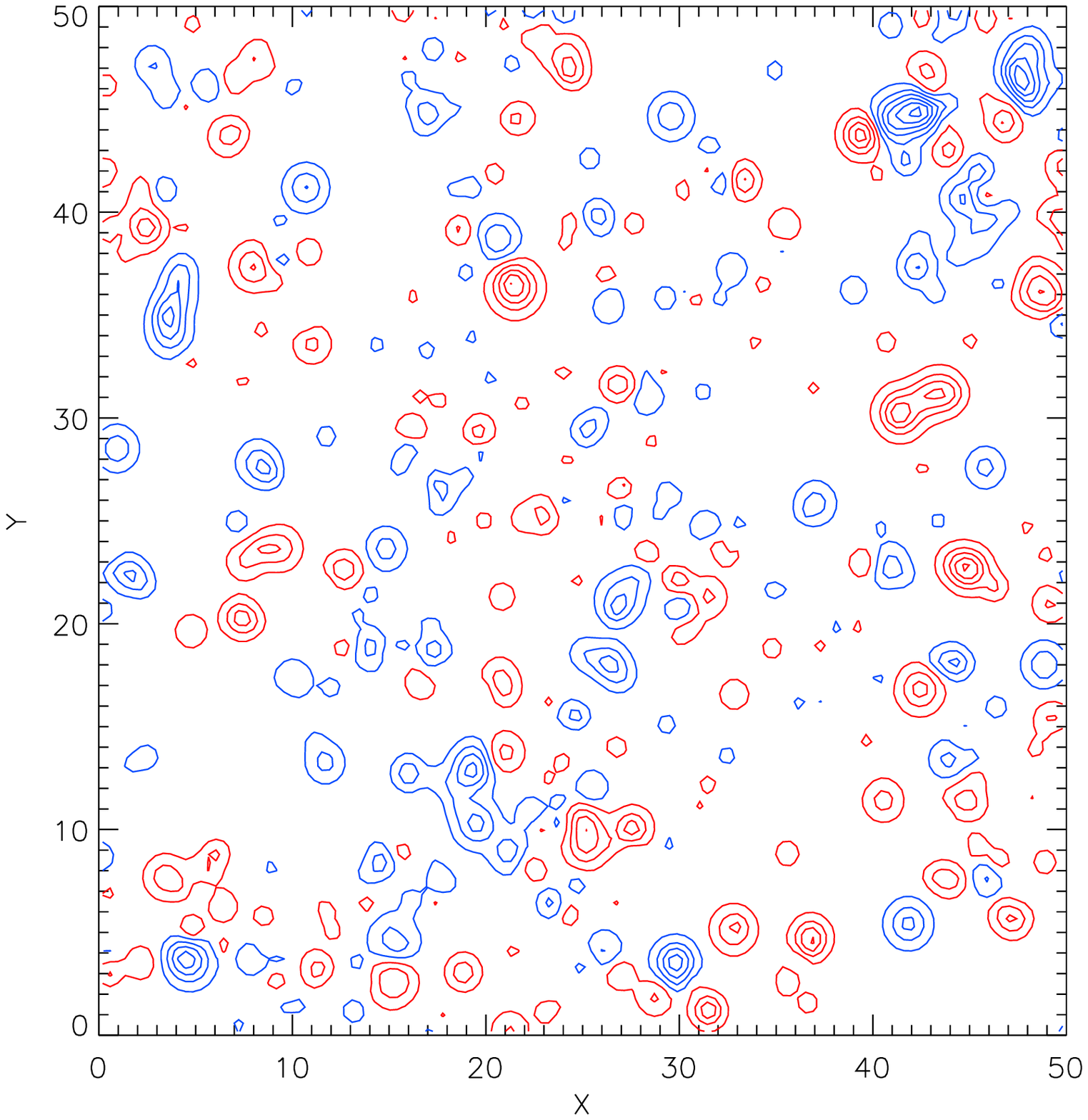}
              }
     \vspace{-0.48\textwidth}   
     \centerline{ \bf     
      \hspace{-0.02 \textwidth}  \color{black}{(a)}
      \hspace{0.44\textwidth}  \color{black}{(b)}
         \hfill}
     \vspace{0.43\textwidth}    
   \centerline{\hspace*{0.015\textwidth}
               \includegraphics[width=0.515\textwidth,clip=]{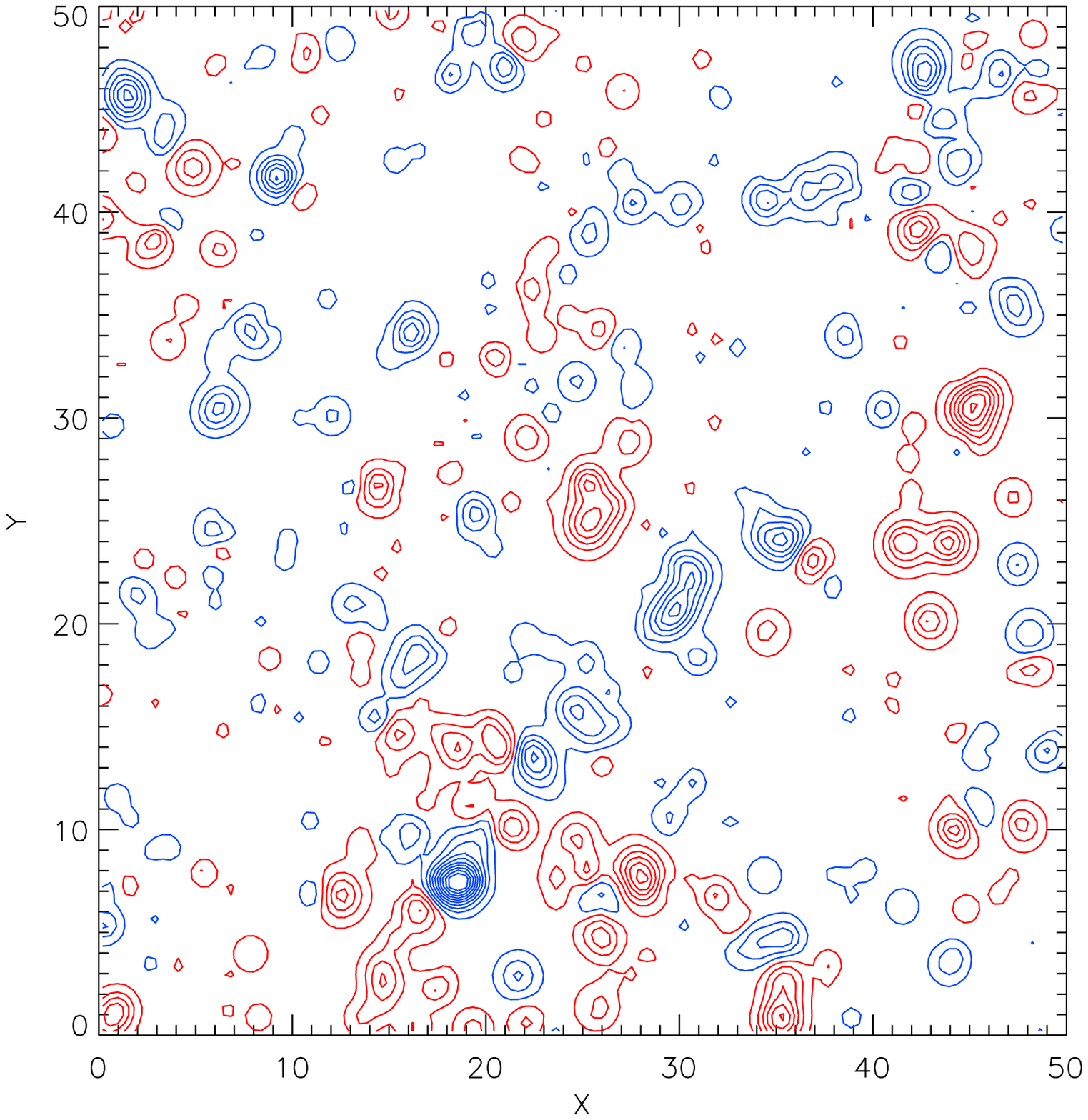}
               \hspace*{-0.03\textwidth}
               \includegraphics[width=0.515\textwidth,clip=]{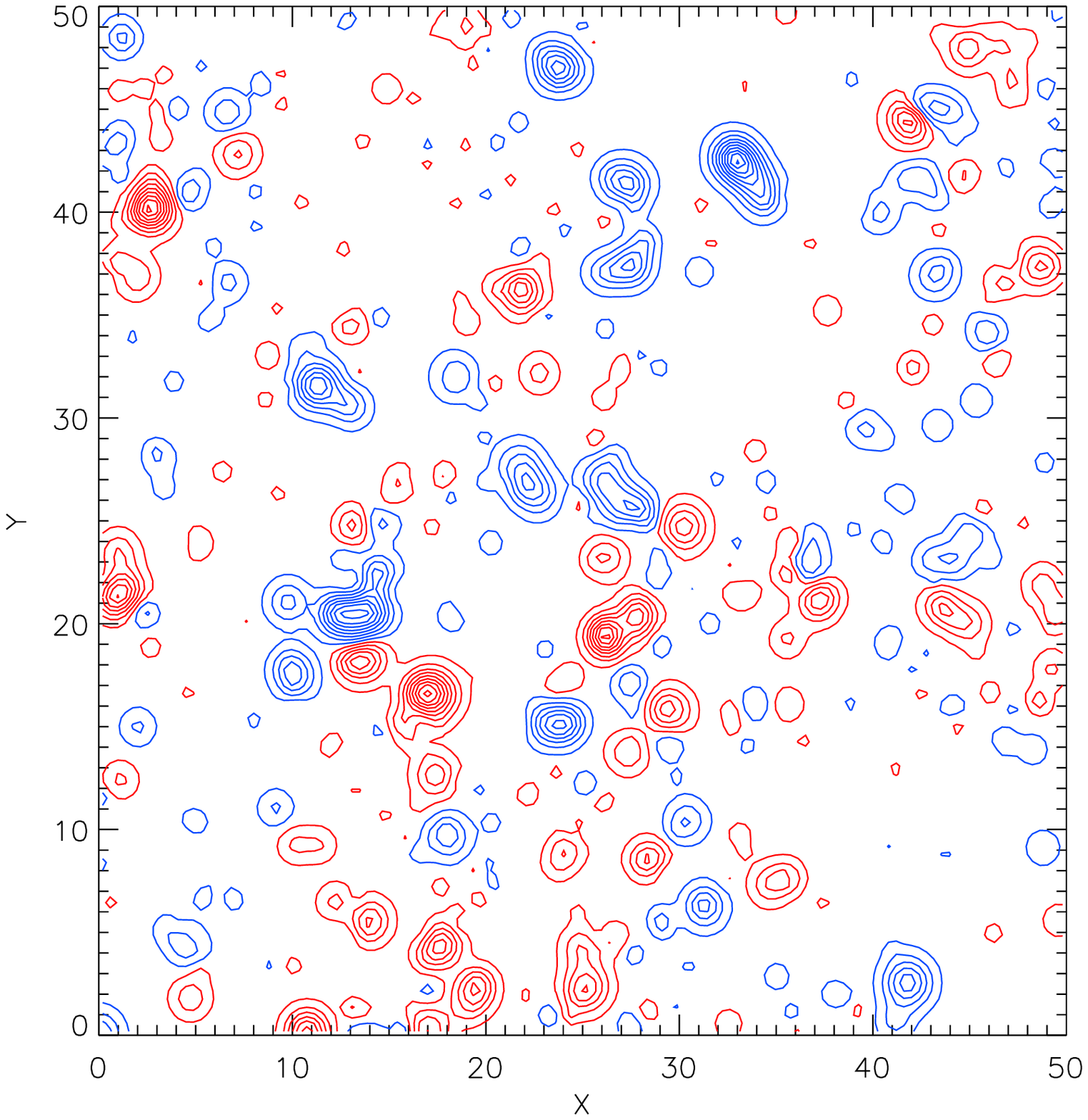}
              }
     \vspace{-0.48\textwidth}   
     \centerline{ \bf     
      \hspace{-0.02 \textwidth}  \color{black}{(c)}
      \hspace{0.44\textwidth}  \color{black}{(d)}
         \hfill}
     \vspace{0.43\textwidth}    
   \centerline{\hspace*{0.015\textwidth}
               \includegraphics[width=0.515\textwidth,clip=]{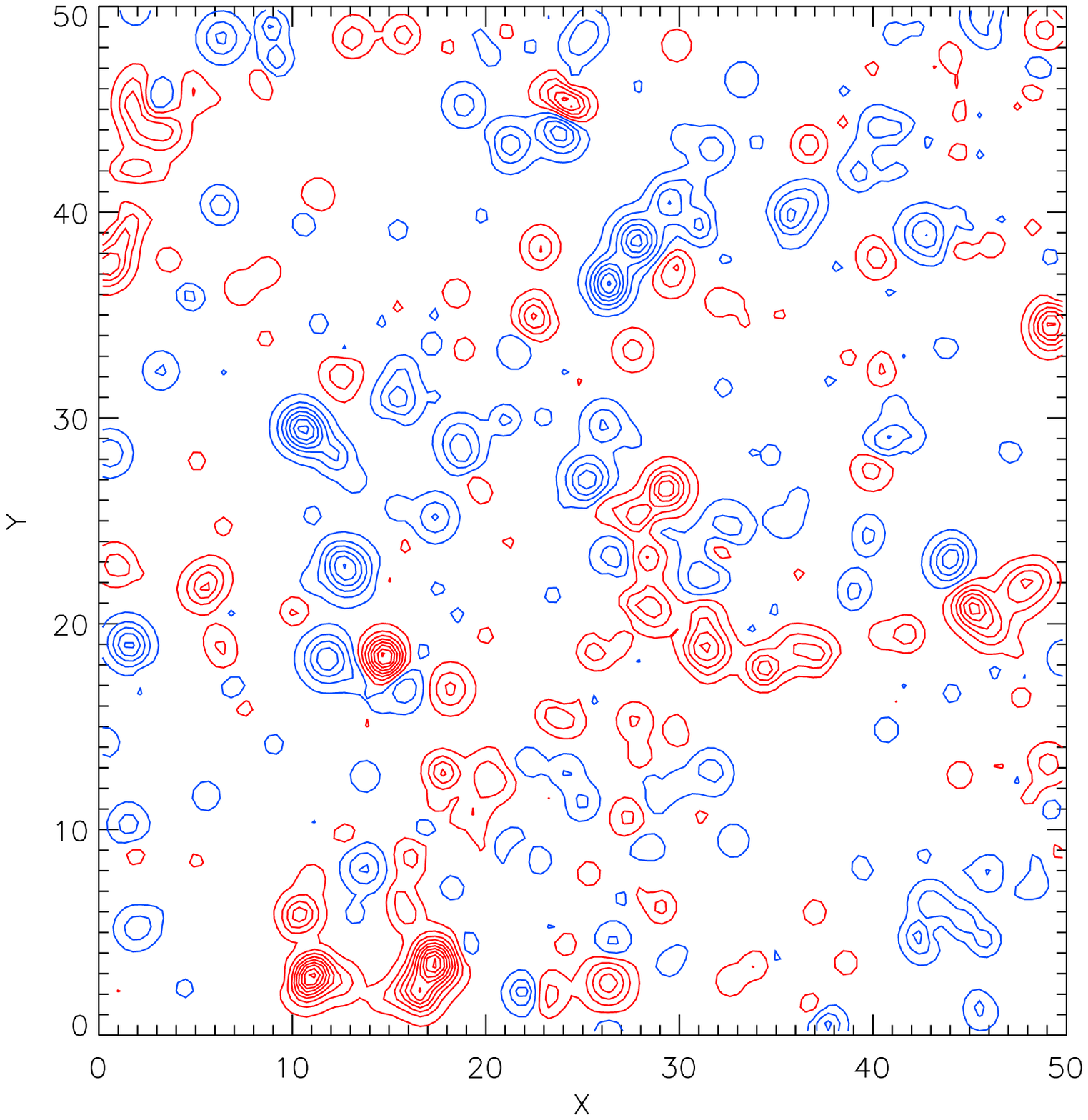}
               \hspace*{-0.03\textwidth}
               \includegraphics[width=0.515\textwidth,clip=]{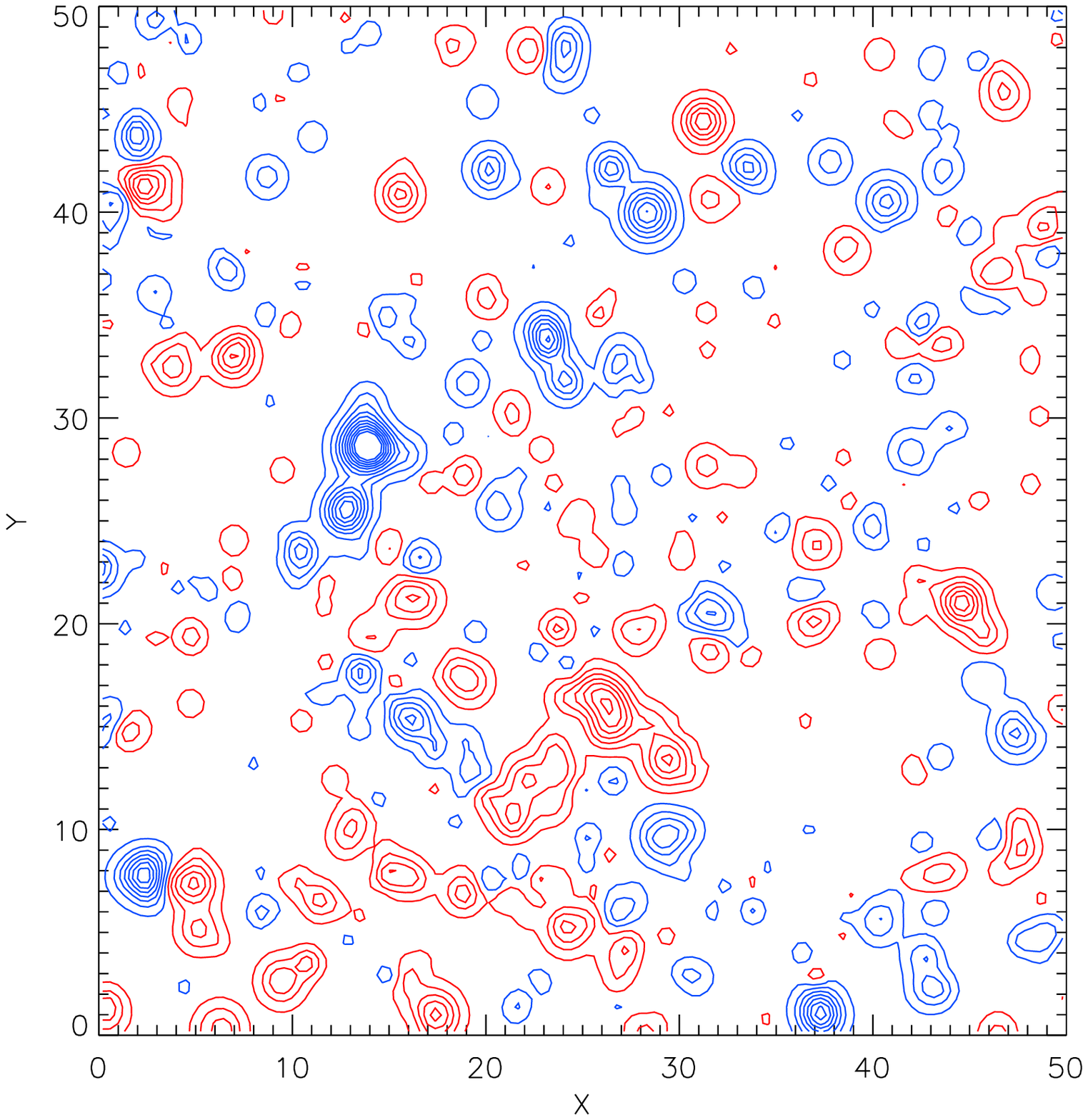}
              }
     \vspace{-0.48\textwidth}   
     \centerline{ \bf     
      \hspace{-0.02 \textwidth}  \color{black}{(e)}
      \hspace{0.44\textwidth}  \color{black}{(f)}
         \hfill}
     \vspace{0.43\textwidth}    

\caption{Synthetic magnetograms for a simulation with a flux emergence range of $8\times 10^{16}$ - $1\times 10^{19} $ Mx. In each frame, red contours represent positive magnetic field, blue contours represent negative magnetic field, where 10 contour levels are shown for each polarity with an absolute peak value of 70 G. The time in hours at which each image is taken and number of individual elements (ns) composing the magnetogram are as follows: (a) $t=0.1$, ns=102, (b) $t=10$, ns=550, (c) $t=20$, ns=607, (d) $t=225$, ns=595, (e) $t=230$, ns=565 and (f) $t=249$, ns=603.}\label{fig:mag}
   \end{figure}

Figure~\ref{fig:mag} shows contour plots of $B_{z}$ for a simulation in which newly emerging bipoles have a total flux in the range $8\times 10^{16}$ - $1\times 10^{19} $ Mx. The images are taken at $t=0.1$, $t=10$, $t=20$, $t=225$, $t=230$ and $t=249$ h. These images may be compared with the flow profile in Figure~\ref{fig:sgflow}(b) to compare the location of the magnetic elements with respect to the underlying supergranule flow pattern.

At $t=0$ the box is empty of magnetic flux. As the simulation progresses, bipoles are allowed to emerge following a supergranular cell pattern. After 6 mins (image (a)) small magnetic elements have begun to appear within the supergranular cells. At (b) $t=10$ h, many of the elements have begun to encounter one another and interact at the cell edges, and by (c) $t=20$ h a magnetic network of larger elements has formed along the supergranule boundaries. By this stage, the model has reached a steady state in which there is a balance between the rates of emerging and cancelling flux. As a result, a similar network of magnetic flux exists in each of the later images, (d), (e) and (f).
Since the supergranular flow profile is steady throughout the simulation, the spatial location of the magnetic network does not vary greatly; however the exact distribution of magnetic flux elements significantly changes. From this it can be seen that the steady flow profile does not lead to the formation of unphysically large magnetic elements. However, it would not lead to the random walk of magnetic elements across the solar surface as proposed by \inlinecite{leighton1964}. Since we are presently only considering a small, localised area, such a random walk is not important. Comparing images (d) and (e) at $t=225$ and $t=230$ h, it can be seen that the distribution of magnetic elements varies significantly between the two images, even though they are only spaced 5 h apart. This implies that the photospheric recycle time of the simulated magnetograms is short, as desired. A plot of total absolute flux versus time is given in Figure~\ref{fig:absflux}(b) and discussed in Section~\ref{subsec:absflux}.

\subsection{Total Absolute Flux and Mean Magnetic Field}\label{subsec:absflux}

  \begin{figure}
   \centerline{\hspace*{0.015\textwidth}
               \includegraphics[width=0.515\textwidth,clip=]{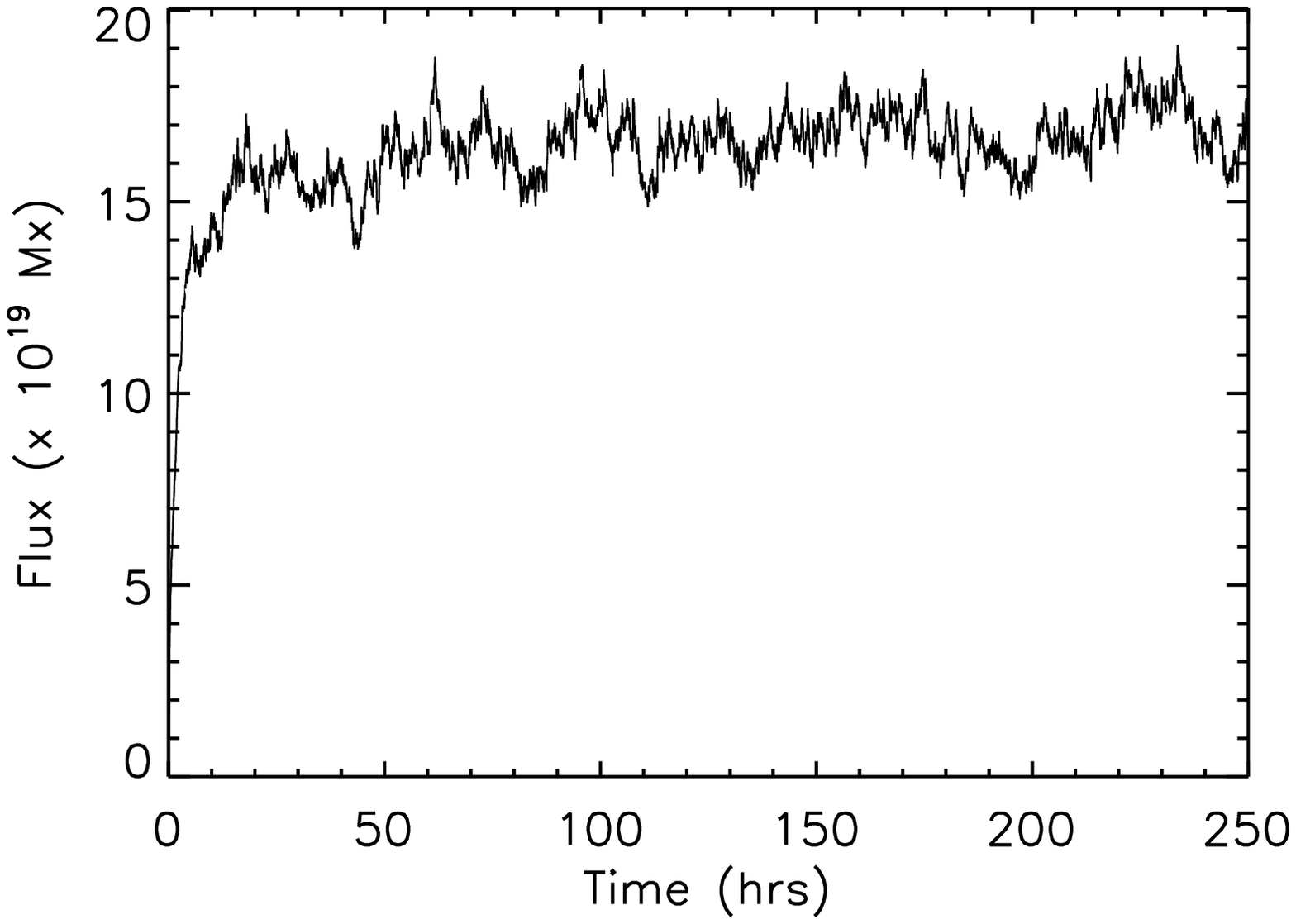}
               \hspace*{-0.03\textwidth}
               \includegraphics[width=0.515\textwidth,clip=]{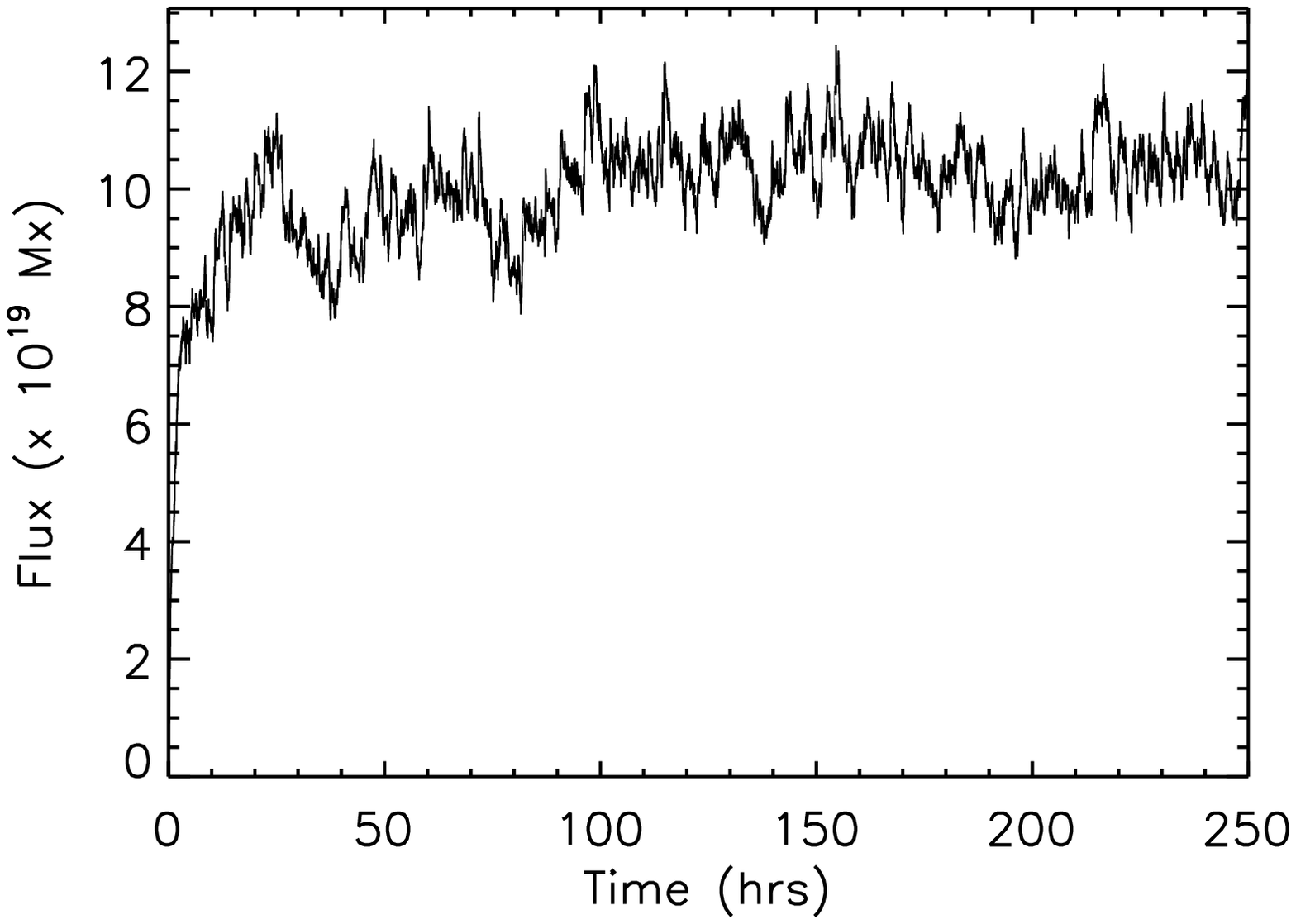}
              }
     \vspace{-0.39\textwidth}   
     \centerline{ \bf     
      \hspace{-0.02 \textwidth}  \color{black}{(a)}
      \hspace{0.44\textwidth}  \color{black}{(b)}
         \hfill}
     \vspace{0.34\textwidth}    
   \centerline{\hspace*{0.015\textwidth}
               \includegraphics[width=0.515\textwidth,clip=]{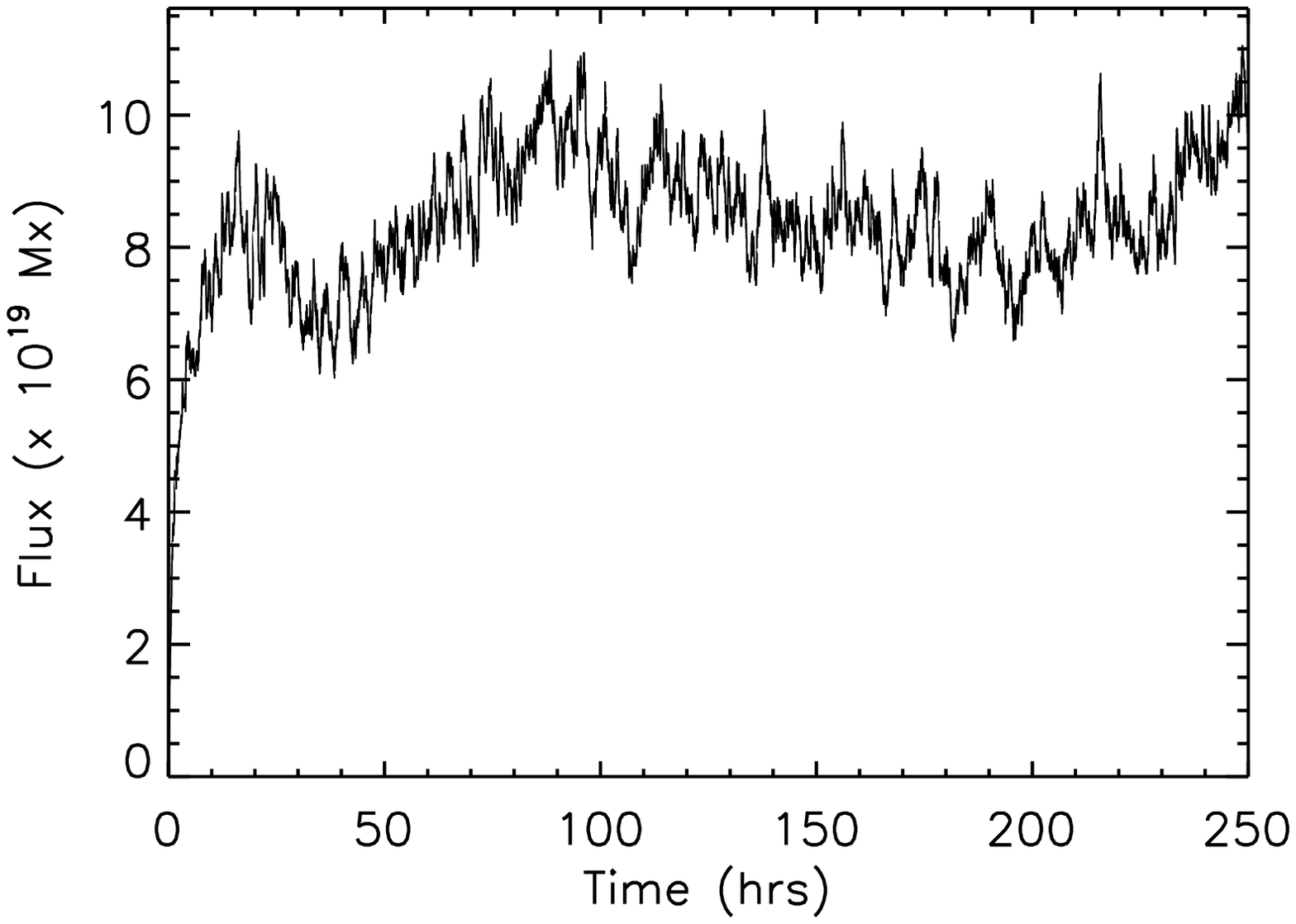}
               \hspace*{-0.03\textwidth}
               \includegraphics[width=0.515\textwidth,clip=]{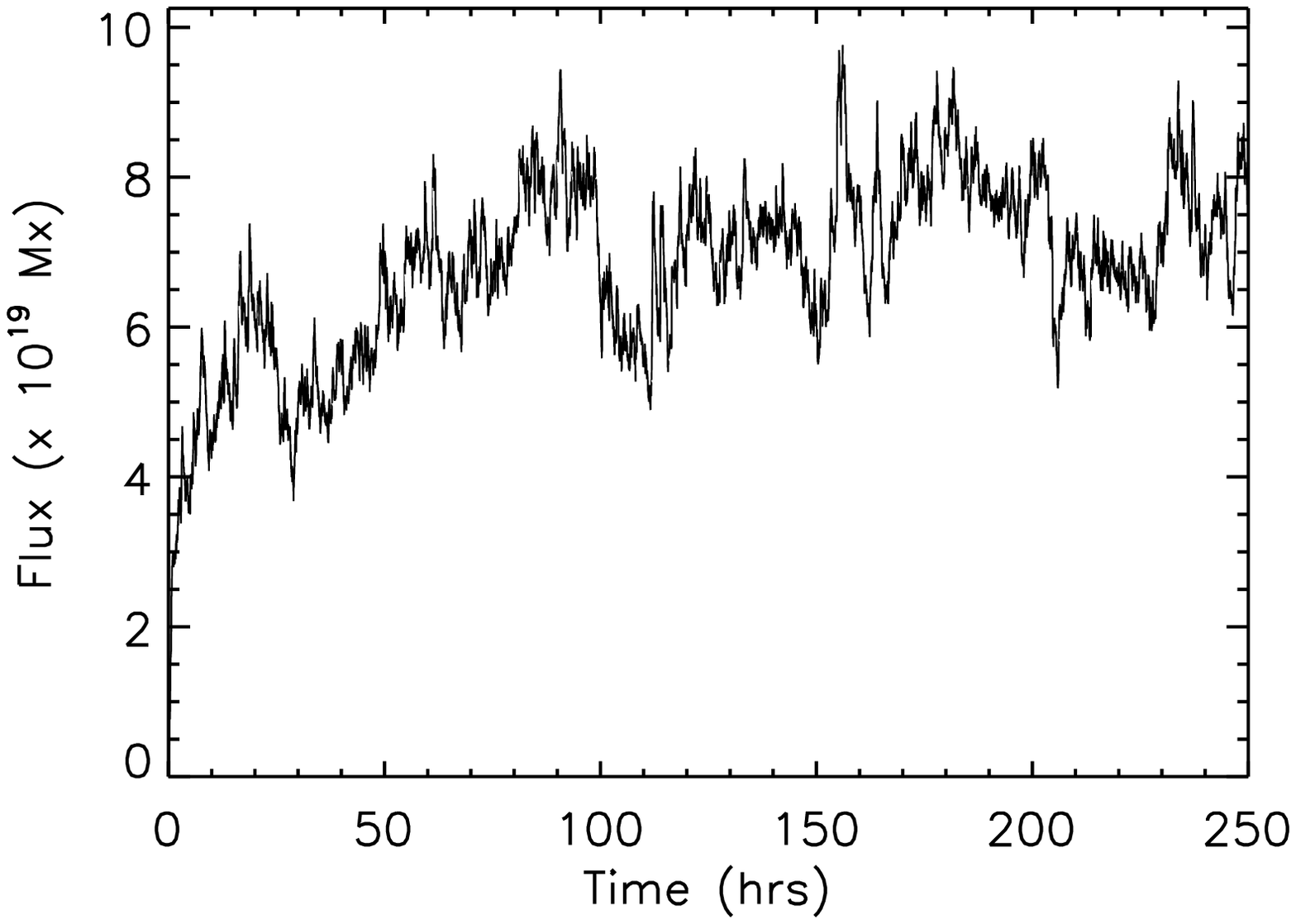}
              }
     \vspace{-0.39\textwidth}   
     \centerline{ \bf     
      \hspace{-0.02 \textwidth} \color{black}{(c)}
      \hspace{0.45\textwidth}  \color{black}{(d)}
         \hfill}
     \vspace{0.34\textwidth}    

\caption{Total absolute flux within the computational box versus time for $\phi_{\textrm{\tiny min}}$ of (a) $4\times 10^{16}$ Mx, (b) $8\times 10^{16}$ Mx, (c) $1\times 10^{17}$ Mx and (d) $2\times 10^{17}$ Mx .}\label{fig:absflux}
   \end{figure}

For all simulations, the computational box is initially empty of magnetic flux. As the simulation progresses, new magnetic bipoles emerge within the supergranular cells and magnetic elements interact with one another. It is important to verify that the model reaches a steady state in which the absolute flux density oscillates about an average value. Plots of the total absolute flux within the box against time are shown in Figure~\ref{fig:absflux} for the four lowest values of $\phi_{\textrm{\small min}}$. Clearly, the lower the value of $\phi_{\textrm{\small min}}$, the larger the range of flux values for emerging bipoles, and thus the greater the rate of flux emergence. This in turn leads to the total absolute flux for the simulation levelling off at a higher value.

It can be seen from Equation~(\ref{eqn:nall}) that $N(\phi)$ will be greater for smaller values of $\phi$. Therefore a lower value of $\phi_{\textrm{\small min}}$ results in a large number of small bipoles emerging. For the case where $\phi_{\textrm{\small min}}=4\times 10^{16}$ Mx (Figure~\ref{fig:absflux}(a)), the absolute flux levels off after approximately 24 h. After this it oscillates between $1.58 \times 10^{20}$ Mx and $1.74 \times 10^{20}$ Mx. This variation is around 5\% from the mean value of $1.66 \times 10^{20}$ Mx. When $\phi_{\textrm{\small min}}=8\times 10^{16}$ Mx, the absolute flux oscillates slightly more than this, but the total flux still becomes roughly steady after 24 h. For higher $\phi_{\textrm{\small min}}$, and hence a smaller flux emergence range, the variation of the absolute flux is much more erratic. For these cases, small numbers of large elements emerge, which have a significant effect on the value of the overall absolute flux within the box at any one instant in time. When $\phi_{\textrm{\small min}}=2\times 10^{17}$ Mx (Figure~\ref{fig:absflux}(d)) there are several points where the total flux value spikes due to a large bipole emerging, or drops due to the lack of emergence of new bipoles and the disappearance of two or more large flux elements through cancellation. However, it is clear from Figure~\ref{fig:absflux} that if $\phi_{\textrm{\small min}}$ is chosen low enough, a steady state is soon found.

In several of the simulations with $\phi_{\textrm{\small min}}$ larger than $2\times 10^{17}$ Mx, the total absolute flux within the box steadily increases throughout the 250 h period, and hence a steady state is not reached. The absolute flux for these simulations may level off if the simulation were run for longer, but there would still be a large scale flux variation about this value. The mean values for the total absolute flux averaged over each simulation, and their percentage variation are shown in Table~\ref{tab:sim}. The means are computed for values taken after 30 h of the simulation have elapsed, so that the model has had a chance to reach a steady state. The highest flux values and lowest variation are found for the lowest value of $\phi_{\textrm{\small min}}$.

\begin{table}
\begin{center}
\begin{tabular}{|ccccc|}
\hline
$\phi_{\textrm{\small min}}$ & Mean total & Percentage & Mean absolute & Mean number \\
 ($\times 10^{16}$ Mx)  & absolute flux & variation of total & flux density (G) & of elements \\
   & ($\times 10^{19}$ Mx) & absolute flux &  & per frame \\
\hline
4 & 16.58 & 4.9 & 6.6 & 1497 \\
8 & 10.14 & 7.4 & 4.1 & 570 \\
10 & 8.50 & 10.7 & 3.4 & 430  \\
20 & 7.07 & 13.4 & 2.8 & 211 \\
30 & 9.03 & 19.1 & 3.6 & 153 \\
40 & 5.93 & 18.4 & 2.4 & 120 \\
60 & 6.35 & 15.2 & 2.5 & 88 \\
80 & 5.88 & 24.7 & 2.4 & 74 \\
100 & 5.03 & 15.8 & 2.0 & 63 \\
\hline
\end{tabular}
\caption{Mean values for each simulation for the total absolute flux, absolute flux density and number of magnetic elements within the box.}\label{tab:sim}
\end{center}
\end{table}

  \begin{figure}
   \centerline{\hspace*{0.015\textwidth}
               \includegraphics[width=0.515\textwidth,clip=]{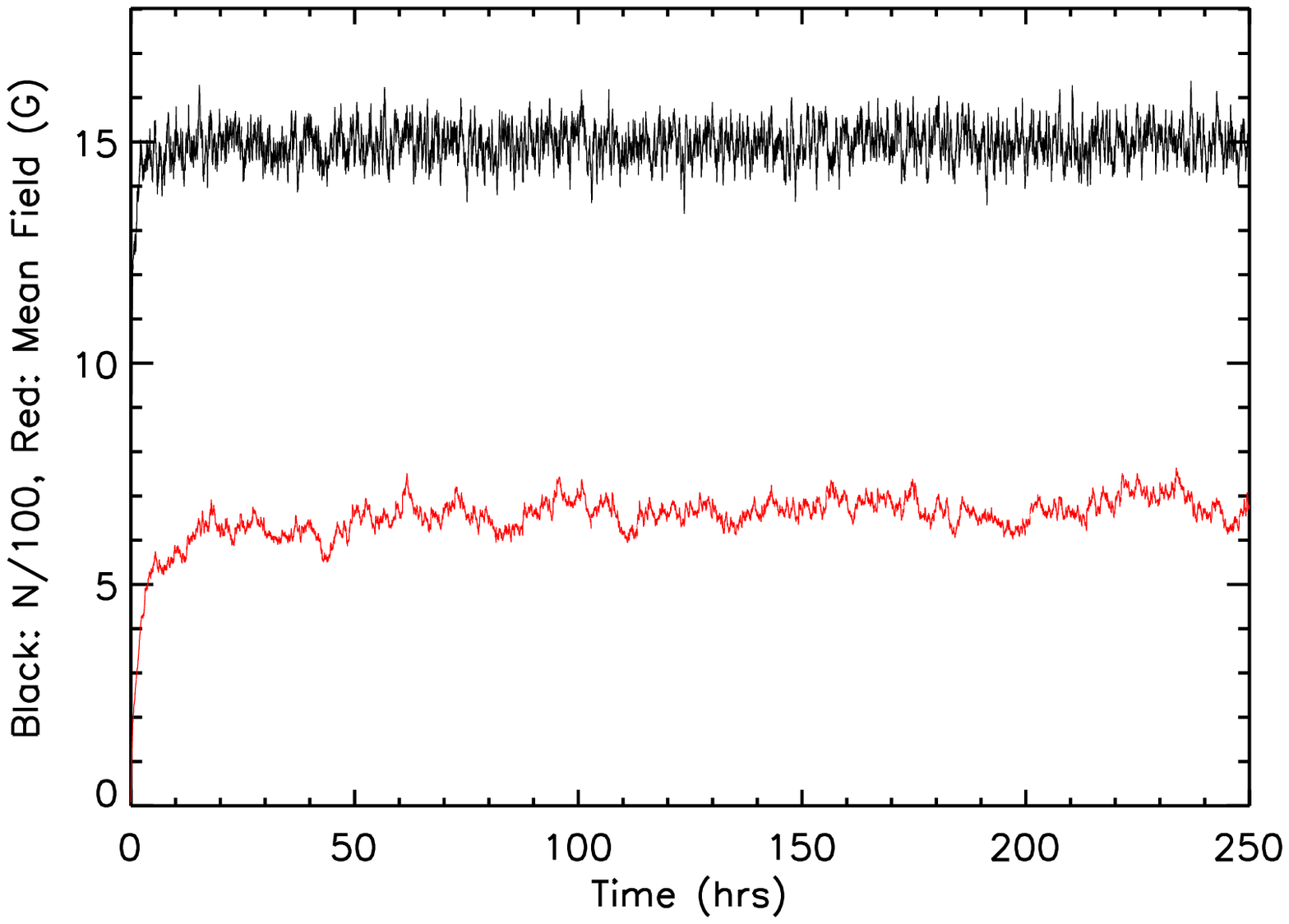}
               \hspace*{-0.03\textwidth}
               \includegraphics[width=0.515\textwidth,clip=]{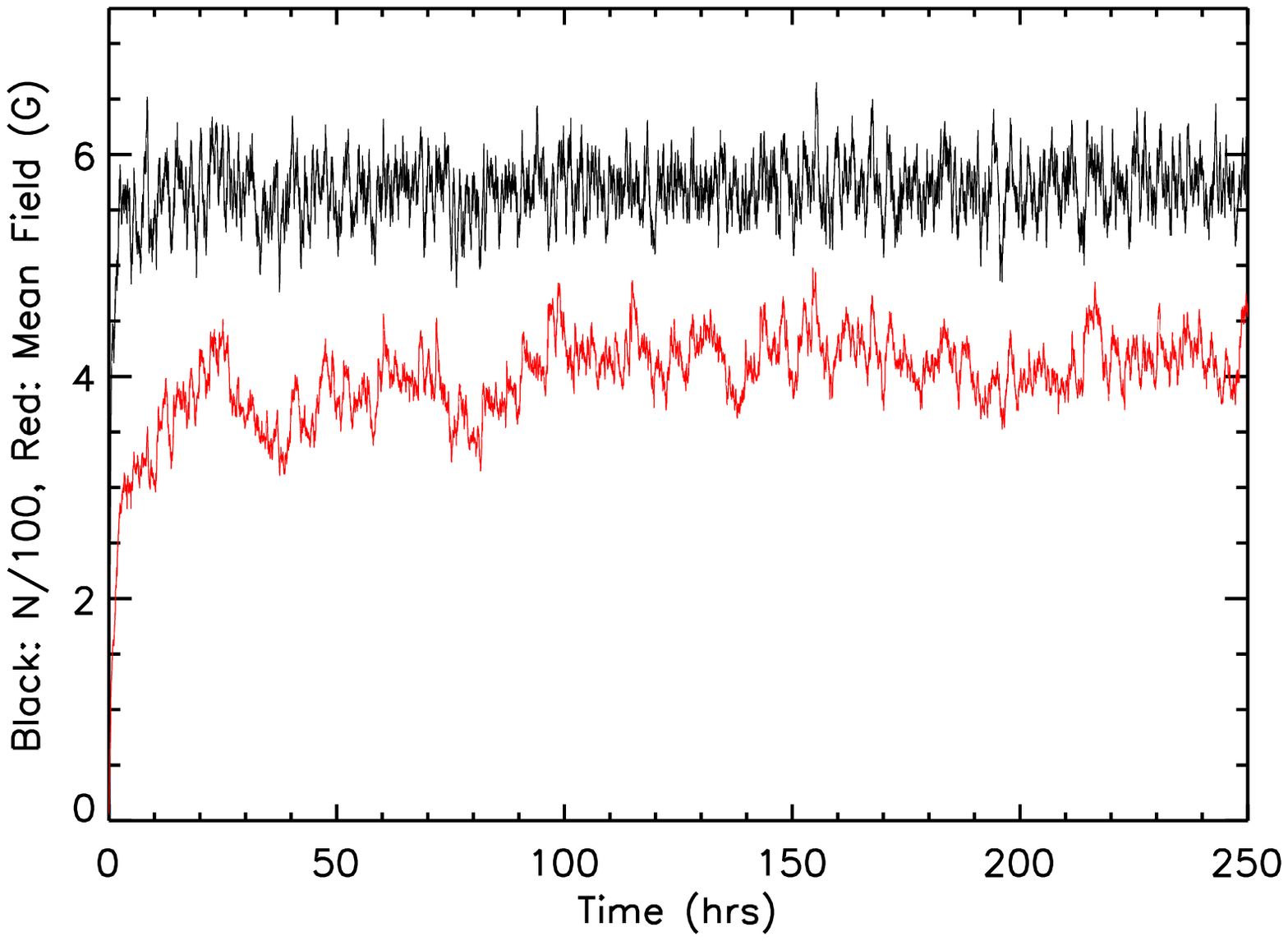}
              }
     \vspace{-0.38\textwidth}   
     \centerline{ \bf     
      \hspace{-0.02 \textwidth}  \color{black}{(a)}
      \hspace{0.45\textwidth}  \color{black}{(b)}
         \hfill}
     \vspace{0.33\textwidth}    
   \centerline{\hspace*{0.015\textwidth}
               \includegraphics[width=0.515\textwidth,clip=]{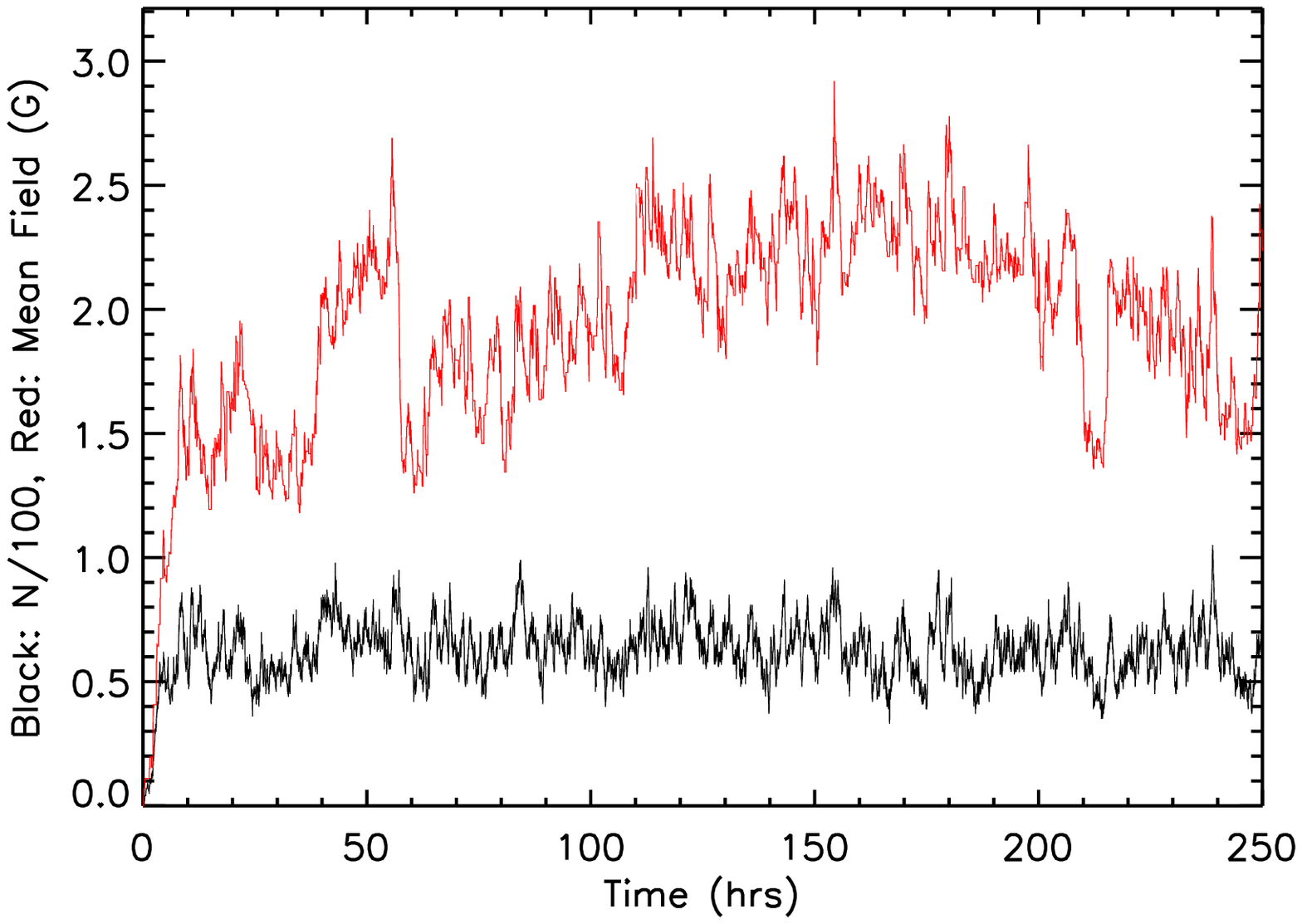}
               \hspace*{-0.03\textwidth}
               \includegraphics[width=0.515\textwidth,clip=]{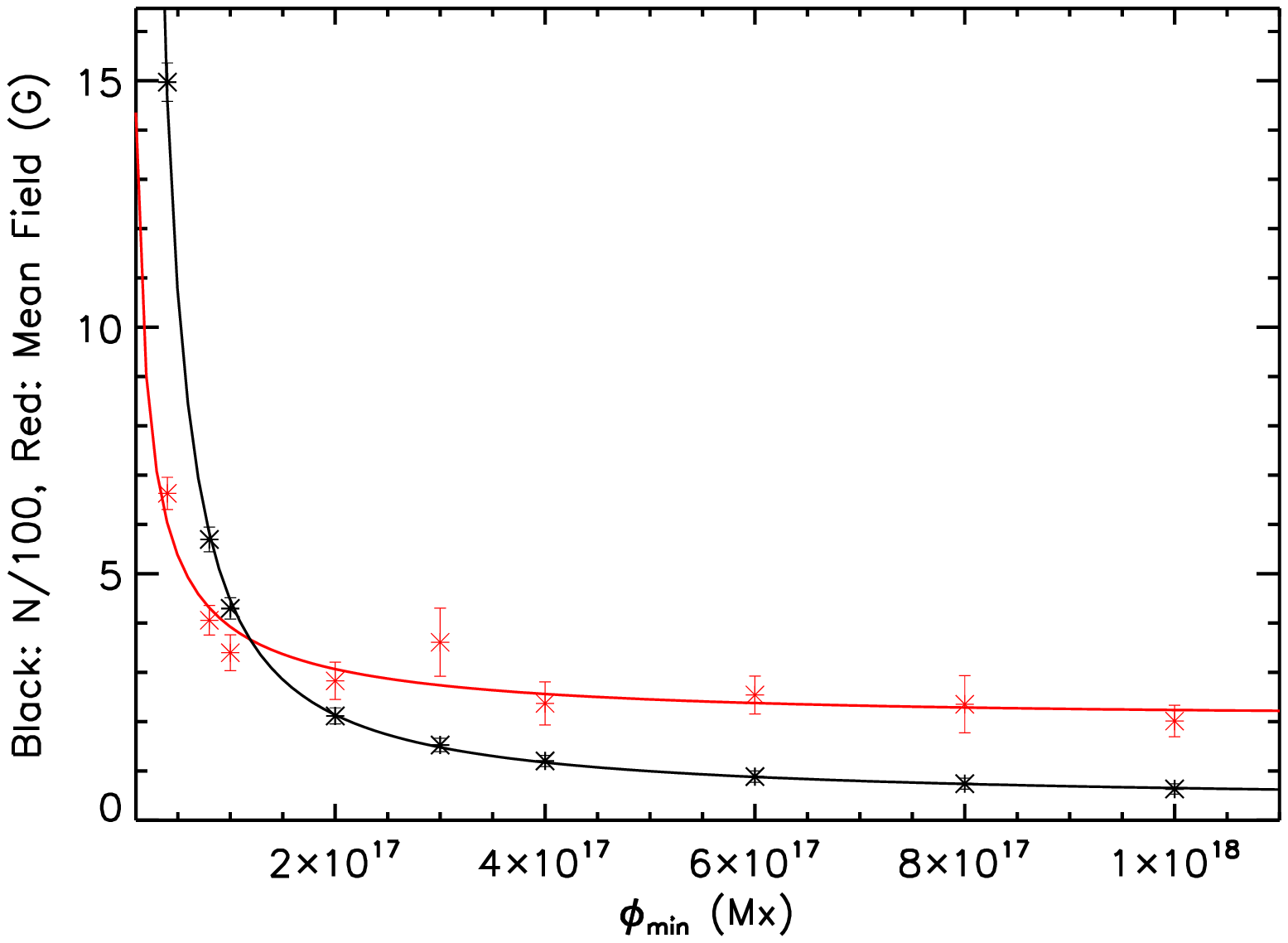}
              }
     \vspace{-0.38\textwidth}   
     \centerline{ \bf     
      \hspace{-0.02 \textwidth} \color{black}{(c)}
      \hspace{0.45\textwidth}  \color{black}{(d)}
         \hfill}
     \vspace{0.33\textwidth}    
              
\caption{(a)-(c) Red line: mean magnetic field (G) and black line: total number of magnetic elements ($N/100$) versus time for $\phi_{\textrm{\tiny min}}$ of (a) $4\times 10^{16}$ Mx, (b) $8\times 10^{16}$ Mx and (c) $1\times 10^{18}$ Mx. (d) Red fitted line: time averaged mean magnetic field (G) and black fitted line: average number of magnetic elements per frame ($N/100$), for each simulation}\label{fig:meanfield}
   \end{figure}

Graphs of the mean magnetic field strength (red line) versus time are shown in Figure~\ref{fig:meanfield} for three of the simulations: (a) $\phi_{\textrm{\small min}}=4\times 10^{16}$ Mx, (b) $\phi_{\textrm{\small min}}=8\times 10^{16}$ Mx and (c) $\phi_{\textrm{\small min}}=1\times 10^{18}$ Mx. Note that in each graph a different scale is used for the vertical axis. Over-plotted on each graph is the number of magnetic elements per frame (black line), divided by 100 so that the curve is in the same range as the mean field. As one would expect, the mean magnetic field strength is higher in simulations where emergence occurs over a larger range. The time averaged values for the mean magnetic field in each simulation are given in Table~\ref{tab:sim} along with the mean number of magnetic elements. The mean values for the simulations with a $\phi_{\textrm{\small min}}<3\times 10^{17}$ are within the observed range of $3-10$ G (Gauss) \cite{cranmer2010}. The mean flux densities calculated for $\phi_{\textrm{\small min}}\geq 4\times 10^{17}$ Mx are slightly lower than observed values, but this is to be expected since realistically, small-scale flux emergence on the Sun is not restricted to such a limited range.

The number of magnetic elements that exist within the box per frame reaches a steady state very rapidly in all nine simulations, levelling off after just a few hours. This occurs shortly after the first magnetic elements within the simulation reach the network between supergranules and begin to coalesce and cancel. The average supergranular crossing time in the model, the time taken for a magnetic element to reach the edge of the supergranule once it has emerged, is around 3-4 h. A typical distance that a bipole might emerge from the edge of a supergranule is 7 Mm, if we assume an average velocity of 0.5 km s$^{-1}$ then each polarity would reach the boundary after 3.9 h. The levelling off of the total absolute flux and number of magnetic elements is therefore determined by the time scale of the flow profile.

Figure~\ref{fig:meanfield}(d) shows the mean field and number of elements ($/100$) averaged over the simulation versus $\phi_{\textrm{\small min}}$. Initially, both the mean field and number of elements drop very rapidly with increasing $\phi_{\textrm{\small min}}$, then level off around $\phi_{\textrm{\small min}}=2\times 10^{17}$ Mx.

\subsection{Rates and Frequencies}\label{subsec:rate}

  \begin{figure}
   \centerline{\hspace*{0.015\textwidth}
               \includegraphics[width=0.515\textwidth,clip=]{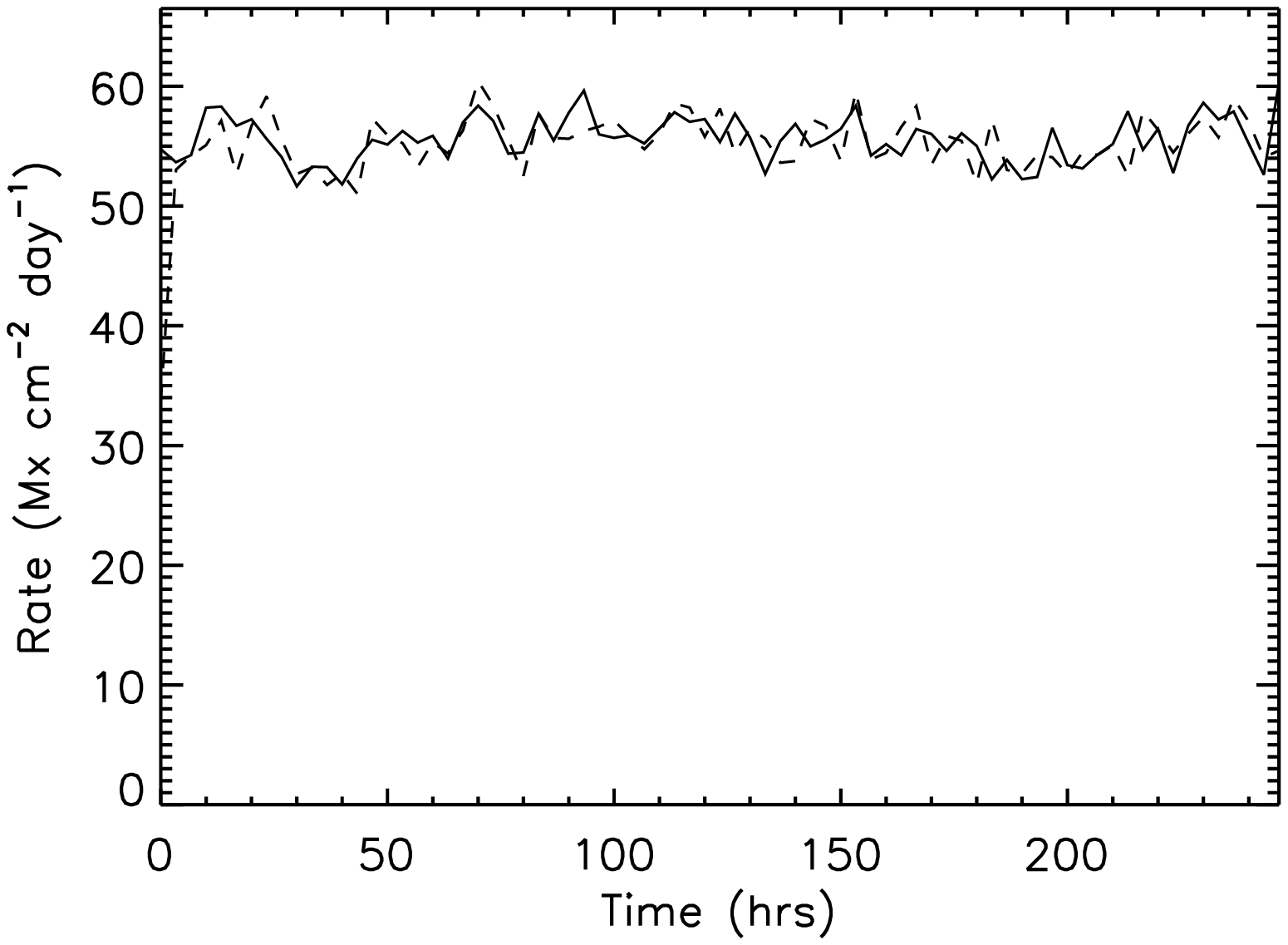}
               \hspace*{-0.03\textwidth}
               \includegraphics[width=0.515\textwidth,clip=]{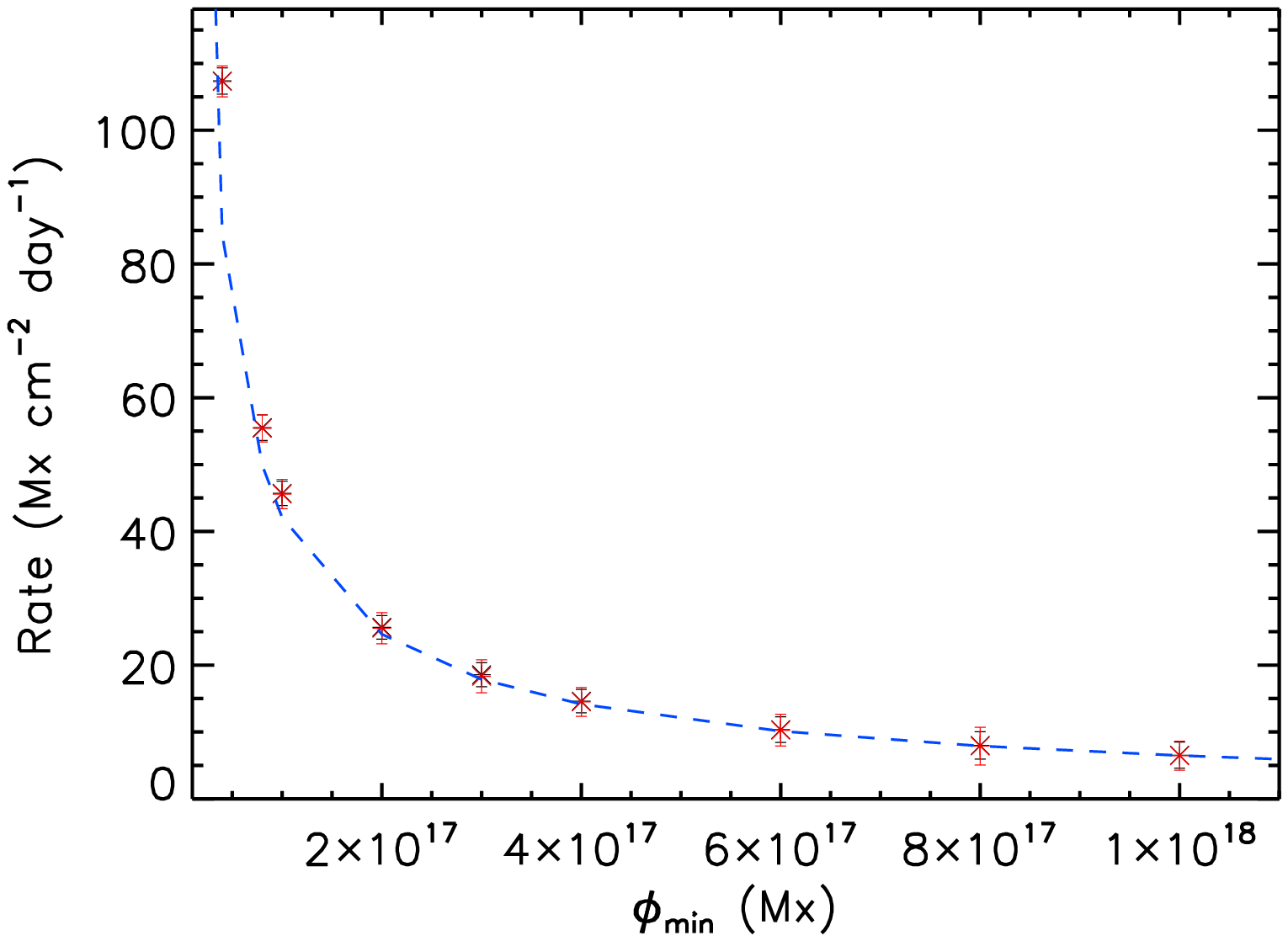}
              }
     \vspace{-0.38\textwidth}   
     \centerline{ \bf     
      \hspace{-0.02 \textwidth}  \color{black}{(a)}
      \hspace{0.45\textwidth}  \color{black}{(b)}
         \hfill}
     \vspace{0.35\textwidth}    

   \centerline{\hspace*{0.015\textwidth}
               \includegraphics[width=0.515\textwidth,clip=]{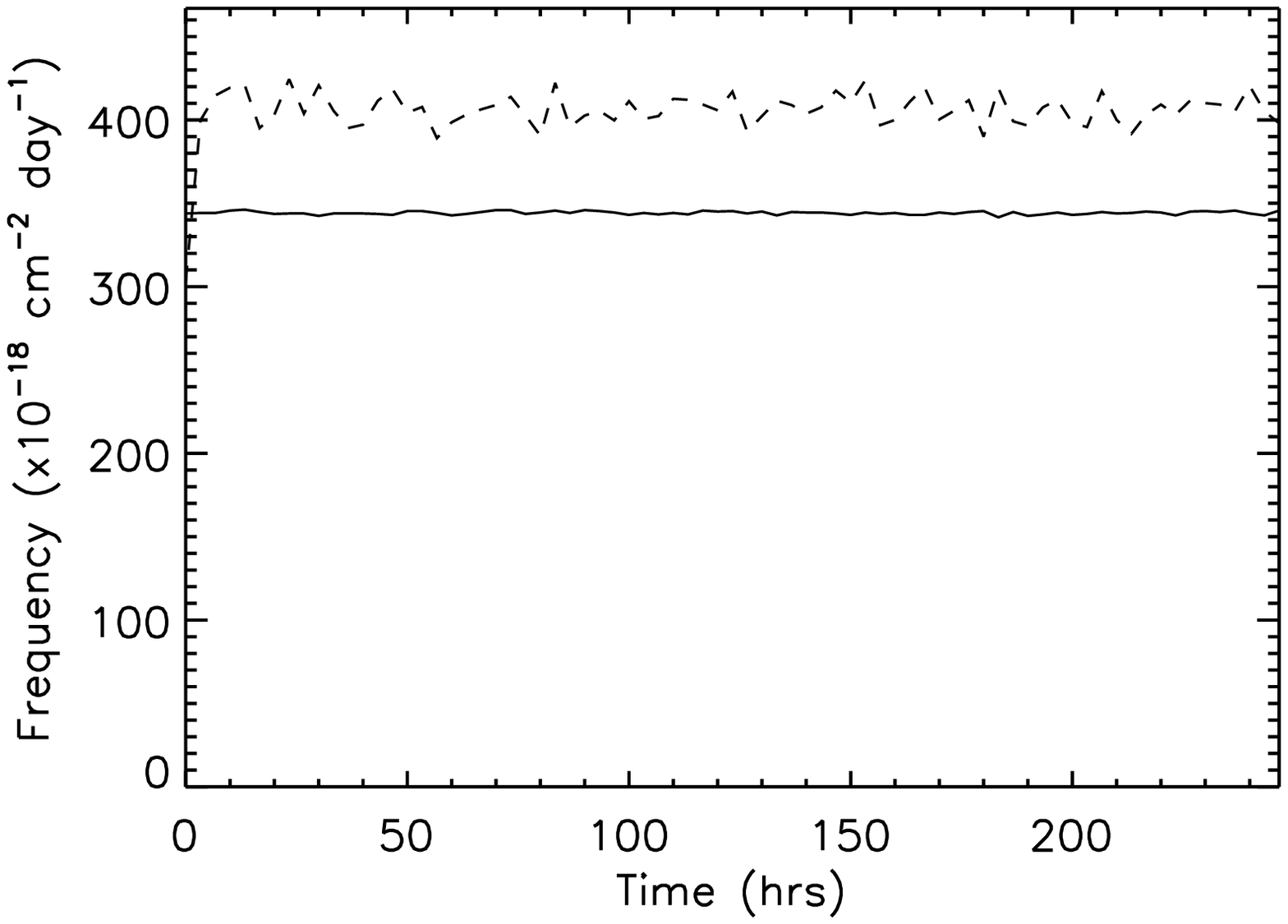}
               \hspace*{-0.03\textwidth}
               \includegraphics[width=0.515\textwidth,clip=]{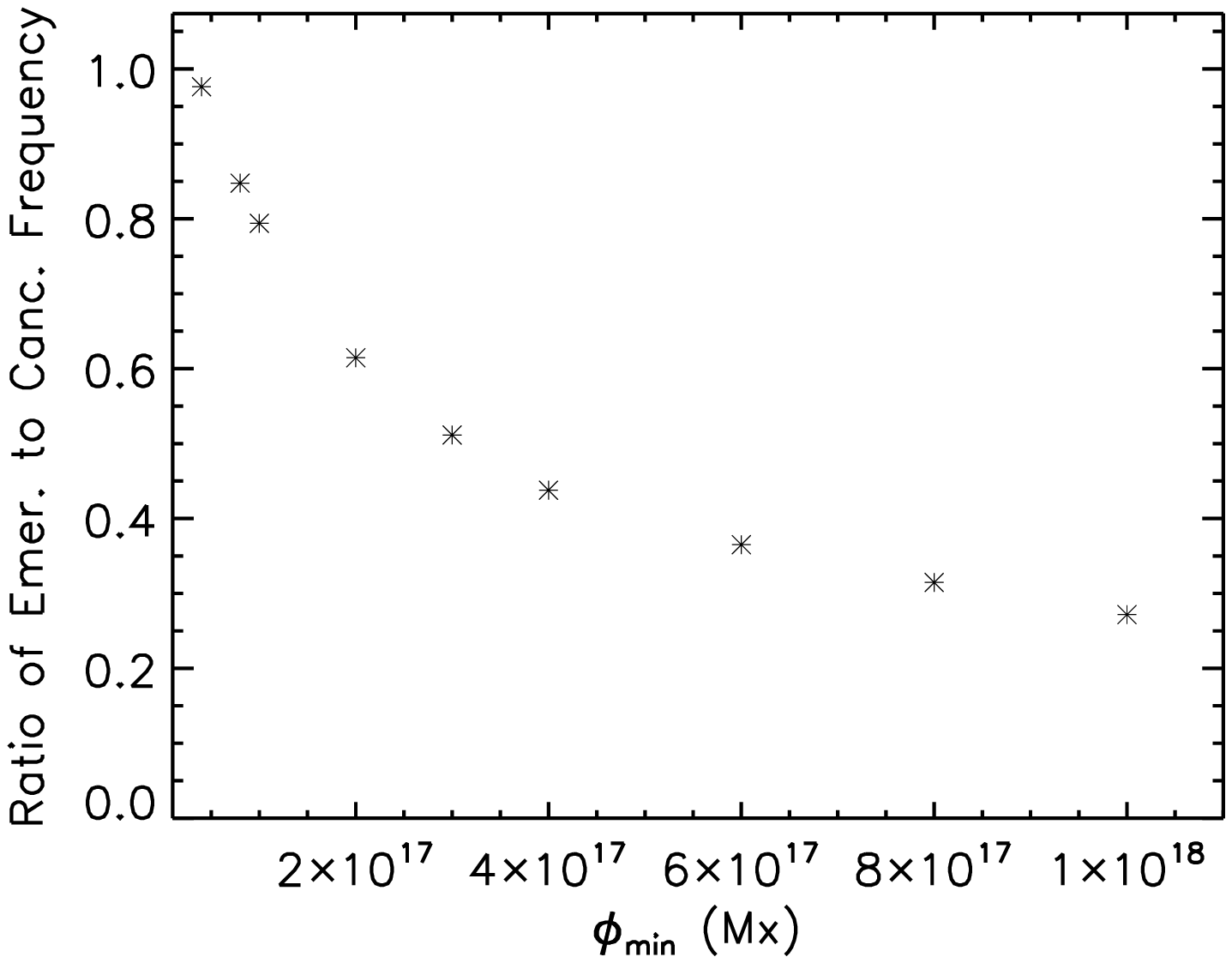}
              }
     \vspace{-0.38\textwidth}   
     \centerline{ \bf     
      \hspace{-0.02 \textwidth} \color{black}{(c)}
      \hspace{0.45\textwidth}  \color{black}{(d)}
         \hfill}
     \vspace{0.33\textwidth}    

\caption{(a) Emergence (solid line) and cancellation (dashed line) rates for the simulation with $\phi_{\textrm{\tiny min}}=8\times 10^{16}$ Mx. (b) Averaged emergence (black) and cancellation (red) rates for all nine simulations. The purple dashed line represents the emergence rates as calculated by Equation~(\ref{eqn:fluxemerge}). (c) Emergence (solid line) and cancellation (dashed line) frequencies for the simulation with $\phi_{\textrm{\tiny min}}=8\times 10^{16}$ Mx. (d) Ratio of average emergence to average cancellation frequencies for each simulation.}\label{fig:ratefreq}
   \end{figure}

Within this paper, the \emph{rates} of magnetic flux emergence and cancellation are defined in terms of Mx cm$^{-2}$ day$^{-1}$. The \emph{frequency} of magnetic flux emergence and cancellation is defined to be the number of occurrences in cm$^{-2}$ day$^{-1}$.

Figures~\ref{fig:ratefreq}(a) shows a plot of the variation of the rates of magnetic flux emergence (solid line) and magnetic flux cancellation (dashed line) during the $\phi_{\textrm{\small min}}=8\times 10^{16}$ Mx simulation, in Mx cm$^{-2}$ day$^{-1}$. In all cases we find that the emergence and cancellation rates quickly become approximately equal, confirming that the model has reached a steady state. When $\phi_{\textrm{\small min}}=8\times 10^{16}$ Mx or less, the plots are steady, as a large number of small magnetic elements emerge and subsequently cancel, helping to keep the emergence and cancellation rates steady. As $\phi_{\textrm{\small min}}$ increases these curves become less steady, but the time averaged emergence and cancellation rates for each simulation are all of similar values.

In simulations with higher $\phi_{\textrm{\small min}}$ one would expect the average emergence rate to be slightly higher than the average cancellation rate, since the total absolute flux in several of these cases increases slightly throughout the simulation. This can be seen in Table~\ref{tab:sim2}, in which the average emergence and cancellation rates for each simulation are listed.
Figure~\ref{fig:ratefreq}(b) shows a plot of the average emergence and cancellation rates for each simulation. The analytical solution for the emergence rate calculated from Equation~(\ref{eqn:fluxemerge}) is over-plotted in purple.

\begin{table}
\begin{center}
\begin{tabular}{|ccccc|}
\hline
                       & Flux               & Flux               & Emergence             & Cancellation \\
 $\phi_{\textrm{\small min}}$ & emergence          & cancellation       & frequency             & frequency \\
 ($\times 10^{16}$ Mx) & rate (Mx cm$^{-2}$ & rate (Mx cm$^{-2}$ & ($\times 10^{-18}$    & ($\times 10^{-18}$ \\
                       & day$^{-1}$)        & day$^{-1}$)        & cm$^{-2}$ day$^{-1}$) & cm$^{-2}$ day$^{-1}$) \\
\hline
4 & 107.37 & 107.30 & 1503.9 & 1540.6  \\
8 & 55.52 & 55.41 & 344.2 & 406.0  \\
10 & 45.69 & 45.59 & 222.1 & 279.7  \\
20 & 25.66 & 25.52 & 60.4 & 98.3  \\
30 & 18.56 & 18.32 & 29.1 & 56.9  \\
40 & 14.61 & 14.50 & 17.4 & 39.6  \\
60 & 10.34 & 10.27 & 8.5 & 23.2  \\
80 & 8.00 & 7.89 & 5.1 & 16.2  \\
100 & 6.53 & 6.45 & 3.4 & 12.5  \\
\hline
\end{tabular}
\caption{Time averaged emergence and cancellation rates and frequencies for each simulation.}\label{tab:sim2}
\end{center}
\end{table}

Figure~\ref{fig:ratefreq} (c) shows a plot of the variation of the emergence and cancellation frequencies throughout the simulation for $\phi_{\textrm{\small min}}=8\times 10^{16}$ Mx. The solid line represents emergence, the dashed line represents cancellation. It can be seen that cancellation events occur more often than emergence events even though the two processes have similar rates. This is because large flux elements may fragment, which leads to more elements that may cancel. As with emergence and cancellation rates, the curves are steadier when $\phi_{\textrm{\small min}}$ is lower.
The average emergence and cancellation frequencies for each simulation are given in Table~\ref{tab:sim2}. In agreement with \inlinecite{parnell2001}, the cancellation frequency is always greater than the emergence frequency.

The difference between the frequencies of emergence and cancellation increases with increasing $\phi_{\textrm{\small min}}$. This is particularly apparent for larger values of $\phi_{\textrm{\small min}}$, where emergence only produces a relatively small number of large elements. These elements may then break apart multiple times to produce large numbers of small continually cancelling and coalescing elements, resulting in a significantly higher frequency of cancellation than emergence. This occurs because emergence within each simulation is restricted to a narrow range of flux values, whereas cancellation may occur on any scale. Figure~\ref{fig:ratefreq}(d) shows the ratio of emergence to cancellation frequencies. Note that even though the cancellation frequency (in cm$^{-2}$ day$^{-1}$) may be much greater than the emergence frequency, the rates of emergence and cancellation (in Mx cm$^{-2}$ day$^{-1}$) are still roughly equal. The same quantity of flux emerges and disappears throughout the simulation, only the size of the magnetic elements involved in these processes changes.

\subsection{Distribution of Flux}\label{subsec:dist}

  \begin{figure}
   \centerline{\hspace*{0.015\textwidth}
               \includegraphics[width=0.515\textwidth,clip=]{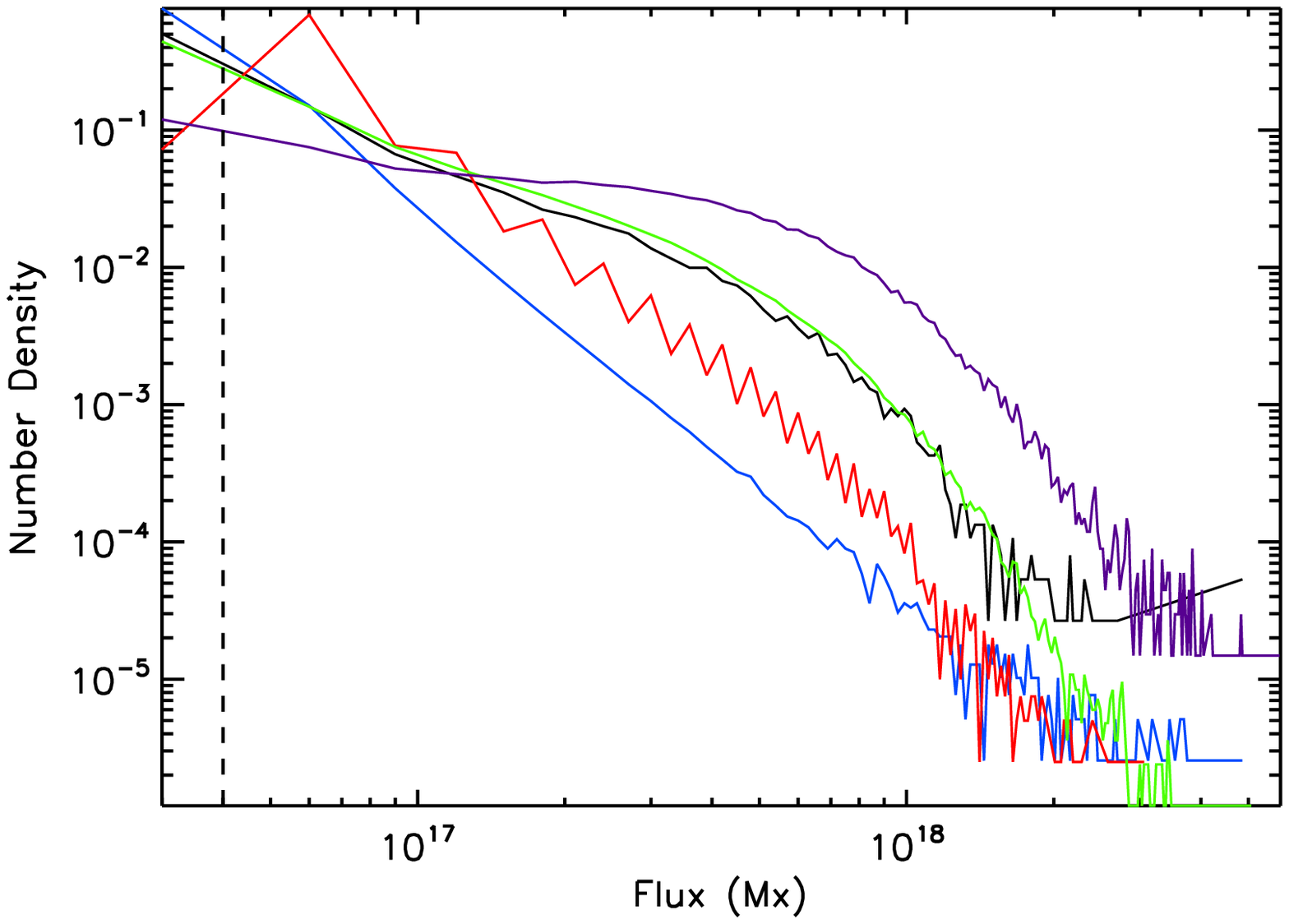}
               \hspace*{-0.03\textwidth}
               \includegraphics[width=0.515\textwidth,clip=]{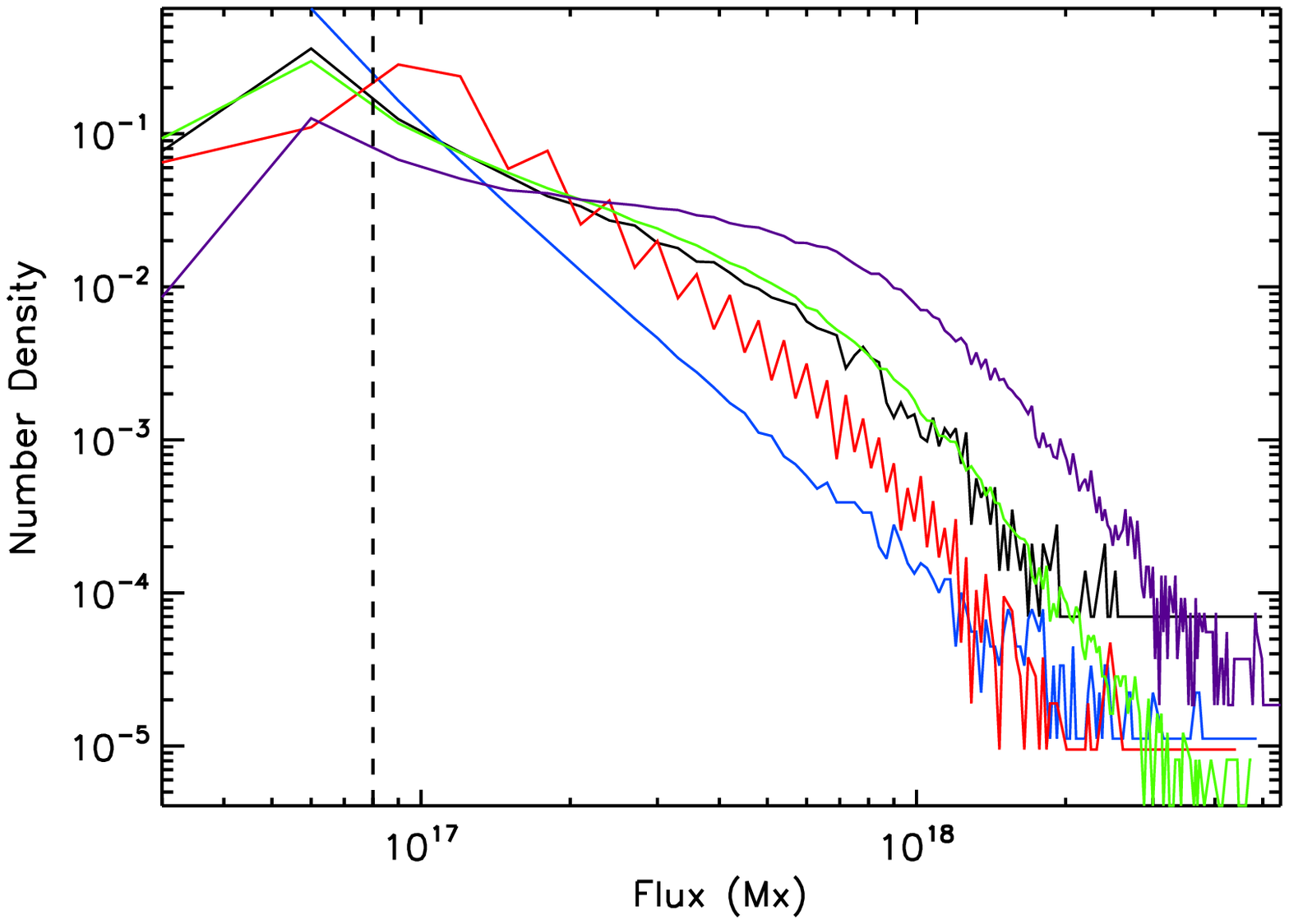}
              }
     \vspace{-0.38\textwidth}   
     \centerline{ \bf     
      \hspace{-0.02 \textwidth}  \color{black}{(a)}
      \hspace{0.45\textwidth}  \color{black}{(b)}
         \hfill}
     \vspace{0.33\textwidth}    

\caption{Number densities for (a) $\phi_{\textrm{\tiny min}}=4\times 10^{16}$ Mx and (b) $\phi_{\textrm{\tiny min}}=8\times 10^{16}$ Mx simulations: the black line represents the number density of flux elements versus absolute flux content. The blue line illustrates number of emergence events versus absolute flux emerging in an event. The red line illustrates number of cancellation events versus absolute flux lost in a cancellation event. The green line illustrates number of coalescence events versus absolute flux in a coalescence. The purple line illustrates number of fragmentation events versus absolute flux of the original element that fragmented. The black dashed line indicates the value of $\phi_{\textrm{\tiny min}}$.}\label{fig:dist1}
   \end{figure}

Figures~\ref{fig:dist1}(a) and (b) show number density plots for the number of magnetic elements as a function of their flux for the simulations with $\phi_{\textrm{\small min}}=4\times 10^{16}$ Mx and $\phi_{\textrm{\small min}}=8\times 10^{16}$ Mx. In each case the black line represents the number density of magnetic elements versus their absolute flux value. This curve is produced by analysing the distribution every 600 time steps, which is equivalent to every 10 h, and summing up all the flux distributions taken for a particular simulation. The length of time between samples was chosen so that the flux within the model will have recycled from one sample to the next. This prevents elements from being counted more than once.

The other four lines on each plot represent the occurrence of each of the four flux evolution processes described in Section~\ref{model}. In the simulations it is possible to keep track of every event that occurs, so the sample size is a lot larger than the flux distribution sample size, as any one element may undergo a number of processes. The blue line corresponds to the number density of emergences versus the total absolute flux of the emerging bipole, while the red line represents the number density of cancellations versus flux removed during the cancellation event. It can be seen that the number density of cancellation events is greater than the number density of emergence events until $\phi>10^{18}$ Mx. This fits with previous results that show many more cancellation events occur than emergence events, where the events tend to occur between smaller magnetic elements. Above a flux of $10^{18}$ Mx, fragmentation (purple) is so strong that it affects the elements before they can cancel (red). Since the rates of emergence and cancellation are roughly equal in each simulation, more large elements must emerge than cancel.

The green line represents the number density of coalescence events versus the absolute flux of the two coalescing elements. In all nine simulations (only two are illustrated here) this curve appears to follow the flux distribution curve (black) very closely, although the fit is not so good at higher $\phi$ (larger elements do not exist for very long within the simulation, so many are missed when sampling the flux distribution only once every 600 time steps). This supports the suggestion of \inlinecite{thornton2011}, that coalescence is the dominant process for small-scale elements. They come to this conclusion because this is one of the reasons why fewer small-scale elements are observed on the solar photosphere than are found to emerge. It can be seen from the plots in Figures~\ref{fig:dist1}(a) and (b) that the number density of emerging events (blue) is greater than the flux distribution (black) for small values of $\phi$, supporting the fact that coalescence is dominant for small-scale fields. In order for our model to be more conclusive on this matter however, it is necessary to allow the existence and emergence of even smaller flux elements than $\phi_{0}=10^{16}$ Mx. Since the green (coalescence) line follows the flux distribution (black) so closely, this implies that any and all magnetic elements may undergo coalescence, it is not flux dependent.

Fragmentation is defined in our simulations to be strongly flux dependent. The purple line represents the number density of fragmentation events versus the flux of the original magnetic element. The curve for fragmentation is lower than those of all of the other processes for small $\phi$. However, it can be seen that as an element's absolute flux increases the fragmentation process becomes dominant.

  \begin{figure}

   \centerline{\hspace*{0.015\textwidth}
               \includegraphics[width=0.515\textwidth,clip=]{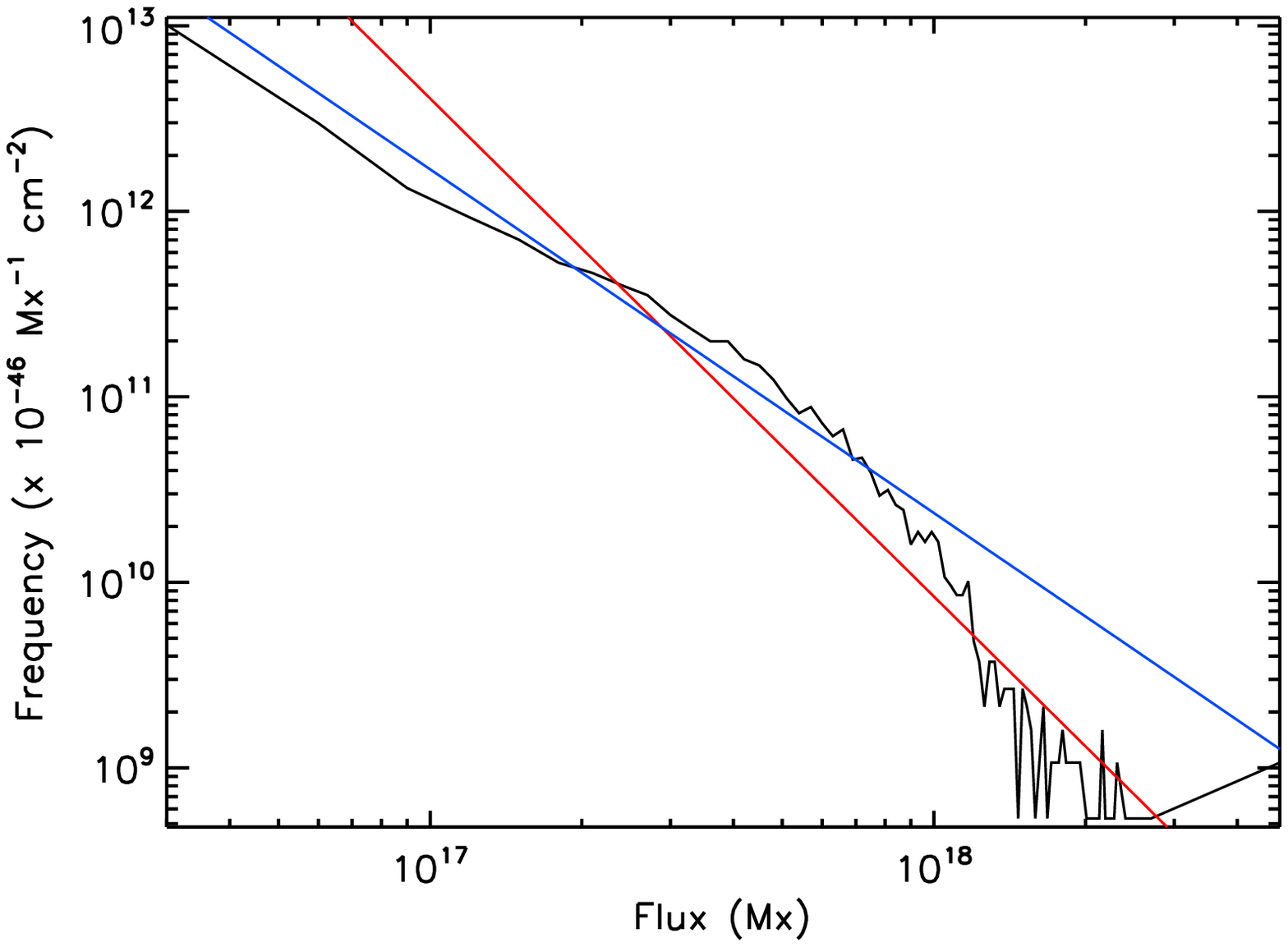}
               \hspace*{-0.03\textwidth}
               \includegraphics[width=0.515\textwidth,clip=]{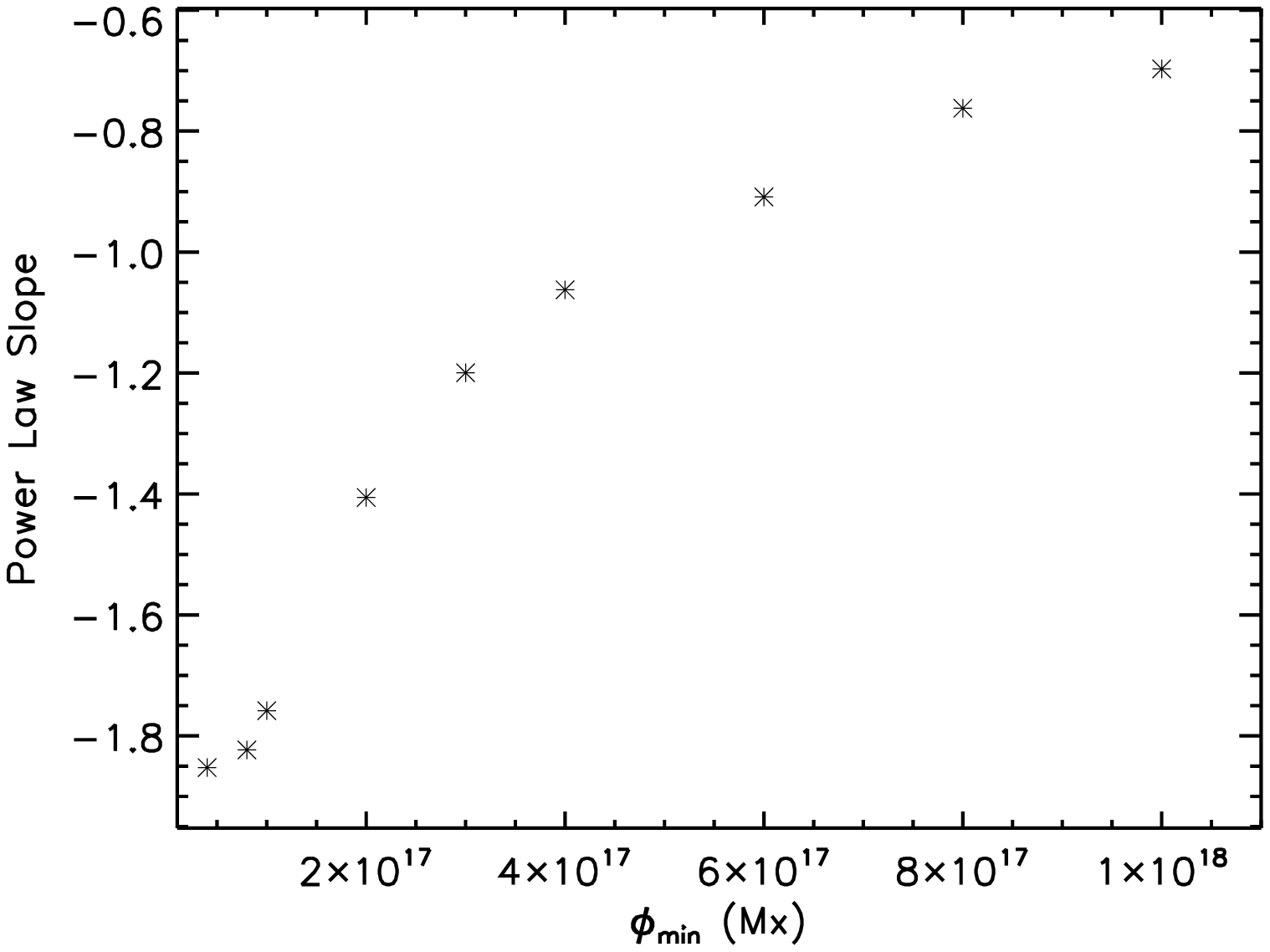}
              }
     \vspace{-0.38\textwidth}   
     \centerline{ \bf     
      \hspace{-0.02 \textwidth} \color{black}{(a)}
      \hspace{0.45\textwidth}  \color{black}{(b)}
         \hfill}
     \vspace{0.33\textwidth}    

\caption{(a) For the simulation with $\phi_{\textrm{\tiny min}}$ of $4\times 10^{16}$ Mx. The black line represents flux density: log plot of frequency of occurrence versus absolute flux of magnetic element. The red line represents the fitted line for the full range of flux values, slope=-2.68. The blue line represents the fitted line for the range $\phi=[10^{16},10^{18}]$ Mx, slope=-1.85. (d) Slopes for flux distribution in the range $\phi=[10^{16},10^{18}]$ Mx for each simulation.}\label{fig:dist2}
   \end{figure}

Figure~\ref{fig:dist2}(a) is a plot of the frequency of occurrence of flux elements versus their flux content, for the simulation with $\phi_{\textrm{\small min}}=4\times 10^{16}$ (black line). Units of $\times 10^{-46}$ Mx$^{-1}$ cm$^{-2}$ have been chosen for the y-axis in order to compare the plot with Figure 5 of \inlinecite{parnell2009}. Two straight lines have been fitted to the data. We note that our data do not span three orders of magnitude, as is technically required to compute a power law relation. However, power laws have been used for observational data so we have fitted one for comparison. The red line is fitted using all of the data, and has a slope of -2.68. The blue line is fitted for elements in the range $10^{16}-10^{18}$ Mx, therefore missing the low emergence upper flux range, as the sample size for our model is much better for smaller elements. Note that as we are considering an isolated region of the solar surface, the upper bound is solely dependent on emergence and coalescence. No elements may enter the domain due to the dispersal of magnetic flux from active regions. As a result, we find that our distribution of elements falls off faster than is observed at higher values of flux. Hence we restrict the power law index to those flux elements that we adequately model. The fit of the blue line is good in this range. It has a slope of -1.85, which is in agreement with the findings of \inlinecite{parnell2009}. They fit a line with a slope of -1.85 to data for flux elements observed by SOHO/MDI and \textit{Hinode}/SOT that spans the range $10^{16}-10^{23}$ Mx, therefore our model, which only models the lower range of this, fits this well. The method that they use to detect magnetic flux features also differs from the way that we define them. They use a `clumping' method, whereas our definition of a magnetic element is more suited to the `downhill' method of feature tracking (De Forest \etal ~\citeyear{deforest2007}; Lamb \etal ,~\citeyear{lamb2008,lamb2010}). The downhill method is good at picking out individual peaks in magnetic flux. Particularly for larger magnetic regions, several peaks may be detected by the downhill method and counted as separate magnetic features. However, the same region may be counted as a single feature by the clumping method (see Figures 1 and 7 of \inlinecite{deforest2007}). A flux distribution detected using downhill would result in a `tail' towards higher fluxes, as is seen in our Figure~\ref{fig:dist2}(a). A feature tracking study of our synthetic magnetograms would be of interest, using a clumping method such as is described by \inlinecite{thornton2011}.

The equation of \inlinecite{parnell2009} describing the frequency of occurrence of elements with an absolute flux of $\phi$ has the form
\begin{equation}
 N(\phi)=\frac{N_{\textrm{\small f}}}{\phi_0}\bigg(\frac{\phi}{\phi_0}\bigg)^{-1.85}\textrm{ Mx}^{-1}\textrm{ cm}^{-2},
\end{equation}
where $N_{\textrm{\small f}}=3.6\times 10^{-17}$ is the value obtained using a clumping method of feature tracking\footnote{C. E. Parnell (private communication).}. In our simulations, we obtain $N_{\textrm{\small f}}=1.2\times 10^{-16}$, which is 3.3 times larger. A possible reason for this is the difference between our definition of a flux element, and the way that features are defined by \inlinecite{parnell2009}. As can be seen in their Figure 4(a), the clumping method they use produces relatively large, irregularly shaped `flux massifs'. Since we define magnetic elements to be compact and circular, such a feature in our model would be composed of many elements. For this reason we require many more small magnetic elements to describe the flux distribution of the quiet Sun, hence the larger value of $N_{\textrm{\small f}}$ in the number density equation. It should be noted that while our definition of a magnetic element is different from that of \inlinecite{parnell2009} after it has emerged, this does not affect the emergence power law \cite{thornton2011}, as during the process of emergence our definitions of magnetic elements agree.

Our reasoning for the difference in values of $N_{\textrm{\small f}}$ is supported by some simple calculations. If $\alpha=1.85$, then the average flux density ($B_{\textrm{\small avg}}$) and average flux of a magnetic element ($\phi_{\textrm{\small avg}}$) are given by
\begin{equation}
B_{\textrm{\small avg}}=\int_{\phi_{\textrm{\tiny min}}}^{\phi_{\textrm{\tiny max}}}N(\phi)\phi d\phi = \frac{N_{\textrm{\small f}}\phi_0}{2-\alpha}\bigg[ \bigg(\frac{\phi}{\phi_0}\bigg)^{2-\alpha}\bigg]_{\phi_{\textrm{\tiny min}}}^{\phi_{\textrm{\tiny max}}}
\end{equation}
and
\begin{equation}
  \phi_{\textrm{\small avg}}=\frac{B_{\textrm{\small avg}}}{N_{\textrm{\small tot}}},
\end{equation}
where
\begin{displaymath}
 N_{\textrm{\small tot}}=\int_{\phi_{\textrm{\tiny min}}}^{\phi_{\textrm{\tiny max}}}N(\phi) d\phi = \frac{N_{\textrm{\small f}}}{1-\alpha}\bigg[ \bigg(\frac{\phi}{\phi_0}\bigg)^{1-\alpha}\bigg]_{\phi_{\textrm{\tiny min}}}^{\phi_{\textrm{\tiny max}}}.
\end{displaymath}
Taking our values of $N_{\textrm{\small f}}=1.2\times 10^{-16}$ cm$^{-2}$, $\phi_{\textrm{\small min}}=4\times 10^{16}$ Mx and $\phi_{\textrm{\small max}}=10^{18}$ Mx we find $\phi_{\textrm{\small avg}}=1.5\times 10^{17}$ Mx and $B_{\textrm{\small avg}}=6.11$ Mx cm$^{-2}$. Our $B_{\textrm{\small avg}}$ is consistent with the values given in Table~\ref{tab:sim}, and is realistic for the quiet Sun. If we now consider the $N_{\textrm{\small f}}=3.6\times 10^{-17}$ cm$^{-2}$ of \inlinecite{parnell2009}, but limit the range of flux values considered to be consistent with the quiet Sun ($4\times 10^{16}-10^{20}$ Mx), we get $\phi_{\textrm{\small avg}}=5.0\times 10^{17}$ Mx and $B_{\textrm{\small avg}}=6.6$ Mx cm$^{-2}$. Our absolute flux density is very similar to theirs, but our average flux element is much smaller. We also find that we have around three times more magnetic elements per unit area than \inlinecite{parnell2009} ($N_{\textrm{\small tot}}=4.06\times 10^{-17}$ cm$^{-2}$ compared with $N_{\textrm{\small tot}}=1.30\times 10^{-17}$ cm$^{-2}$). Therefore the difference in the parameters of the power law is due to our varying definition of `magnetic elements' to `magnetic features'.

The slope of the fitted line in the range $10^{16}-10^{18}$ Mx for each simulation is plotted versus $\phi_{\textrm{\small min}}$ in Figure~\ref{fig:dist2}(b). One can see that as $\phi_{\textrm{\small min}}$ increases the slope becomes less steep. This is because as only larger elements emerge, these large elements have more of an impact on the number density. Such a limited range for emerging bipoles is less realistic, so we would expect this to be less of a match with the results obtained through observations. When the lower bound of emerging flux is less than $10^{17}$ Mx, the power law index converges around -1.8.

\subsection{Lifetime of Magnetic Elements}\label{subsec:life}

  \begin{figure}
   \centerline{\hspace*{0.015\textwidth}
               \includegraphics[width=0.515\textwidth,clip=]{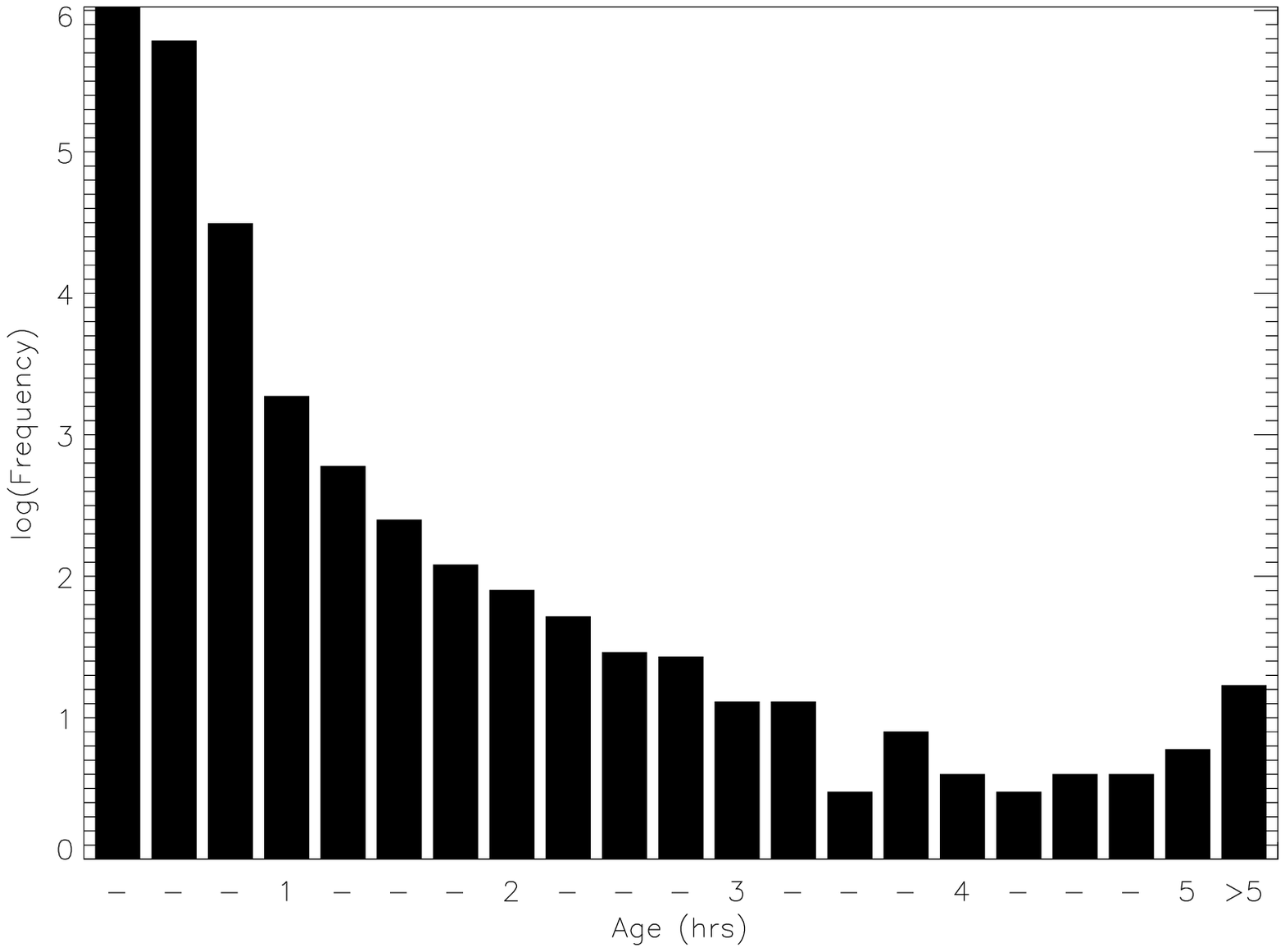}
               \hspace*{-0.03\textwidth}
               \includegraphics[width=0.515\textwidth,clip=]{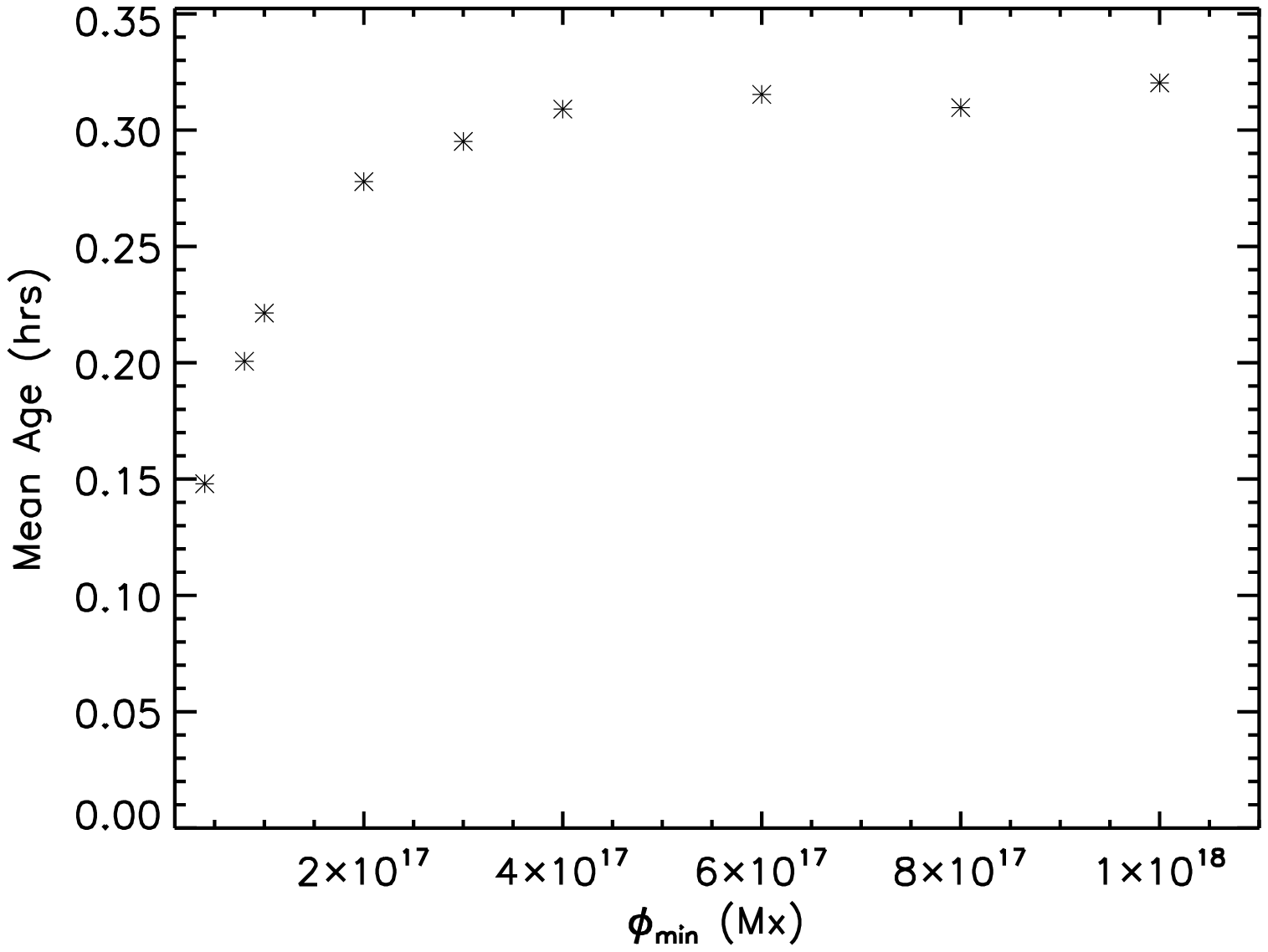}
              }
     \vspace{-0.38\textwidth}   
     \centerline{ \bf     
      \hspace{-0.02 \textwidth}  \color{black}{(a)}
      \hspace{0.45\textwidth}  \color{black}{(b)}
         \hfill}
     \vspace{0.33\textwidth}    
   \centerline{\hspace*{0.015\textwidth}
               \includegraphics[width=0.515\textwidth,clip=]{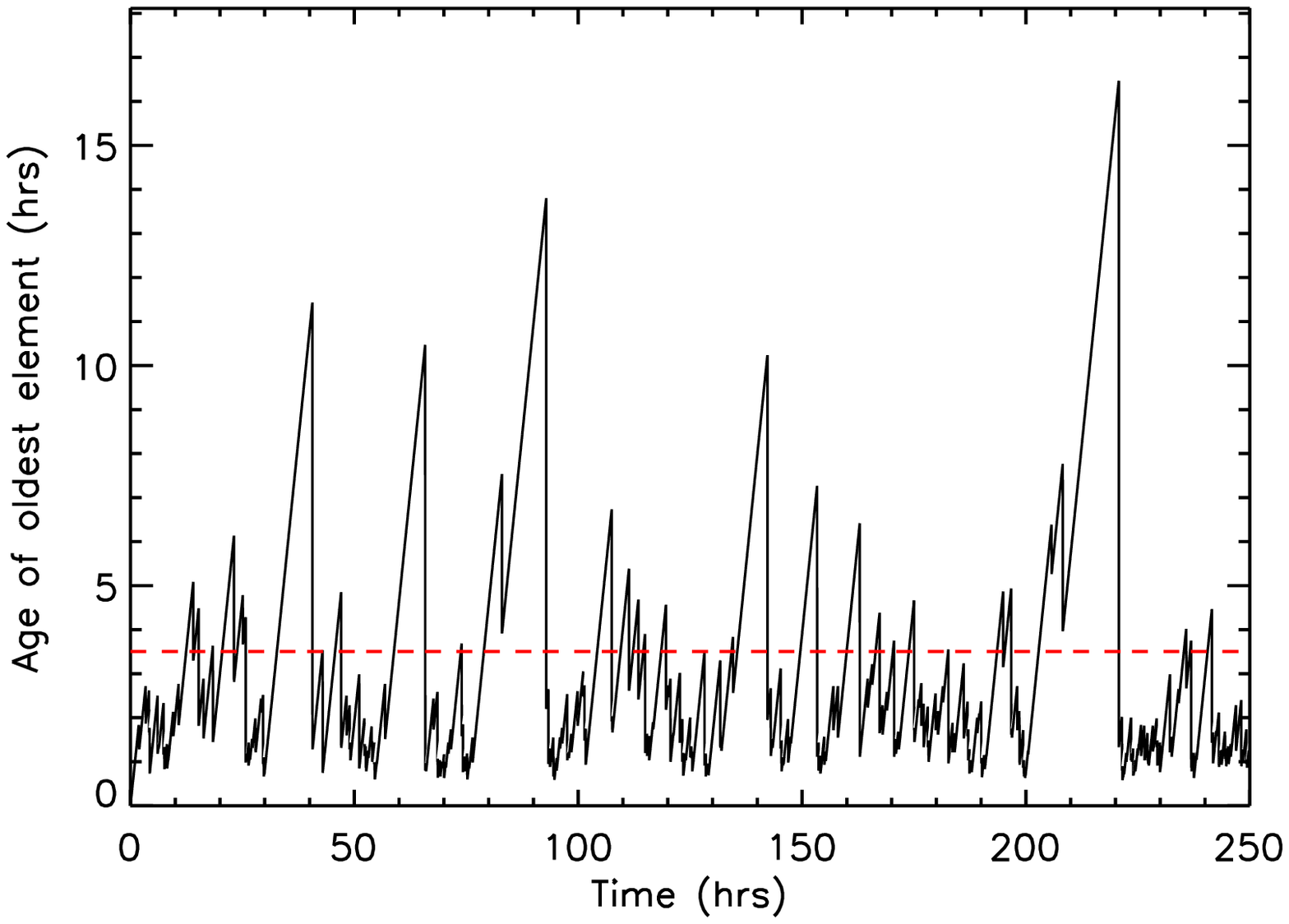}
               \hspace*{-0.03\textwidth}
               \includegraphics[width=0.515\textwidth,clip=]{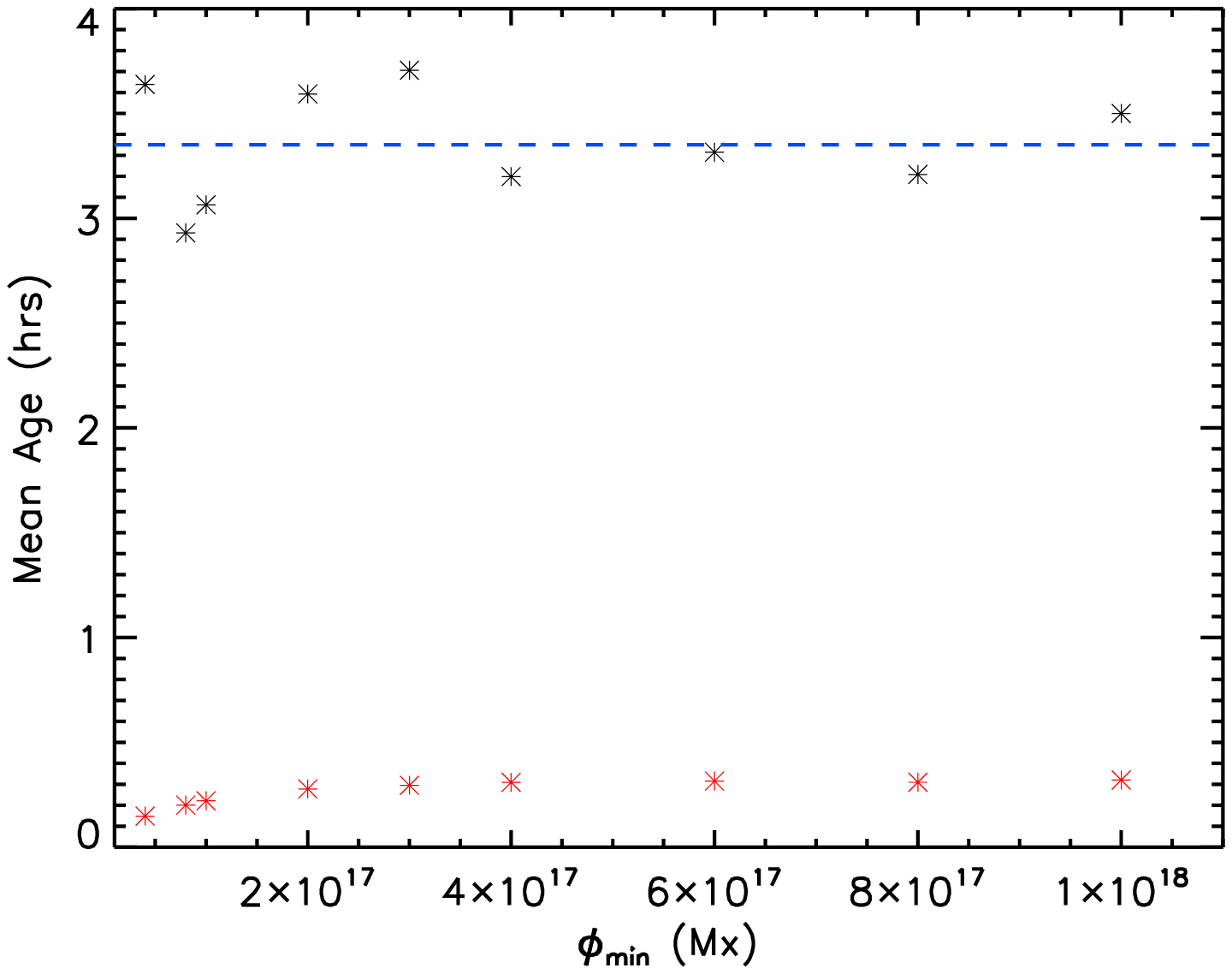}
              }
     \vspace{-0.38\textwidth}   
     \centerline{ \bf     
      \hspace{-0.02 \textwidth} \color{black}{(c)}
      \hspace{0.45\textwidth}  \color{black}{(d)}
         \hfill}
     \vspace{0.33\textwidth}    
              
\caption{(a) Frequency of occurrence of magnetic elements with a lifespan of $t$ hours, within the $\phi_{\textrm{\tiny min}}=4\times 10^{16}$ Mx simulation. (b) Mean lifespan of a magnetic element in hours for each of the simulations. (c) The age of the oldest existing magnetic element throughout the simulation, for $\phi_{\textrm{\tiny min}}=4\times 10^{16}$ Mx. The red dashed line represents the mean age of the oldest element averaged over the simulation. (d) Black stars: mean maximum age of an element averaged over each simulation. Blue line: mean maximum age of an element averaged over all simulations. Red stars: mean lifespan of a magnetic element within each simulation.}\label{fig:life}
  \end{figure}

The plots in Figure~\ref{fig:life} relate to the \emph{lifespan} of magnetic elements within each simulation. Within our model, an element is defined to `die' when its flux changes, either by fragmenting into two new elements or cancelling or coalescing with another element. An element begins its life when it newly emerges; has just split from another as a result of fragmentation; or is produced by two separate elements cancelling or coalescing together.

Our definition of an element's lifespan likely differs somewhat from that of an observer studying magnetogram data, as does our counting of magnetic elements. It is easy to count and keep track of the processes that magnetic elements undergo within our model because they are treated as individual point sources. However, when the synthetic magnetogram is created using the method described in Section~\ref{subsec:magn}, many of these elements overlap to produce fewer, larger magnetic elements. In addition, the resolution and cadence of real data are not always high enough to be able to detect the smallest and fastest evolving elements. This is another reason why further study of the photospheric model using the synthetic magnetogram series as input into the variety of feature tracking techniques that have been produced would be of interest.

Figure~\ref{fig:life}(a) shows the log frequency of occurrence of magnetic elements with a lifespan of $t$ h, for the simulation in which $\phi_{\textrm{\small min}}=4\times 10^{16}$ Mx. The mean lifespan of all elements is plotted for each simulation in Figure~\ref{fig:life}(b). It can be seen that most elements do not even `live' for one hour before fragmenting, cancelling or coalescing with another. The mean lifespan of an element is only around 9-20 mins. Observationally, it makes more sense that larger, more isolated magnetic elements would be long-lived. However, since fragmentation is so frequent within our model, this is not the case for our simulations. The longest lived elements most likely occur at the beginning of the simulation before the number of magnetic elements increases to the point where they interact frequently; or are isolated small elements that do not fragment because their flux is so low. Within the synthetic magnetograms, large, irregularly shaped features form where several individual elements lie close together but have not yet coalesced. These elements tend to appear in the network between supergranules. Examples of these can be seen in the six images shown in Figure~\ref{fig:mag}.

Figure~\ref{fig:life}(c) shows a plot of the age of the oldest magnetic element existing within the model versus time, within the $\phi_{\textrm{\small min}}=4\times 10^{16}$ Mx simulation. The dashed red line represents the mean maximum age averaged over the simulation. The large spikes represent occasional long-lived magnetic elements, but in general the maximum age of an element throughout the simulation remains at roughly 3-4 h. In every case the mean maximum age computed for the whole simulation is around this value, as can be seen in Figure~\ref{fig:life}(d). This is similar to the lifetime of ephemeral regions, determined by \inlinecite{harvey1973} to be around 4.4 h. \inlinecite{zhou2010} find the lifespan of internetwork features to be between 1 and 20 mins, with a mean of 2.9$\pm$2.0 mins. The mean lifespan for magnetic elements within our model is 9-20 mins.

The \emph{photospheric recycle time} is calculated by dividing the mean field by the emergence rate. For our two most realistic simulations, $\phi_{\textrm{\small min}}=4\times 10^{16}$ Mx and $\phi_{\textrm{\small min}}=8\times 10^{16}$ Mx, we find the recycle time to be 1.48 h and 1.75 h respectively. This is in excellent agreement with \inlinecite{hagenaar2008}'s recycle time of 1-2 h.

The next section considers one of many possibilities for future studies using this theoretical model for the magnetic carpet, in this example, emergence is switched off midway through the simulation.

\subsection{Switching Off Emergence}\label{subsec:noem}

  \begin{figure}
   \centerline{\hspace*{0.015\textwidth}
               \includegraphics[width=0.515\textwidth,clip=]{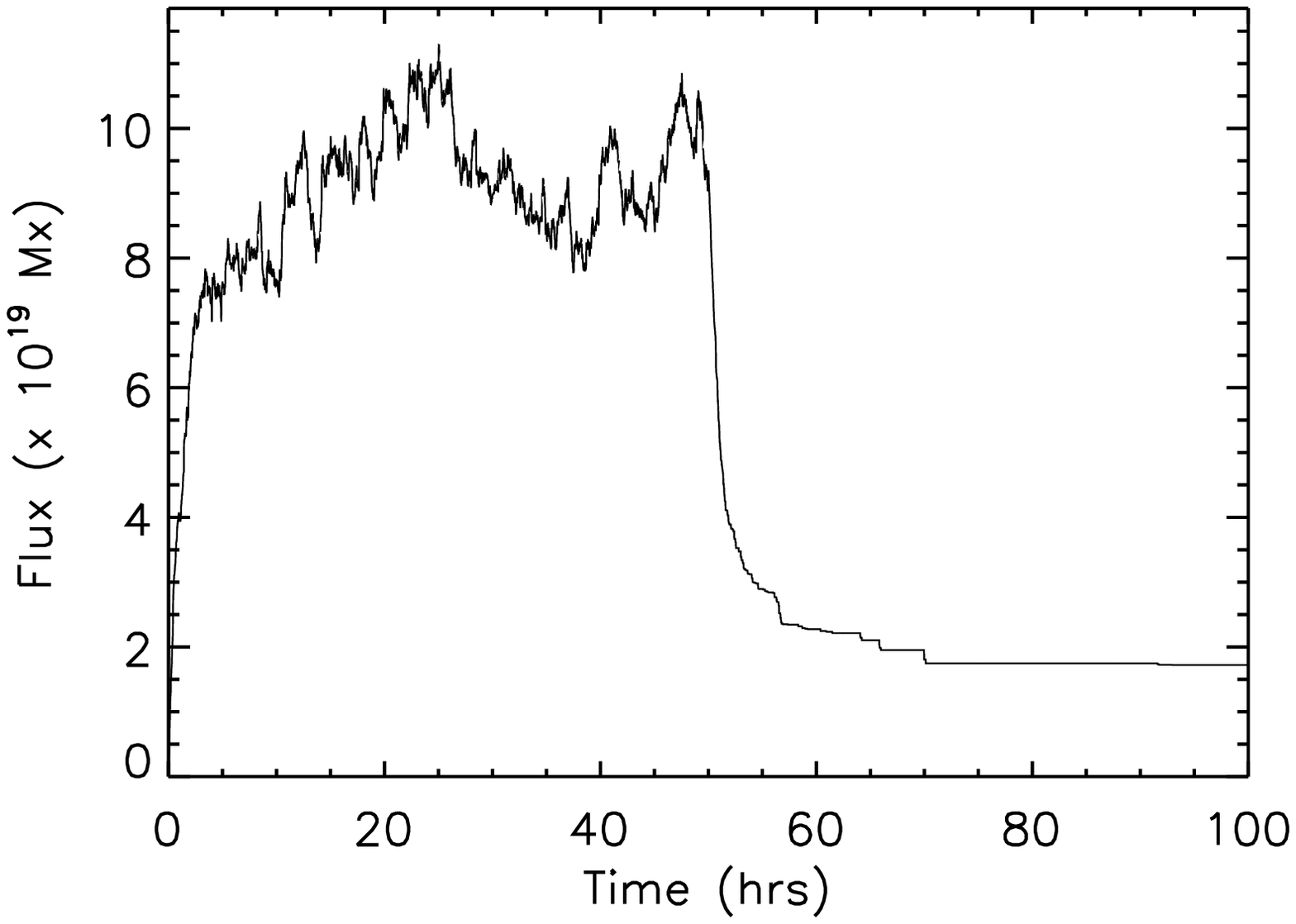}
               \hspace*{-0.03\textwidth}
               \includegraphics[width=0.515\textwidth,clip=]{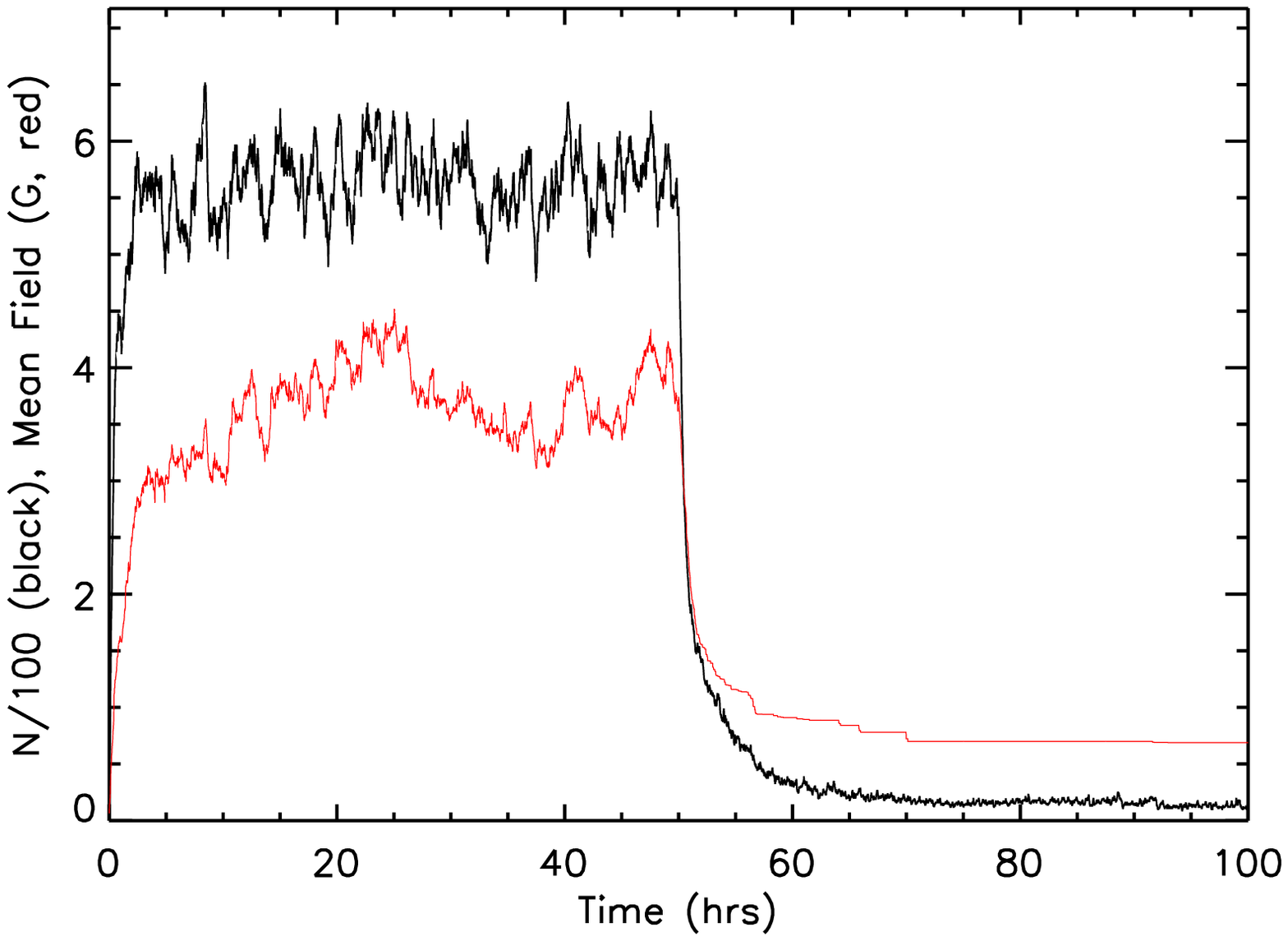}
              }
     \vspace{-0.35\textwidth}   
     \centerline{ \bf     
      \hspace{-0.02 \textwidth} \color{black}{(a)}
      \hspace{0.44\textwidth}  \color{black}{(b)}
         \hfill}
     \vspace{0.33\textwidth}    
   \centerline{\hspace*{0.015\textwidth}
               \includegraphics[width=0.515\textwidth,clip=]{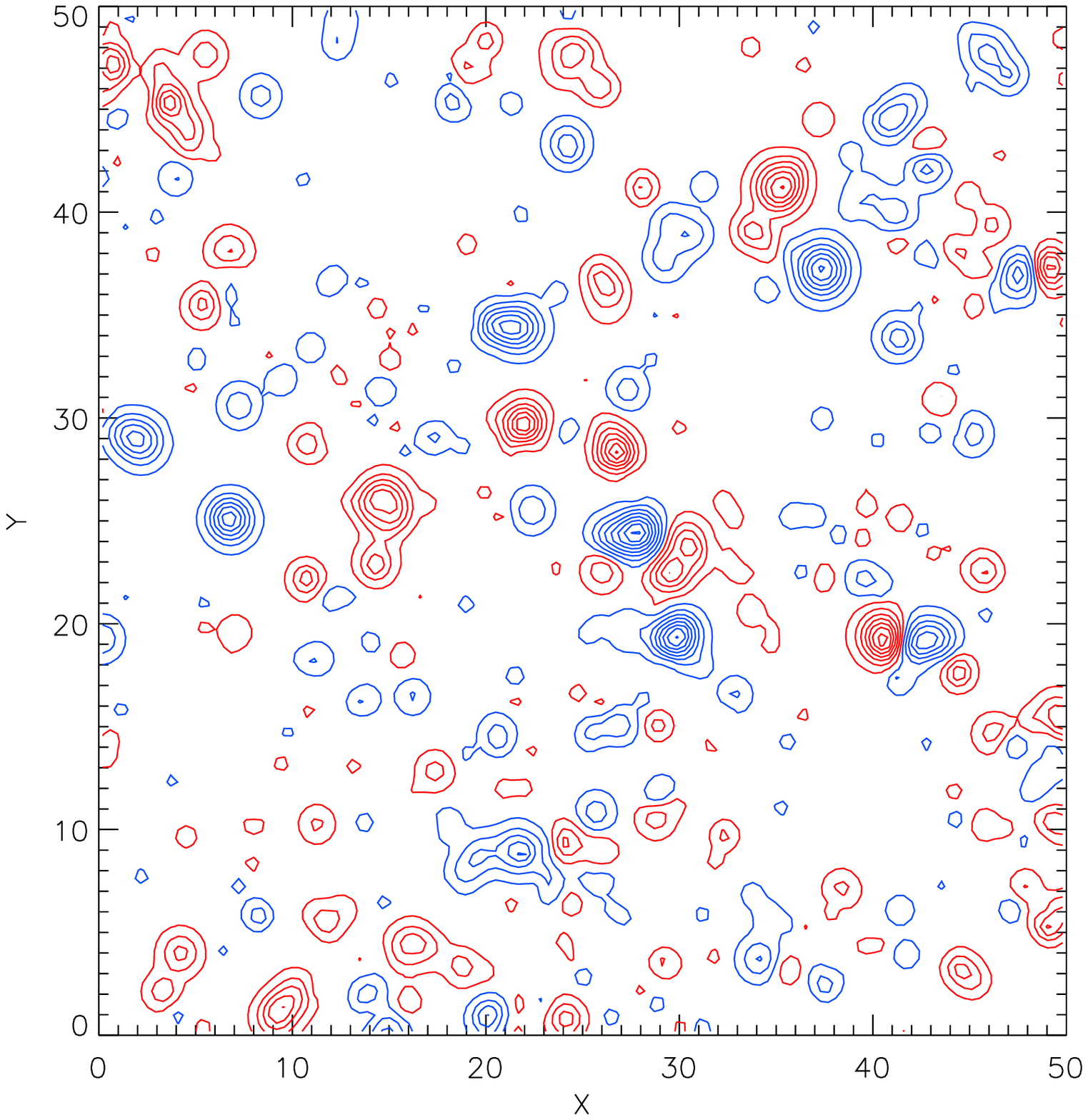}
               \hspace*{-0.03\textwidth}
               \includegraphics[width=0.515\textwidth,clip=]{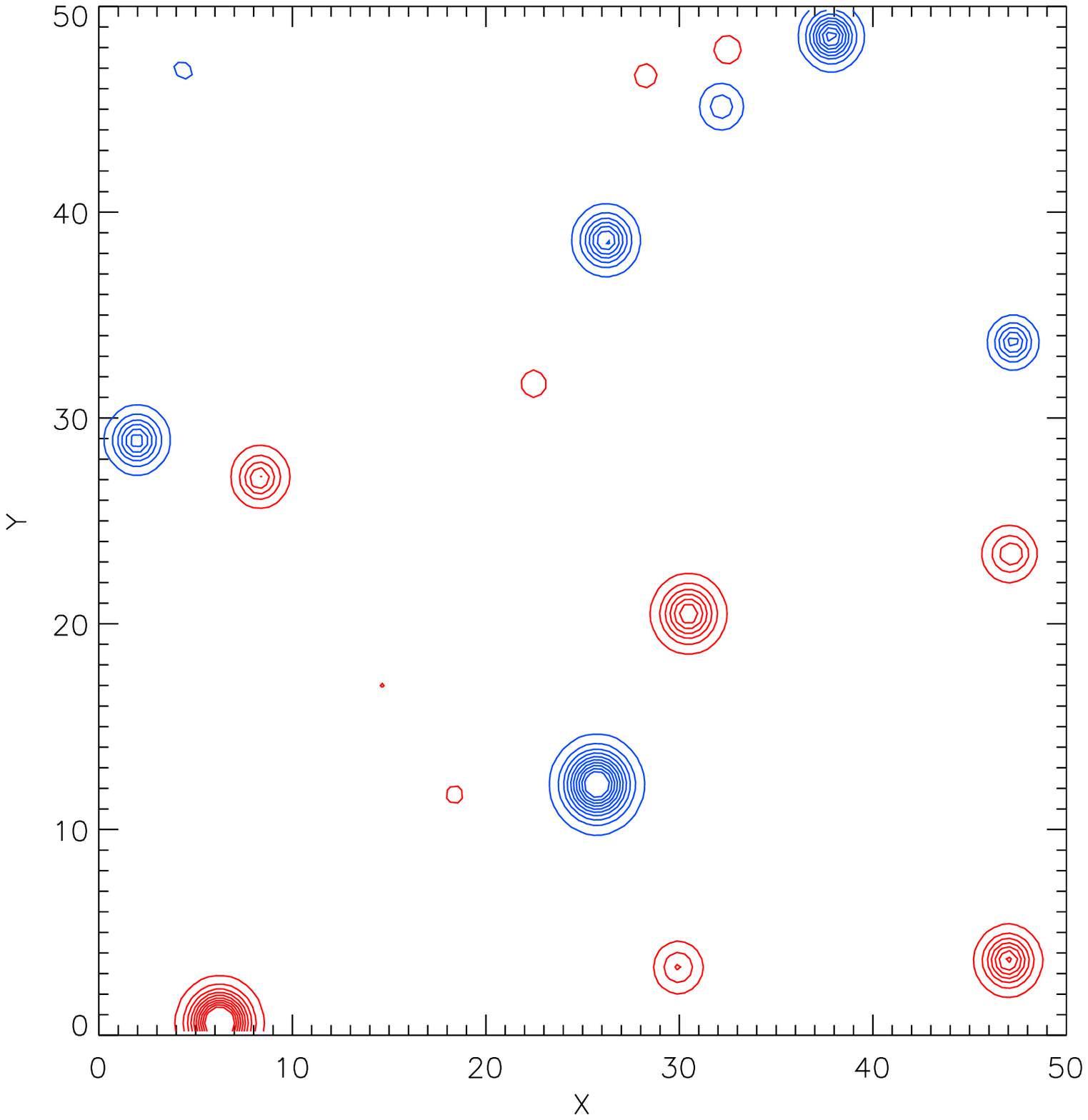}
              }

     \vspace{-0.48\textwidth}   
     \centerline{ \bf     
      \hspace{-0.02 \textwidth}  \color{black}{(c)}
      \hspace{0.44\textwidth}  \color{black}{(d)}
         \hfill}
     \vspace{0.43\textwidth}    

\caption{A simulation with emergence of bipoles in the range $8\times 10^{16}-10^{19}$ Mx. Emergence is switched off after $50$ h (3\,000 time steps), and the simulation is allowed to run for a further $50$ h. (a) Total absolute flux within the simulation versus time. (b) Mean field and number of magnetic elements ($N/100$) versus time, in red and black respectively. (c) and (d) show the synthetic magnetograms produced at $t=50$ h and $t=60$ h.}\label{fig:em}
  \end{figure}

Clearly, many other studies may be undertaken using our theoretical model simply by varying different input parameters to determine their effect upon photospheric evolution. One interesting test is to `switch off' emergence at some stage in the simulation to observe how rapidly flux disappears. An example of such a simulation was run with identical parameters to those described in Section~\ref{results}. The flux emergence range for newly appearing bipoles is $8\times 10^{16}-1\times 10^{19}$ Mx. Emergence is switched off at $t=50$ h, and the simulation is allowed to run for a further $50$ h. Results from this experiment are shown in Figure~\ref{fig:em}. Plot (a) shows the total absolute flux and (b) shows both the mean field and the number of elements ($N/100$) throughout the simulation. It can be seen that all three of these quantities rapidly decrease as soon as emergence is switched off after $50$ h, levelling off and becoming steady at roughly $t=60$ h. The total number of magnetic elements becomes very small as all remaining flux within the computational box rapidly cancels and coalesces together at the supergranule boundaries. The synthetic magnetogram images (c) and (d) are taken at $t=50$ and $t=60$ h. In image (d) one can see the few remaining magnetic elements. They are spaced far from one another where the supergranular velocities are small at the boundaries, so they do not encounter one another before the simulation ends. If the supergranular flow profile was also allowed to evolve in time, this would likely aid the process by increasing and randomising further the motion of the remaining magnetic elements, eventually removing them completely from the simulation.

\section{Discussion, Conclusions and Future Work} \label{conclusion}

The aim of this paper is to construct a realistic model for the photospheric evolution of the solar magnetic carpet, which we intend to use as a lower boundary condition of 3D non-potential modelling. We have built into this model the processes of flux emergence, cancellation, coalescence and fragmentation, as well as a steady supergranular flow profile that influences the motion of magnetic elements. Many parameters for the model are taken from studies of observational solar data, such as the probability distribution for newly emerging bipoles \cite{thornton2011} and the peak radial velocity of a supergranule \cite{simon1964,paniveni2004}.

A series of nine simulations of length 250 h were run, keeping all parameters fixed apart from the minimum value of total absolute flux, $\phi_{\textrm{\small min}}$, that a newly emerging bipole may take. The maximum value for newly emerging bipoles was fixed at $\phi_{\textrm{\small max}}=10^{19}$ Mx. The lower the value of $\phi_{\textrm{\small min}}$ is taken to be, the larger the range of emerging flux elements is, and hence the more flux is emerged into the system. The upper bound produces a cap on the size of large elements, and we assume that we are considering a quiet area of the Sun, with no input from active regions. A larger flux emergence range is more realistic so as one would expect, the lower the value of $\phi_{\textrm{\small min}}$, the closer the simulation results agree with solar observations.

Particularly for $\phi_{\textrm{\small min}}=4\times 10^{16}$ Mx and $\phi_{\textrm{\small min}}=8\times 10^{16}$ Mx, the total absolute flux within the simulation quickly reaches a steady state in which the rates of emergence and cancellation are roughly equal. These simulations also result in a mean magnetic field that is within the range determined from observations. For the less realistic simulations with the highest $\phi_{\textrm{\small min}}$ ($\geq3\times 10^{17}$ Mx) the total absolute flux takes longer to reach a steady state illustrated by the fact that the mean emergence rate is slightly higher than the mean cancellation rate. For low $\phi_{\textrm{\small min}}$, the mean field is 4-6 G, which fits observational data.

The number of magnetic elements within each simulation very quickly reaches a steady state in all cases however. This occurs shortly after the first magnetic elements reach the boundaries of the supergranules and begin to interact with one another.

Although the cancellation and emergence rates, defined in Mx cm$^{-2}$ day$^{-1}$, become roughly equal for all simulations, it is also the case that the cancellation frequency, in cm$^{-2}$ day$^{-1}$, is always greater than the emergence frequency. This is in rough agreement with the theoretical magnetic carpet model of \inlinecite{parnell2001}, who attribute the difference to large numbers of cancellations between small magnetic elements arising through fragmentation. They suggest that the energy release from so many small cancellation events could significantly contribute to the energy required to heat the solar corona. In future in our 3D simulations, we will consider the energy build up at such locations.

The flux distribution resulting from the simulations that include a larger range of flux emergence are in agreement with the power-law distribution of \inlinecite{parnell2009}. The slope of their fitted line is -1.85, while the slopes produced by our simulations with $\phi_{\textrm{\small min}}=4\times 10^{16}$ Mx and $\phi_{\textrm{\small min}}=8\times 10^{16}$ Mx are -1.85 and -1.82 respectively. Our definition of magnetic elements means that our flux distribution would fit better with a downhill method of feature tracking. A feature tracking study of our synthetic magnetograms using a clumping method would be useful to check the power law, in addition to other values such as lifetimes and physical extent of magnetic elements.

Our model produces a highly dynamic small-scale photosphere as desired, with the mean lifespan of a magnetic element in any of the simulations being just 9-20 mins. We also find a photospheric recycle time of just 1.48 h, this is in good agreement with \inlinecite{hagenaar2008}'s recycle time time of 1-2 h.

As suggested in Section~\ref{subsec:noem}, an evolving supergranular flow profile is one improvement that could be made to the model. \inlinecite{simon1964} determine the average lifespan of a supergranule to be 20 h, whereas \inlinecite{wang1989} find their lifetime to be $\geq$ 50 h. \inlinecite{simon2001} assign lifetimes of between 18 and 42 h to the supergranules within their model. In any case, these studies show that a magnetic carpet model that runs for longer than around a day should consider a time evolving flow pattern. The flow profile evolution could either be built into the model or taken from observational data using a method such as described by \inlinecite{potts2004}. They track solar photospheric flows by using a ball tracking technique.

The fragmentation process is limitation of our model. Rather than allowing fragmentation to arise naturally as a consequence of photospheric flows, we currently impose it. This is another feature of the model that could be improved in future.

The next stage in our process of constructing a non-potential coronal model for the solar magnetic carpet is the theoretical study of small-scale basic interactions between magnetic flux elements, using the technique of \inlinecite{vanballegooijen2000}. This will produce a continuously evolving non-linear force-free magnetic field in response to photospheric motions \cite{meyer2011}. This method has recently been applied to the evolving structure of magnetic filaments \cite{mackay2009}, and a decaying active region observed by SOHO/MDI \cite{mackay2011}. It is useful to understand these simple interactions before moving onto more complex cases. We will later insert the synthetic magnetograms produced by this 2D model into the 3D coronal model as lower boundary data.

We will study many aspects of the resultant coronal magnetic field, such as locations of electric currents, free magnetic energy, and coronal null points, as these are associated with the problem of coronal heating. It would also be of interest to conduct this study side by side with an identical study of SDO magnetogram data, for which the coronal magnetic field could also be modelled. Since we know exactly which processes are occurring where within the simulated synthetic magnetograms illustrated here, we would be able to relate these with events occurring in the real magnetogram data, and study the simulated coronal field in both cases.

We conclude that we have successfully produced a realistic model for the photospheric evolution of the solar magnetic carpet that reproduces many observational properties. This will now be used to produce a 3D model for the small-scale coronal magnetic field under controlled circumstances.

\begin{acks}
KAM would like to thank ISSI (Bern) for their support of the team ``Solar small-scale transient phenomena
and their role in coronal heating'', and teammates for their encouragement and support of her as a young researcher. She would also like to thank the Harvard-Smithsonian Center for Astrophysics (Boston, MA) for their warm hospitality and helpful discussions during her six week stay in summer 2010. KAM and DHM gratefully acknowledge the support of the STFC. DHM would like to thank the Royal Society for their support through the Research Grant Scheme. DHM and CEP acknowledge the support of the EU-FP7 project SWIFF. Finally, we thank the anonymous referee, who's helpful comments and suggestions have improved the paper.
\end{acks}

\end{article} 

\end{document}